\documentclass[pra,twocolumn,nofootinbib,showpacs]{revtex4}
\usepackage{amsmath}
\usepackage{graphicx}

\def\>{\rangle}
\def\<{\langle}

\def\be{\begin{equation}}
\def\ee{\end{equation}}

\begin{document}

\author{Robert W. Spekkens}
\address{Perimeter Institute for Theoretical Physics,\\
31 Caroline St.~North, Waterloo, Canada N2L 2Y5}
\title{In defense of the epistemic view of quantum states: a toy theory}
\date{\today}

\begin{abstract}
We present a toy theory that is based on a simple principle: the
number of questions about the physical state of a system that are
answered must always be equal to the number that are unanswered in
a state of maximal knowledge. A wide variety of quantum phenomena
are found to have analogues within this toy theory. Such phenomena
include: the noncommutativity of measurements, interference, the
multiplicity of convex decompositions of a mixed state, the
impossibility of discriminating nonorthogonal states, the
impossibility of a universal state inverter, the distinction
between bi-partite and tri-partite entanglement, the monogamy of
pure entanglement, no cloning, no broadcasting, remote steering,
teleportation, dense coding, mutually unbiased bases, and many
others. The diversity and quality of these analogies is taken as
evidence for the view that quantum states are states of incomplete
knowledge rather than states of reality. A consideration of the
phenomena that the toy theory fails to reproduce, notably,
violations of Bell inequalities and the existence of a
Kochen-Specker theorem, provides clues for how to proceed with
this research program.
\end{abstract}

\maketitle
\tableofcontents

\section{Introduction}

\label{intro}

In this article, we introduce a simple toy theory based on a principle that
restricts the amount of knowledge an observer can have about reality.
Although not \emph{equivalent} to quantum theory nor even competitive as an
explanation of empirical phenomena, this theory bears an uncanny resemblance
to the latter insofar as it reproduces in detail a large number of phenomena
that are typically taken to be characteristically quantum. This, and the
fact that the object analogous to the quantum state in the toy theory is a
state of incomplete knowledge, are the grounds upon which we argue for our
thesis, that quantum states are also states of incomplete knowledge.

We begin by clarifying the dichotomy between states of reality and states of
knowledge. To be able to refer to this distinction conveniently, we
introduce the qualifiers \emph{ontic}, (from the Greek \emph{ontos, }meaning
``to be'') and \emph{epistemic} (from the Greek \emph{epist\={e}m\={e}, }%
meaning ``knowledge''). An \emph{ontic state} is a state of reality, whereas
an \emph{epistemic state} is a state of knowledge. To understand the content
of the distinction, it is useful to study how it arises in an
uncontroversial context: that of classical physics.

The first notion of state that a student typically encounters in their study
of classical physics is the one associated with a point in phase space. This
state provides a complete specification of all the properties of the system
(in particle mechanics, such a state is sometimes called a ``Newtonian
state''). It is an ontic state. On the other hand, when a student learns
classical \emph{statistical} mechanics, a new kind of state is introduced,
corresponding to a probability distribution over the phase space (sometimes
called a ``Liouville state''). This is an epistemic state. The critical
difference between a point in phase space and a probability distribution
over phase space is not that the latter is a function. An electromagnetic
field configuration is a function over three-dimensional space, but is
nonetheless an ontic state. What is critical about a probability
distribution is that the relative height of the function at two different
points is not a property of the system (unlike the relative height of an
electromagnetic field at two points in space). Rather this relative height
represents the relative likelihood that some agent assigns to the two ontic
states associated with those points of the phase space. The distribution
describes only what this agent knows about the system.

There is one case wherein the distinction between an ontic state and an
epistemic state breaks down, and that is for epistemic states describing
complete knowledge, since the latter also contain a complete specification
of a system's properties. For example, states of complete knowledge in a
classical theory are represented by Dirac-delta functions on phase space,
and these are associated one-to-one with the points of phase space. The
epistemic states with which we shall be interested in this paper -- the ones
with which we hope to associate quantum states -- are those describing \emph{%
incomplete} knowledge.

The standard view among physicists and philosophers of physics is that pure
quantum states are ontic states. Only mixed quantum states are taken to be
epistemic states, specifically, states of incomplete knowledge about which
pure quantum state is really occurrent. In a variant of this view, even the
mixed quantum states are interpreted as ontic (this approach is motivated by
the fact that a mixed state may be expressed as a convex sum of pure states
in many different ways). We shall describe proponents of both of these
viewpoints as proponents of the \emph{ontic view} of quantum states. In
contrast, the thesis we wish to defend is that \emph{all }quantum states,
mixed and pure, are states of incomplete knowledge. This viewpoint will be
referred to as the \emph{epistemic view} of quantum states.

The ontic view of quantum states has a long history in the interpretation of
quantum mechanics. Schr\"{o}dinger initially interpreted the quantum state
as a physical wave, and never wholly abandoned this view. In the classic
textbooks of Dirac \cite{Dirac} and of von Neumann \cite{vonNeumann}, the
quantum state is taken to provide a complete specification of the properties
of a system. This is also true of both collapse theories \cite%
{GRW,Pearle,review} and Everett-type interpretations \cite{Everett,Barrett}.
Even within the popular hidden variable theories, such as the deBroglie-Bohm
theory \cite{BohmHiley,Holland,Valentini} and the modal interpretation \cite%
{KDH,DieksVermaas,Bacciagaluppi}, although the quantum state has an
epistemic role to play in specifying the probability distribution over
hidden variables, it is fundamentally an ontic state insofar as it acts as a
guiding wave, causally influencing the dynamics of the hidden variables. The
tension between the epistemic and ontic roles of the quantum state in these
interpretations has understandably troubled many authors, and although
efforts have been made to reduce the tension \cite{Valentini}, these have
tended to assign \emph{less} rather than more epistemic significance to the
quantum state.

The epistemic view, although less common than the ontic view, also
has a long tradition. As we shall see in detail further on,
Einstein's argument for the incompleteness of quantum mechanics
(which is most clear in his correspondence with Schr\"{o}dinger
\cite{Einsteinletters} but was made famous in the EPR paper
\cite{EPR}) is an argument for an epistemic view of quantum
states. The work of Ballentine on the statistical interpretation
\cite{BallentineRMP,Ballentine} can be interpreted as a defense of
the epistemic view, as can that of Emerson \cite{Emerson}. Peierls
was also an early advocate of this interpretation of the quantum
state \cite{Peierls}. It is only recently, with the advent of
quantum information theory, that the epistemic view has become
more widespread, with the most convincing and eloquent advocate of
the approach being Fuchs \cite{Fuchs}. Indeed, our work owes much
of its inspiration to Fuchs's research program, in particular, the
idea of deriving quantum phenomena from a principle that maximal
information is incomplete and cannot be completed
\cite{Fuchscloning,Fuchssamizdat}.

Despite the fact that the epistemic view appears to be on the rise, our
impression is that many would-be supporters have failed to completely
abandon their ontic preconceptions, perhaps due to the ubiquity of ontic
language in the literature and perhaps due to a vague feeling that the
epistemic path is one that has been shown to be inconsistent. We hope
through this article to correct some of these misconceptions and to increase
the respectability of this viewpoint.

We shall argue for the superiority of the epistemic view over the ontic view
by demonstrating how a great number of quantum phenomena that are mysterious
from the ontic viewpoint, appear natural from the epistemic viewpoint. These
phenomena include interference, noncommutativity, entanglement, no cloning,
teleportation, and many others. Note that the distinction we are emphasizing
is whether the phenomena can be understood \emph{conceptually}, not whether
they can be understood as mathematical consequences of the formalism, since
the latter type of understanding is possible regardless of one's
interpretation of the formalism. The greater the number of phenomena that
appear mysterious from an ontic perspective but natural from an epistemic
perspective, the more convincing the latter viewpoint becomes. For this
reason, the article devotes much space to elaborating on such phenomena.

Of course, a proponent of the ontic view might argue that the phenomena in
question are not mysterious \emph{if} one abandons certain preconceived
notions about physical reality. The challenge we offer to such a person is
to present a few simple \emph{physical} principles by the light of which all
of these phenomena become conceptually intuitive (and not merely
mathematical consequences of the formalism) within a framework wherein the
quantum state is an ontic state. Our impression is that this challenge
cannot be met. By contrast, a single information-theoretic principle, which
imposes a constraint on the amount of knowledge one can have about any
system, is sufficient to derive all of these phenomena in the context of a
simple toy theory, as we shall demonstrate.

A few words are in order about the motivation for such a principle. In
Liouville mechanics, states of incomplete knowledge exhibit phenomena
analogous to those exhibited by pure quantum states. Among these are the
existence of a no-cloning theorem for such states~\cite%
{Fuchscloning,classicalnocloning}, the impossibility of discriminating such
states with certainty~\cite{Fuchscloning,RudolphSpekkens03}, the lack of
exponential divergence of such states (in the space of epistemic states)
under chaotic evolution~\cite{Ballentinechaos}, and, for correlated states,
many of the features of entanglement~\cite{Col01}. On the other hand, states
of \emph{complete} knowledge do not exhibit these phenomena. This suggests
that one would obtain a better analogy with quantum theory if states of
complete knowledge were somehow impossible to achieve, that is, if somehow
\emph{maximal} knowledge was always\emph{\ incomplete} knowledge~\cite%
{Fuchscloning,FCS,Fuchssamizdat}. This idea is borne out by the results of
this paper. In fact, the toy theory suggests that the restriction on
knowledge should take a particular form, namely, that one's knowledge be
quantitatively equal to one's ignorance in a state of maximal knowledge.

It is important to bear in mind that one cannot derive quantum theory from
the toy theory, nor from any simple modification thereof. The problem is
that the toy theory is a theory of incomplete knowledge about local and
noncontextual hidden variables, and it is well known that quantum theory
cannot be understood in this way \cite{Bell,KochenSpecker,Bell2}. This
prompts the obvious question: if a quantum state is a state of knowledge,
and it is not knowledge of local and noncontextual hidden variables, then
what is it knowledge about? We do not at present have a good answer to this
question. We shall therefore remain completely agnostic about the nature of
the reality to which the knowledge represented by quantum states pertains.
This is not to say that the question is not important. Rather, we see the
epistemic approach as an unfinished project, and this question as the
central obstacle to its completion. Nonetheless, we argue that even in the
absence of an answer to this question, a case can be made for the epistemic
view. The key is that one can hope to identify phenomena that are
characteristic of states of incomplete knowledge regardless of what this
knowledge is about.

The outline of the paper is as follows. In Sec.~\ref{principle}, we
introduce our foundational principle -- that there is a balance between
knowledge and ignorance in a state of maximal knowledge -- and define our
measures of knowledge and ignorance. From this starting point, and a few
other assumptions, we derive the toy theory. We begin in Sec.~\ref{1system}
by considering the simplest possible system that can satisfy the principle.
In Sec.~\ref{2systems} we consider pairs of these systems, and in Sec.~\ref%
{3systems}, triplets. For each of these cases, we determine the epistemic
states, measurements and transformations that are allowed by the principle,
as well as the manner in which epistemic states must be updated after a
measurement. Along the way, we draw attention to various analogues of
quantum phenomena. Some additional analogues are enumerated in Sec.~\ref%
{otheranalogues}, while in Sec.~\ref{phenomenathatdonotarise} we
identify some quantum phenomena that are \emph{not} reproduced by
the toy theory and consider what these teach us about how to
proceed with the epistemic research program. In Sec.~\ref
{relatedwork}, we discuss related work, specifically,
Kirkpatrick's playing card model~\cite{Kirkpatrick}, Hardy's toy
theory~\cite{Hardydisentangling}, Smolin's
"lockboxes"~\cite{Smolin}, Zeilinger's foundational
approach~\cite{Zeilinger} and Wootters' discrete Wigner
function~\cite{WoottersWignerfunctions}. We conclude in
Sec.\ref{conclusions} with some questions for future research.
Some additional material is presented in the appendices, namely, a
discussion of why the toy theory for $N$ elementary systems cannot
be understood as a restriction upon quantum theory for $N$ qubits,
and of the significance of our results for information-theoretic
derivations of quantum theory.


\section{The knowledge balance principle}

\label{principle}

The toy theory is built on the following foundational principle:

\begin{quotation}
If one has maximal knowledge, then for every system, at every time, the
amount of knowledge one possesses about the ontic state of the system at
that time must equal the amount of knowledge one lacks.
\end{quotation}

We call this the \emph{knowledge balance principle}. As stated, it is not
sufficiently explicit, because the manner of quantifying the amount of
knowledge one possesses and the amount one lacks has yet to be specified.
Although the measure of knowledge that we adopt is very simple, it is not a
conventional one, and consequently we must define it carefully.

We begin by introducing the notion of a \emph{canonical} set of yes/no
questions. This is a set of yes/no questions that is sufficient to fully
specify the ontic state, and that has a minimal number of elements. To
clarify this notion, consider a situation wherein there are four possible
ontic states. A set of questions that will determine which of the four
applies is: ``Is it 1, or not?'', ``Is it 2, or not?'', ``Is it 3, or not?''
and ``Is it 4, or not?''. This questioning scheme is inefficient however. A
more efficient scheme divides the set of possibilities into two with every
question. Indeed, one can fully specify the ontic state with just two
questions, for instance: ``Is it in the set \{1,2\}, or not?'' and ``Is it
in the set \{1,3\}, or not?''. As there are four answers to two yes/no
questions, two is the minimal number of questions that can possibly specify
which of four states applies. So the two questions just described form a
canonical set. Note also that there can be many canonical sets of questions.
For instance, a different pair of questions, namely, ``Is it in the set
\{1,2\}, or not?'' and ``Is it in the set \{2,3\}, or not?'' also form a
canonical set.

With the notion of a canonical set in hand, we can define our measure of
knowledge. It is simply the maximum number of questions for which the answer
is known, in a variation over all canonical sets of questions. Our measure
of ignorance is simply the difference between this number and the total
number of questions in the canonical set.

The knowledge balance principle, made specific with our measure of
knowledge, is the starting point of the toy theory. There are, however, a
few other assumptions that shall go into its derivation, to which we now
turn.

We assume that all physical systems are such that there can be a balance of
knowledge about them. This implies that the number of yes/no questions in a
canonical set must be a multiple of two, because if this were not the case,
one couldn't have an equality between the number of questions answered and
the number of questions unanswered. The simplest possible case is to have
just two questions in the canonical set. Because a canonical set is, by
definition, the minimal sufficient set of questions required to specify the
ontic state, it follows that for two questions there are four possible ontic
states. Thus the simplest possible system in the toy theory has four ontic
states. We call this an \emph{elementary system.}

We shall also assume that every system is built of elementary systems. For a
pair of elementary systems, there are four questions in the canonical set,
and sixteen possible ontic states in all. For $N$ systems, there are $2N$
questions in the canonical set and $2^{2N}$ possible ontic states. This
``reductionist'' assumption will have very significant consequences in the
development of the toy theory, as the knowledge balance principle will yield
more constraints for composite systems: not only must there be a balance of
knowledge and ignorance for the whole, but for every part of the whole,
right down to the smallest subsystems.

The motional degree of freedom for all systems is treated classically, a
background of flat space-time is assumed and every elementary system is
taken to exist at a point in space.

We also assume that the outcome of a reproducible measurement depends only
on the ontic state of the system being measured. Moreover, we assume that a
transformation applied to one system can only affect the ontic state of that
system, and not the ontic state of others. If the systems are spatially
separated, this amounts to an assumption of locality. Further, we shall
assume that an observer's state of knowledge about a system does not dictate
what can and can't be done to the system, nor does it ever determine the
change that occurs in the system's ontic state during a measurement. This
assumption is motivated by the implausibility of there being a causal
relation between the mental state of the agent and the ontic state of the
apparatus or the system.

Finally, we assume that information gain about a system is always possible.
This will allow us to infer the existence of a disturbance when a
reproducible measurement is performed, rather than inferring the
impossibility of reproducible measurements.

\section{Elementary systems}

\label{1system}

\subsection{Epistemic states}

\label{epistemicstates1}

\strut An elementary system is one for which the number of questions in the
canonical set is two, and consequently the number of ontic states is four$.$
Although it takes two yes/no questions to specify the ontic state, the
answer to only one of these can be known according to the knowledge balance
principle. Thus, the epistemic states for which the balance occurs are those
which identify the ontic state of the system to be one of two possibilities.
Denoting the four ontic states by `1',`2',`3' and `4', and disjunction by
the symbol `$\vee $' (read as `or'), we can specify the possible epistemic
states as disjunctions of the ontic states. In all, there are six states of
maximal knowledge, namely,
\begin{eqnarray}
&&1\vee 2,  \notag \\
&&3\vee 4,  \notag \\
&&1\vee 3,  \notag \\
&&2\vee 4,  \notag \\
&&2\vee 3,  \notag \\
&&1\vee 4.  \label{epistemicstates}
\end{eqnarray}
It is useful to represent these graphically as follows:
\begin{eqnarray}
&&\includegraphics[width=15mm]{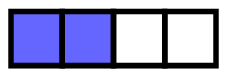}  \notag \\
&&\includegraphics[width=15mm]{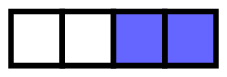}  \notag \\
&&\includegraphics[width=15mm]{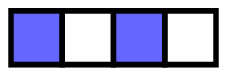}  \notag \\
&&\includegraphics[width=15mm]{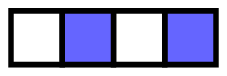}  \notag \\
&&\includegraphics[width=15mm]{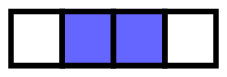}  \notag \\
&&\includegraphics[width=15mm]{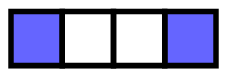}
\end{eqnarray}
with the understanding that the four cells represent the four ontic states,
and the filled cells denote the set in which the actual ontic state of the
system is known to lie.

For a single elementary system, the only way to have less than maximal
knowledge is for both questions in the canonical set to be unanswered. This
corresponds to the epistemic state
\begin{equation}
1\vee 2\vee 3\vee 4.
\end{equation}%
It is denoted pictorially by
\begin{equation}
\begin{array}{l}
\includegraphics[width=15mm]{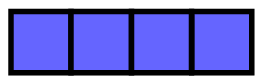}%
\end{array}%
.
\end{equation}

A single elementary system in the toy theory is analogous to a system
described by a two-dimensional Hilbert space in quantum theory, called a
\emph{qubit} in quantum information theory. In particular, the six epistemic
states describing maximal knowledge of a single elementary system are
analogous to the following six pure qubit states
\begin{eqnarray}
&&1\vee 2\Leftrightarrow \left| 0\right\rangle  \notag \\
&&3\vee 4\Leftrightarrow \left| 1\right\rangle  \notag \\
&&1\vee 3\Leftrightarrow \left| +\right\rangle  \notag \\
&&2\vee 4\Leftrightarrow \left| -\right\rangle  \notag \\
&&2\vee 3\Leftrightarrow \left| +i\right\rangle  \notag \\
&&1\vee 4\Leftrightarrow \left| -i\right\rangle  \label{epistemicstatesmap}
\end{eqnarray}
where $\left| \pm \right\rangle =\frac{1}{\sqrt{2}}\left| 0\right\rangle \pm
\left| 1\right\rangle ,$ and $\left| \pm i\right\rangle =\frac{1}{\sqrt{2}}%
\left| 0\right\rangle \pm i\left| 1\right\rangle ,$ while the single state
of nonmaximal knowledge is analogous to the completely mixed state for a
qubit, that is,
\begin{equation}
1\vee 2\vee 3\vee 4\Leftrightarrow I/2
\end{equation}
where $I$ is the identity operator on the 2-dimensional Hilbert space. The
rest of this section will demonstrate the extent of this analogy. Note,
however, that the choice of which three epistemic states to associate with $%
\left| 0\right\rangle ,\left| +\right\rangle $ and $\left| +i\right\rangle $
is simply a convention.

\textbf{Disjointness.} It is useful to define the \emph{ontic base }of an
epistemic state to be the set of ontic states which are consistent with it.
For instance, the ontic base of $1\vee 2$ is the set $\{1,2\}.$ If the
intersection of the ontic bases of a pair of epistemic states is empty, then
those states are said to be \emph{disjoint. }A set of epistemic states are
said to be disjoint if they are pairwise disjoint. The relation of
disjointness is analogous to the relation of orthogonality among quantum
states. The fact that there are pairs of epistemic states which are
nondisjoint demonstrates that there exists an analogue of nonorthogonality
in the toy theory.

\textbf{Fidelity.} One can also introduce a measure of the degree of
nondisjointness, equivalently, a measure of the distance between a pair of
epistemic states in the space of such states. A standard measure of distance
between two probability distributions, $\mathbf{p}=(p_{k})_{k}$ and $\mathbf{%
q}=(q_{k})_{k}\mathbf{,}$ is the \emph{classical fidelity}, defined by $F(%
\mathbf{p,q})=\sum_{k}\sqrt{p_{k}}\sqrt{q_{k}}.$ If we treat the epistemic
states of the toy theory as uniform probability distributions, for instance,
associating the distribution $(1/2,1/2,0,0)$ with $1\vee 2,$ and $%
(1/4,1/4,1/4,1/4)$ with $1\vee 2\vee 3\vee 4,$ then we can use the classical
fidelity as a measure of distance. For the epistemic states of a single
elementary system, the fidelity between a pair takes one of four values: the
value 0 if they are disjoint, such as $1\vee 2$ and $3\vee 4;$ the value $%
1/2 $ if they are nondisjoint states of maximal knowledge, such as $1\vee 2$
and $1\vee 3$; the value $1/\sqrt{2}$ if one is a state of maximal
knowledge, the other not, such as $1\vee 2$ and $1\vee 2\vee 3\vee 4;$ and
the value $1$ if the elements of the pair are identical. The analogous
measure of distance between quantum states is the quantum fidelity \cite%
{NielsenChuang}, which is defined for a pair of density operators, $\rho $
and $\sigma $, as $\mathrm{Tr}\sqrt{\rho }\sqrt{\sigma }.$ In the case of a
pair of pure states, $\left| \psi \right\rangle $ and $\left| \chi
\right\rangle ,$ the fidelity is simply the inner product squared, $\left|
\left\langle \psi |\chi \right\rangle \right| ^{2}.$ It turns out that the
classical fidelities between pairs of epistemic states are precisely equal
to the quantum fidelities for the analogous pairs of quantum states under
the mapping of Eq.~(\ref{epistemicstatesmap}). For instance, the quantum
fidelity between $\left| 0 \right\rangle$ and $\left| 1 \right\rangle$ is 0,
between $\left| 0 \right\rangle$ and $\left| + \right\rangle$ is $1/2$,
between $\left| 0 \right\rangle$ and $I/2$ is $1/\sqrt{2}$, and between any
state and itself is $1$.

\textbf{Compatibility.} Another useful relation to introduce is that of
compatibility. Two epistemic states are said to be \emph{compatible} if the
intersection of their ontic bases is the ontic base of a valid epistemic
state. Thus, the epistemic states $1\vee 2$ and $1\vee 2\vee 3\vee 4$ have
the ontic states $1$ and $2$ in common, and are therefore compatible, while $%
1\vee 2$ and $2\vee 3$ have only the ontic state $2$ in common, and are
therefore incompatible. Whenever two observers are describing the same
system, their epistemic states must be compatible. This follows from the
fact that if these individuals pool their information they will rule out any
ontic state that either one of them rules out, which is equivalent to taking
the intersection of the ontic bases of their epistemic states. If their
epistemic states were incompatible, this would result in a final epistemic
state that violated the knowledge balance principle. Note that this implies
that if two observers both have maximal knowledge of a system, then their
states of knowledge must be identical; there is always inter-subjective
agreement among maximally informed observers. This relation of compatibility
is analogous to the Brun-Finkelstein-Mermin compatibility relation for
quantum states, according to which two states are compatible whenever the
intersection of their supports (in Hilbert space) is not null \cite{BFM}.

\textbf{Convex combination.} We now introduce a way of combining epistemic
states that is analogous to the convex combination (or incoherent
superposition) of quantum states. A pair of epistemic states in the toy
theory must satisfy two conditions for the convex combination to be defined.
The first condition is that they be disjoint. The second condition is that
the union of their ontic bases must form the ontic base of a valid epistemic
state. If both conditions are met, then the epistemic state that results by
taking the union of the ontic bases of the pair is defined to be the convex
combination of that pair. Thus, the convex combination of $1\vee 2$ and $%
3\vee 4$ is $1\vee 2\vee 3\vee 4$, while the convex combination of $1\vee 2$
and $1\vee 3$ is undefined, as is the convex combination of $1\vee 2$ and $%
1\vee 2\vee 3\vee 4$. The convex combination of a larger set of epistemic
states is defined similarly.

Note that in addition to being sometimes undefined, the convex combination
of a set of epistemic states in the toy theory also differs from the convex
combination of a set of quantum states in there being nothing analogous to a
convex sum with unequal weights.

It is useful to introduce the terms \emph{mixed }and \emph{pure }to specify
whether or not an epistemic state can be obtained as a convex combination of
distinct epistemic states or not. For a single elementary system, the
epistemic states $1\vee 2,$ $3\vee 4,$ $1\vee 3,$ $2\vee 4,$ $1\vee 4,$ and $%
2\vee 3$ are pure, while the epistemic state $1\vee 2\vee 3\vee 4$ is mixed.
There are in fact many convex decompositions of $1\vee 2\vee 3\vee 4$.
Denoting convex combination by the symbol `$+_{\text{cx}}$', we have
\begin{eqnarray}
1\vee 2\vee 3\vee 4 &=&(1\vee 2)+_{\text{cx}}(3\vee 4)  \notag \\
&=&(1\vee 3)+_{\text{cx}}(2\vee 4)  \notag \\
&=&(2\vee 3)+_{\text{cx}}(1\vee 4),
\end{eqnarray}
Graphically,
\begin{eqnarray}
\begin{array}{l}
\includegraphics[width=15mm]{1or2or3or4.eps}%
\end{array}
&=&
\begin{array}{l}
\includegraphics[width=15mm]{1or2.eps}%
\end{array}
+_{\text{cx}}
\begin{array}{l}
\includegraphics[width=15mm]{3or4.eps}%
\end{array}
\notag \\
&=&
\begin{array}{l}
\includegraphics[width=15mm]{1or3.eps}%
\end{array}
+_{\text{cx}}
\begin{array}{l}
\includegraphics[width=15mm]{2or4.eps}%
\end{array}
\notag \\
&=&
\begin{array}{l}
\includegraphics[width=15mm]{2or3.eps}%
\end{array}
+_{\text{cx}}
\begin{array}{l}
\includegraphics[width=15mm]{1or4.eps}%
\end{array}
\notag \\
\end{eqnarray}
This is analogous to the fact that in quantum theory, the completely mixed
state of a qubit, $I/2,$ has convex decompositions
\begin{eqnarray}
I/2 &=&\frac{1}{2}\left| 0\right\rangle \left\langle 0\right| +\frac{1}{2}
\left| 1\right\rangle \left\langle 1\right|  \notag \\
&=&\frac{1}{2}\left| +\right\rangle \left\langle +\right| +\frac{1}{2}\left|
-\right\rangle \left\langle -\right|  \notag \\
&=&\frac{1}{2}\left| +i\right\rangle \left\langle +i\right| +\frac{1}{2}
\left| -i\right\rangle \left\langle -i\right| .
\end{eqnarray}

Thus, the toy theory mirrors quantum theory in admitting multiple convex
decompositions of a mixed state into pure states. This multiplicity is a
direct result of the fact that in the toy theory, pure epistemic states are
states of incomplete knowledge.

\begin{figure}[h]
\includegraphics[width=60mm]{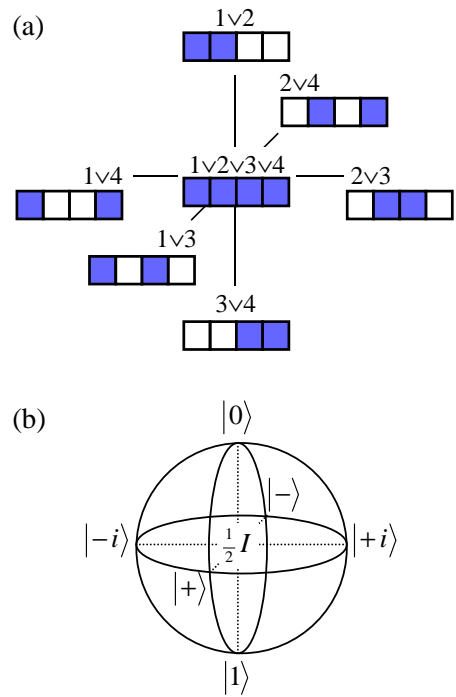}
\caption{(a) A representation of the space of epistemic states in the toy
theory, and (b) the Bloch ball representation of the states in quantum
theory.}
\label{Blochsphere}
\end{figure}

\textbf{A geometric representation of the space of epistemic states.} In
quantum theory, the Bloch sphere (or, more precisely, the Bloch ball) offers
a useful geometric representation of the quantum states of a qubit and the
relations of orthogonality and convex combination that hold among them~\cite%
{NielsenChuang}. Specifically, orthogonal quantum states are represented by
antipodal points on the sphere, and every convex decomposition of a mixed
state is associated with a convex polytope that contains in its interior the
point representing the mixed state, with the vertices of the polytope
representing the elements of the convex decomposition \cite%
{SpekkensRudolphQIC}. Similarly, the epistemic states for an elementary
system in the toy theory can be represented by a subset of the points inside
a unit ball. Disjoint epistemic states are represented by antipodal points,
and convex decompositions of the mixed epistemic state are represented by
line segments, the endpoints of which are the elements of the decomposition.
The two pictures are presented for comparison in Fig.~\ref{Blochsphere}.

\textbf{Coherent superposition.} One can also introduce a way of combining
epistemic states that is analogous to the coherent superposition of quantum
states. What we seek is a binary operation that takes a pair of pure
epistemic state to another pure epistemic state (unlike the operation of
convex combination we have just introduced, which takes a pair of pure
states to a mixed state). Suppose the two epistemic states we seek to
combine are of the form $a\vee b$ and $c\vee d$ (here, of course, $%
a,b,c,d\in \{1,2,3,4\}$ and $a\ne b,c\ne d).$ We assume that they are
disjoint, so that $a,b\ne c,d.$ Moreover, we adopt the convention that $a<b$
and $c<d.$ One can define four new pure epistemic states from these two,
namely, $a\vee c,$ $a\vee d,$ $b\vee c,$ and $b\vee d. $ We can think of
these as the result of applying four distinct binary operations to the
original pair of states. Denoting these four operations by $%
+_{1},+_{2},+_{3},$ and $+_{4},$ we have
\begin{eqnarray}
(a\vee b)+_{1}(c\vee d) &=&a\vee c  \notag \\
(a\vee b)+_{2}(c\vee d) &=&b\vee d  \notag \\
(a\vee b)+_{3}(c\vee d) &=&b\vee c  \notag \\
(a\vee b)+_{4}(c\vee d) &=&a\vee d.
\end{eqnarray}
The first operation can be described as follows: take the ontic state of
lowest index from the first epistemic state, and the ontic state of lowest
index from the second epistemic, then define a new epistemic state in terms
of these. The other three operations can be defined similarly. All that
differs is whether one takes the ontic state with the lowest or highest
index from each epistemic state. The convention we have chosen is that $%
+_{1},+_{2},+_{3}$ and $+_{4}$ are associated with low-low, high-high,
high-low and low-high. We call these \emph{coherent binary operations.}

The four possible coherent binary operations acting on $1\vee 2$ and $3\vee
4 $ yield
\begin{eqnarray}
(1\vee 2)+_{\text{1}}(3\vee 4) &=&1\vee 3,  \label{bo1} \\
(1\vee 2)+_{\text{2}}(3\vee 4) &=&2\vee 4,  \label{bo2} \\
(1\vee 2)+_{\text{3}}(3\vee 4) &=&2\vee 3,  \label{bo3} \\
(1\vee 2)+_{\text{4}}(3\vee 4) &=&1\vee 4,  \label{bo4}
\end{eqnarray}
acting on $1\vee 3$ and $2\vee 4$ they yield
\begin{eqnarray}
(1\vee 3)+_{\text{1}}(2\vee 4) &=&1\vee 2,  \label{bo5} \\
(1\vee 3)+_{\text{2}}(2\vee 4) &=&3\vee 4,  \label{bo6} \\
(1\vee 3)+_{\text{3}}(2\vee 4) &=&2\vee 3,  \label{bo7} \\
(1\vee 3)+_{\text{4}}(2\vee 4) &=&1\vee 4,  \label{bo8}
\end{eqnarray}
and acting on $2\vee 3$ and $1\vee 4,$ they yield
\begin{eqnarray}
(2\vee 3)+_{\text{1}}(1\vee 4) &=&1\vee 2,  \label{bo9} \\
(2\vee 3)+_{\text{2}}(1\vee 4) &=&3\vee 4,  \label{bo10} \\
(2\vee 3)+_{\text{3}}(1\vee 4) &=&1\vee 3,  \label{bo11} \\
(2\vee 3)+_{\text{4}}(1\vee 4) &=&2\vee 4.  \label{bo12}
\end{eqnarray}
These relations should be compared with the following relations among
quantum states:
\begin{eqnarray}
\sqrt{2}^{-1}\left( \left| 0\right\rangle +\left| 1\right\rangle \right)
&=&\left| +\right\rangle  \label{cs1} \\
\sqrt{2}^{-1}\left( \left| 0\right\rangle -\left| 1\right\rangle \right)
&=&\left| -\right\rangle  \label{cs2} \\
\sqrt{2}^{-1}\left( \left| 0\right\rangle +i\left| 1\right\rangle \right)
&=&\left| +i\right\rangle  \label{cs3} \\
\sqrt{2}^{-1}\left( \left| 0\right\rangle -i\left| 1\right\rangle \right)
&=&\left| -i\right\rangle  \label{cs4}
\end{eqnarray}
and
\begin{eqnarray}
\sqrt{2}^{-1}\left( \left| +\right\rangle +\left| -\right\rangle \right)
&=&\left| 0\right\rangle  \label{cs5} \\
\sqrt{2}^{-1}\left( \left| +\right\rangle -\left| -\right\rangle \right)
&=&\left| 1\right\rangle  \label{cs6} \\
\sqrt{2}^{-1}\left( \left| +\right\rangle +i\left| -\right\rangle \right)
&=&e^{i\pi /4}\left| -i\right\rangle  \label{cs7} \\
\sqrt{2}^{-1}\left( \left| +\right\rangle -i\left| -\right\rangle \right)
&=&e^{-i\pi /4}\left| +i\right\rangle  \label{cs8}
\end{eqnarray}
and
\begin{eqnarray}
\sqrt{2}^{-1}\left( \left| +i\right\rangle +\left| -i\right\rangle \right)
&=&\left| 0\right\rangle  \label{cs9} \\
\sqrt{2}^{-1}\left( \left| +i\right\rangle -\left| -i\right\rangle \right)
&=&i\left| 1\right\rangle  \label{cs10} \\
\sqrt{2}^{-1}\left( \left| +i\right\rangle +i\left| -i\right\rangle \right)
&=&e^{i\pi /4}\left| +\right\rangle  \label{cs11} \\
\sqrt{2}^{-1}\left( \left| +i\right\rangle -i\left| -i\right\rangle \right)
&=&e^{-i\pi /4}\left| -\right\rangle  \label{cs12}
\end{eqnarray}

Note that the combinations we have enumerated do not exhaust the
possibilities, since for the operations $+_{3}$ and $+_{4}$, the order of
the arguments in the operation is important. That is, $+_{3}$ and $+_{4}$
are not commutative operations. For instance, $(1\vee 2)+_{3}(3\vee 4)=2\vee
3$ while $(3\vee 4)+_{3}(1\vee 2)=1\vee 4.$ The same sensitivity to ordering
is found in quantum theory for coherent superpositions with relative phases $%
\pi /2$ and $3\pi /2$. For instance, $\sqrt{2}^{-1}\left( \left|
0\right\rangle +i\left| 1\right\rangle \right) =\left| +i\right\rangle $
while $\sqrt{2}^{-1}\left( \left| 1\right\rangle +i\left| 0\right\rangle
\right) =i\left| -i\right\rangle .$

It is natural to associate the operations $+_{1},+_{2},+_{3}$ and $+_{4}$
with coherent superpositions of two quantum states where the relative
weights are equal and the relative phases of the second term to the first
are $0,\pi ,\pi /2$ and $3\pi /2$ respectively,
\begin{eqnarray}
&&+_{1}\Leftrightarrow 0  \notag \\
&&+_{2}\Leftrightarrow \pi  \notag \\
&&+_{3}\Leftrightarrow \pi/2  \notag \\
&&+_{4}\Leftrightarrow 3 \pi/2  \label{binaryoperationsmap}
\end{eqnarray}
Under this association of toy-theoretic operations with quantum operations
and under the association of epistemic states with quantum states expressed
in (\ref{epistemicstatesmap}), the relations (\ref{bo1})-(\ref{bo12})
parallel (modulo global phases) the relations (\ref{cs1})-(\ref{cs12}), with
two notable exceptions. Given the form of the relations (\ref{bo7}) and (\ref%
{bo8}), and the fact that $2\vee 3$ maps to $\left| +i\right\rangle $ and $%
1\vee 4$ maps to $\left| -i\right\rangle $ under (\ref{epistemicstatesmap}),
one would expect the right hand side of (\ref{cs7}) to be proportional to $%
\left| +i\right\rangle $ and the right hand side of (\ref{cs8}) to be
proportional to $\left| -i\right\rangle $ rather than vice-versa. Note that
one cannot achieve a better analogy by modifying the associations adopted in
(\ref{epistemicstatesmap}) and (\ref{binaryoperationsmap}). For instance, by
associating $2\vee 3$ with $\left| -i\right\rangle $ and $1\vee 4$ with $%
\left| +i\right\rangle $, the relations (\ref{bo7}) and (\ref{bo8}) can be
made to parallel the relations (\ref{cs7}) and (\ref{cs8}), however, in this
case the relations (\ref{bo3}) and (\ref{bo4}) fail to parallel (\ref{cs3})
and (\ref{cs4}). This curious failure of the analogy shows that an
elementary system in the toy theory is not simply a constrained version of a
qubit.

There are two other important respects in which our coherent binary
operations for a single elementary system differ from those one finds in
quantum theory for a qubit. First, whereas any pair of quantum states of a
qubit can be coherently superposed, the binary operations in the toy theory
are not defined for arbitrary pairs of epistemic states. Specifically, they
are not defined for nondisjoint epistemic states. Second, whereas there are
a continuum of different types of coherent superposition of a pair of
quantum states of a qubit, corresponding to all possible relative weights
and all possible relative phases, there are only four coherent binary
operations in the toy theory.

\subsection{Transformations}

\label{transformations1}

We now consider the sorts of transformations of the ontic states that are
allowed by the knowledge balance principle. Imagine a transformation that
takes two different ontic states, say $1$ and $2,$ to a single ontic state,
say $3$. If the epistemic state prior to the transformation was $1\vee 2,$
then after the transformation, one would be certain that the ontic state was
$3.$ But such an epistemic state violates the knowledge balance principle,
therefore this transformation is not allowed. A similar example can be
devised for any many-to-one map. Thus, all such maps are ruled out by the
principle.

We are left with the one-to-one maps and the one-to-many maps. We focus on
the former here, since these correspond to the reversible maps. Clearly,
these are simply the set of permutations of the four ontic states.

One can describe permutations in terms of cycles. For instance, the
permutation $a\rightarrow a,b\rightarrow c\rightarrow d\rightarrow b$
involves two cycles: a 1-cycle, $a\rightarrow a,$ and a 3-cycle $%
b\rightarrow c\rightarrow d\rightarrow b.$ In cycle notation, this
permutation is written as $(a)(bcd).$ The set of permutations of $4$
elements is a group, called $S_{4},$ containing $24$ elements. Permutations
with the same number of cycles form a class. We list the elements of $S_{4},$
and their class structure in Table \ref{S_4}. If an element is written
alone, it is its own inverse, whereas elements appearing in pairs are each
other's inverses.
\begin{table}[h]
\par
\begin{tabular}{ccccc}
$(1^{4})$ & $(31)$ & $(21^{2})$ & $(2^{2})$ & $(4)$ \\ \hline
(1)(2)(3)(4) & (234)(1) & (12)(3)(4) & (12)(34) & (1234) \\
& (243)(1) &  &  & (1432) \\
&  & (13)(2)(4) & (13)(24) &  \\
& (134)(2) &  &  & (1243) \\
& (143)(2) & (14)(2)(3) & (14)(23) & (1342) \\
&  &  &  &  \\
& (124)(3) & (23)(1)(4) &  & (1324) \\
& (142)(3) &  &  & (1423) \\
&  & (24)(1)(3) &  &  \\
& (123)(4) &  &  &  \\
& (132)(4) & (34)(1)(2) &  &
\end{tabular}%
\caption{The class structure of the group $S_{4}$ of permutations of four
elements}
\label{S_4}
\end{table}

The valid transformations may be usefully represented graphically by arrows
between the ontic states. For instance,
\begin{eqnarray}
(123)(4) &:&
\begin{array}{l}
\includegraphics[width=15mm]{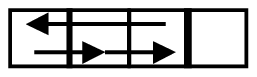}%
\end{array}
\notag \\
(13)(24) &:&
\begin{array}{l}
\includegraphics[width=15mm]{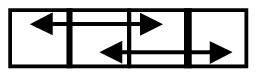}%
\end{array}
\notag \\
(13)(2)(4) &:&
\begin{array}{l}
\includegraphics[width=15mm]{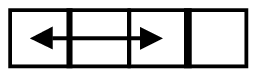}%
\end{array}
\notag \\
(1234) &:&
\begin{array}{l}
\includegraphics[width=15mm]{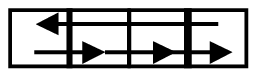}%
\end{array}%
\end{eqnarray}

It is interesting to determine how the set of epistemic states are
transformed under a permutation of the ontic states. For instance, the
permutation $(123)(4)$ leads to the following map on the epistemic states
\begin{eqnarray}
1\vee 2 &\rightarrow & 2\vee 3  \notag \\
3\vee 4 &\rightarrow & 1\vee 4  \notag \\
1\vee 3 &\rightarrow & 1\vee 2  \notag \\
2\vee 4 &\rightarrow & 3\vee 4  \notag \\
2\vee 3 &\rightarrow & 1\vee 3  \notag \\
1\vee 4 &\rightarrow & 2\vee 4
\end{eqnarray}

\begin{figure}[h]
\includegraphics[width=85mm]{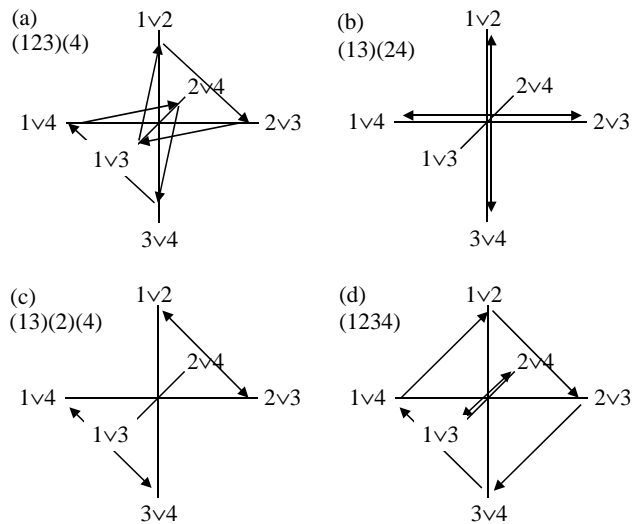}
\caption{How four permutations of the ontic states appear in the Bloch
sphere representation of the space of epistemic states.}
\label{transformations}
\end{figure}

Representing the epistemic states in the ``Bloch sphere'' picture, we see
that this permutation appears as a rotation by 120$^{\circ }$ about the axis
that points in the $\hat{x}+\hat{y}+\hat{z}$ direction, as seen in Fig.~\ref%
{transformations}(a). Similarly, the permutation $(13)(24)$ appears as a
rotation by 180$^{\circ }$ about the $\hat{x}$ axis, as seen in Fig.~\ref%
{transformations}(b). These permutations are analogous to unitary maps in
Hilbert space, which appear as rotations in the Bloch sphere. These two
examples might lead one to think that \emph{all} permutations appear as
rotations in the Bloch sphere picture, but this is not the case. A
permutation such as $(13)(2)(4)$ is a reflection about the plane spanned by $%
\hat{x}$ and $(\hat{y}+\hat{z})$, as seen in Fig.~\ref{transformations}(c),
while $(1234)$ involves a rotation of 90$^{\circ }$ about $\hat{x}$ and a
reflection about the plane spanned by $\hat{y}$ and $\hat{z}$, as seen in
Fig.~\ref{transformations}(d). These are analogous to anti-unitary maps in
Hilbert space. Anti-unitary maps do not represent possible evolutions of a
system in quantum theory because evolution is assumed to be continuous in
time. The fact that transformations analogous to anti-unitary maps arise in
the toy theory is a consequence of the fact that the transformations in the
toy theory are discrete.

Note that the set of all transformations for an elementary system
corresponds to the symmetry group of a tetrahedron the four vertices of
which are located along the $\hat{x}-\hat{y}+\hat{z}$ axis, the $-\hat{x}+%
\hat{y}+\hat{z}$ axis, the $\hat{x}+\hat{y}-\hat{z}$ axis and the $-\hat{x}-%
\hat{y}-\hat{z}$ axis. These vertices are associated with the ontic states $%
1,2,3$, and $4$ respectively.

\subsection{No universal state inverter}

\label{nostateinverter}

Given the aspects of the toy theory developed so far, we can already
demonstrate an analogy to a characteristically quantum phenomenon, namely,
the impossibility of building a universal state inverter. For a single
qubit, a universal state inverter is a device that deterministically maps
every pure quantum state to the orthogonal quantum state, that is,
\begin{equation}
\left| \psi \right\rangle \rightarrow \left| \bar{\psi}\right\rangle \text{
for all }\left| \psi \right\rangle
\end{equation}
where $\left\langle \psi |\bar{\psi}\right\rangle =0.$ Such a map cannot be
physically implemented because it is not unitary \cite{nostateinverter}.

The analogous task in the toy theory is to deterministically map every pure
epistemic state of an elementary system to the one that is disjoint with it.
Thus, we require
\begin{eqnarray}
1\vee 2 &\leftrightarrow &3\vee 4,  \notag \\
1\vee 3 &\leftrightarrow &2\vee 4,  \notag \\
2\vee 3 &\leftrightarrow &1\vee 4.
\end{eqnarray}
But this transformation is impossible since it does not correspond to any
permutation of the ontic states; the first two conditions together imply
that $1\leftrightarrow 4$ and $2\leftrightarrow 3,$ which is in
contradiction with the third condition.

The impossibility of universal state inversion in both quantum theory and
the toy theory can also be seen by noting that it would appear as an
inversion about the origin in the Bloch ball representation, and such an
inversion cannot be achieved by any rotation, nor by any combination of the
rotations and reflections that are allowed in the toy theory.

\subsection{Measurements}

\label{measurements1}

\strut We now turn to the nature of measurements in the toy theory. We shall
here consider only measurements that are \emph{reproducible} in the sense
that if repeated upon the same system, they yield the same outcome. For this
to be possible, the epistemic state after the measurement must rule out all
of the ontic states that are not consistent with the outcome (otherwise, the
epistemic state would not reflect the fact that a different outcome cannot
occur upon repetition).

The knowledge balance principle imposes restrictions on the sort of
reproducible measurement that can be implemented. Again, we start by ruling
out a certain kind of measurement, namely one which identifies whether or
not the ontic state is in a singleton set. To be specific, consider the
measurement which determines whether the ontic state is $1$ or not. The `not
1' outcome identifies the ontic state as being either $2$ or $3$ or $4.$
Now, if in this measurement the outcome $1$ occurs (and nothing prevents it
from occurring when the initial epistemic state deems it to be possible),
then by virtue of the assumed reproducibility of measurements, the epistemic
state after the measurement must rule out the ontic states $2,3$ and $4.$
But this would mean that after the measurement one would be certain that the
ontic state was $1,$ and such a state of knowledge violates the knowledge
balance principle. Thus, the measurement considered is not allowed.

Clearly, the fewest ontic states that can be associated with a single
outcome of a measurement is two. Thus, the only valid reproducible
measurements are those which partition the four ontic states into two sets
of two ontic states. There are only three such partitionings:
\begin{eqnarray}
&&\{1\vee 2,3\vee 4\}  \notag \\
&&\{1\vee 3,2\vee 4\}  \notag \\
&&\{1\vee 4,2\vee 3\}.
\end{eqnarray}

In our pictorial representation, we can represent these as
\begin{eqnarray}
\includegraphics[width=15mm]{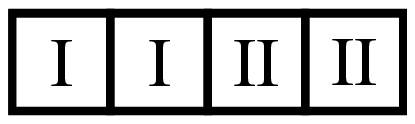}  \notag \\
\includegraphics[width=15mm]{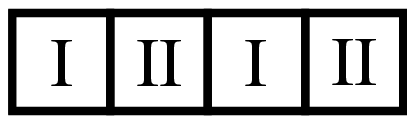}  \notag \\
\includegraphics[width=15mm]{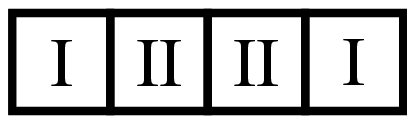}
\end{eqnarray}
where in each case the two sets are distinguished by a roman numeral. These
three partitionings are analogous to the following three bases in quantum
theory:
\begin{eqnarray}
\{1\vee 2,3\vee 4\} &\Leftrightarrow &\{\left| 0\right\rangle ,\left|
1\right\rangle \},  \notag \\
\{1\vee 3,2\vee 4\} &\Leftrightarrow &\{\left| +\right\rangle ,\left|
-\right\rangle \},  \notag \\
\{1\vee 4,2\vee 3\} &\Leftrightarrow &\{\left| +i\right\rangle ,\left|
-i\right\rangle \}.
\end{eqnarray}

We call the set of ontic states associated with a particular outcome the
\emph{ontic base} of that outcome. If the initial epistemic state has its
ontic base inside the ontic base of a particular outcome, then that outcome
is certain to occur, otherwise, the outcome is not determined by the initial
epistemic state. For instance, suppose the epistemic state is $1\vee 2,$ so
that graphically we have
\begin{equation}
\includegraphics[width=15mm]{1or2.eps}.
\end{equation}
If one performs the measurement that distinguishes $1\vee 2$ from $3\vee 4,$
depicted
\begin{equation}
\includegraphics[width=15mm]{12vs34.eps}
\end{equation}
then the first outcome is certain to occur. while if one performs the
measurement that distinguishes $1\vee 3$ from $2\vee 4,$ depicted
\begin{equation}
\includegraphics[width=15mm]{13vs24.eps}
\end{equation}
then the outcome is not determined. In a large ensemble of such experiments,
one expects the two outcomes to occur with equal frequency \footnote{%
This presumes that the relative frequency of different ontic states in the
ensemble is equivalent to the probability distribution defined by the
epistemic state. This assumption can be questioned. See, for instance, the
work of Valentini \cite{Valentini}.}. This is analogous to what occurs in
quantum theory: if the initial quantum state is one of the elements of the
orthogonal basis associated with the measurement, then the outcome
associated with that element is certain to occur, while if it is not, only
the expected relative frequencies of the outcomes are determined by the
quantum state.

\subsection{Measurement update rule}

\label{transformationaspect1}

Suppose the initial epistemic state is $1\vee 2,$ a reproducible measurement
of $1\vee 3$ versus $2\vee 4$ is performed, and the outcome $1\vee 3$
occurs. In this case, one can retrodict that the ontic state of the system
must have been $1$ prior to the measurement. This is not in conflict with
the knowledge balance principle since the latter does not place restrictions
on what one can know, at a given time, about the ontic state at an earlier
time. The principle \emph{does}, however, place restrictions on what one can
know, at a given time, about the ontic state at that time. If it were the
case that the system's ontic state was known to be unaltered in the process
of measurement, then one's description of the system prior to the
measurement would apply also after the measurement. But then, one would know
the system to be in the ontic state $1$ after the measurement, and this
\emph{is} in violation of the knowledge balance principle. Since we assume
that information gain through measurements is always possible, we must
conclude that measurement causes an unknown disturbance to the ontic state
of the system.

In our particular example, the assumption that the measurement is
reproducible implies that the epistemic state after the measurement must
rule out the ontic states $2$ and $4$. Thus, the only final epistemic state
that makes the measurement result reproducible and abides by the knowledge
balance principle is $1\vee 3.$

It follows that the nature of the unknown disturbance must be such that
although one knows that the ontic state that applied prior to the
measurement was $1,$ all one knows about the ontic state that applies after
the measurement is that it is $1$ or $3.$ Thus, the unknown disturbance must
ensure that
\begin{equation}
1\rightarrow 1\vee 3.
\end{equation}

Similarly, if the initial epistemic state was $3\vee 4$ and a measurement of
$1\vee 3$ versus $2\vee 4$ found the outcome $1\vee 3,$ one could infer that
prior to the measurement, the ontic state must have been $3.$ However, in
order to have reproducibility and to abide by the knowledge balance
principle, it must be the case that after the measurement, the ontic state
is only know to be $1$ or $3.$ Thus, the unknown disturbance must ensure
that
\begin{equation}
3\rightarrow 1\vee 3
\end{equation}

These two conditions can be satisfied by assuming that the measurement
induces either the permutation $(1)(2)(3)(4)$ or the permutation $(13)(2)(4)$%
, but that it is not known which. For instance, if the ontic state was $1,$
then either it remains $1$ or it becomes 3. Which of these two possibilities
occurs is unknown, so all that can be said of the ontic state that applies
after the measurement is that it is $1$ or $3.$

This is generalized as follows. In a measurement of $a\vee b$ versus $c\vee
d,$ if the outcome $a\vee b$ occurs, then either the permutation
(a)(b)(c)(d) occurs (i.e. nothing happens to the system) or the permutation
(ab)(c)(d) occurs (if the ontic state is $a$, it becomes $b$ and
vice-versa), but it is unknown which.

\strut Note that the possible permutations resulting from a measurement
depend only on the identity and outcome of the measurement and not on the
initial epistemic state. This is appropriate, since the nature of someone's
knowledge of a system should not influence how the ontic state of the system
evolves during a measurement. By the same token, whether or not the system
is initially correlated with other systems should not influence the nature
of the evolution of the ontic state of the system during a measurement,
because the presence or absence of such correlation is a feature of an
observer's knowledge of the system, not a property of the system itself
\footnote{%
The only way in which the initial epistemic state could influence the
evolution of the ontic state is if there was a physical influence exerted by
the mental state of the observer on the physical system. In our derivation
of the toy theory, we are explicitly rejecting this sort of possibililty.}.
Thus, although we have derived the nature of the unknown disturbance by
considering an example where the system being measured is not correlated
with any other system, the results obtained must also be applicable when
such correlation is present. We will therefore make use of the results
derived above when we consider measurements on one member of a pair of
systems in Sections \ref{epistemicstates2}, \ref{steering} and \ref%
{measurements2}.

In the case considered here, where the system of interest is\emph{\ }%
uncorrelated with all other systems, the nature of the transformation of the
ontic states for reproducible maximally-informative measurements implies a
particularly simple rule for updating the epistemic state. The final
epistemic state is simply the set of ontic states that are associated with
the outcome obtained in the measurement. This is analogous to the update
rule for a reproducible maximally-informative measurement in quantum theory,
where the final quantum state is simply the eigenvector associated with the
outcome obtained in the measurement.

We now consider a few more quantum phenomena for which we can provide an
analogue in the toy theory.

\subsection{Noncommutativity of measurements}

\label{non-commutativity}

In quantum theory, the order in which measurements occur is important for
the outcome that is obtained in these measurements. For instance,
implementing a reproducible measurement of the basis $\{\left|
0\right\rangle ,\left| 1\right\rangle \}$ followed by a reproducible
measurement of the basis $\{\left| +\right\rangle ,\left| -\right\rangle \}$
in general has different results from the case where they are implemented in
the opposite order. Specifically, if the quantum state is $\left|
0\right\rangle $ initially, then if the measurement of $\{\left|
0\right\rangle ,\left| 1\right\rangle \}$ comes first, it will yield the
outcome $\left| 0\right\rangle $ with certainty. On the other hand, if it
comes second, then the outcomes $\left| 0\right\rangle $ and $\left|
1\right\rangle $ will occur with equal probability. The reason is that the
intervening measurement of $\{\left| +\right\rangle ,\left| -\right\rangle
\} $ collapses the quantum state to $\left| +\right\rangle $ or $\left|
-\right\rangle $ with equal probabilities, and the latter states make the
outcome of $\{\left| 0\right\rangle ,\left| 1\right\rangle \}$ completely
unpredictable.

Similarly, the order in which measurements occur in the toy theory also has
a bearing on the outcomes obtained. Indeed, the example just provided has a
perfect analogue in the toy theory. We consider implementing a reproducible
measurement associated with the partitioning $\{1\vee 2,3\vee 4\}$ followed
by the reproducible measurement associated with the partitioning $\{1\vee
3,2\vee 4\},$ and the same measurements in reverse order:
\begin{eqnarray}
&&
\begin{array}{l}
\includegraphics[width=15mm]{12vs34.eps}%
\end{array}
\text{ then }
\begin{array}{l}
\includegraphics[width=15mm]{13vs24.eps}%
\end{array}
\notag \\
&&\text{or}  \notag \\
&&
\begin{array}{l}
\includegraphics[width=15mm]{13vs24.eps}%
\end{array}
\text{ then }
\begin{array}{l}
\includegraphics[width=15mm]{12vs34.eps}%
\end{array}
.
\end{eqnarray}
Suppose that initially the epistemic state is $1\vee 2$,
\begin{equation}
\begin{array}{l}
\includegraphics[width=15mm]{1or2.eps}%
\end{array}
.
\end{equation}
If the measurement of $\{1\vee 2,3\vee 4\}$ comes first, it will yield the
outcome $1\vee 2$ with certainty. On the other hand, if it comes second,
then the outcomes $1\vee 2$ and $3\vee 4$ will occur with equal frequency.
The reason is that the measurement of $\{1\vee 3,2\vee 4\}$ causes the
epistemic state to be updated to $1\vee 3$ or $2\vee 4$ with equal
probabilities, and each of these epistemic states makes the outcome of $%
\{1\vee 2,3\vee 4\}$ completely unpredictable.

\subsection{Interference}

\label{interference}

Another quantum phenomenon that the toy theory reproduces qualitatively is
interference. We offer the following paradigmatic example of interference in
quantum theory. Consider three experiments:
\begin{eqnarray*}
&&\text{(a) Prepare }\left| 0\right\rangle \text{, then measure }\{\left|
+\right\rangle ,\left| -\right\rangle \} \\
&&\text{(b) Prepare }\left| 1\right\rangle \text{, then measure }\{\left|
+\right\rangle ,\left| -\right\rangle \} \\
&&\text{(c) Prepare }\sqrt{2}^{-1}\left( \left| 0\right\rangle +\left|
1\right\rangle \right) \text{, then measure }\{\left| +\right\rangle ,\left|
-\right\rangle \} \text{.}
\end{eqnarray*}
The probability distribution over the outcomes is $(1/2,1/2)$ for (a), $%
(1/2,1/2)$ for (b) and $(1,0)$ for (c). The probability zero for the outcome
$\left| -\right\rangle$ in case (c) is, of course, a result of the
destructive interference between the amplitude for this outcome in states $%
\left| 0\right\rangle$ and $\left| 1\right\rangle$.


Interference is often cited as evidence against the epistemic view of
quantum states. The argument runs as follows. If quantum states are
associated with probability distributions over some hidden reality, then the
only way one could possibly understand a coherent superposition of quantum
states (so the argument goes) is as a convex combination of the associated
probability distributions with weights given by the amplitudes squared. In
particular, the distribution associated with the state $\sqrt{2}^{-1}\left(
\left| 0\right\rangle +\left| 1\right\rangle \right) $ must be a convex sum,
with equal weights, of the distributions associated with $\left|
0\right\rangle $ and $\left| 1\right\rangle .$ But given that in a
measurement of $\{\left| +\right\rangle ,\left| -\right\rangle \}$ the $%
\left| -\right\rangle $ outcome occurs with probability $1/2$ for both $%
\left| 0\right\rangle $ and $\left| 1\right\rangle ,$ if $\sqrt{2}%
^{-1}\left( \left| 0\right\rangle +\left| 1\right\rangle \right) $
corresponded to a convex sum of these possibilities, one would still expect
the $\left| -\right\rangle $ outcome to occur with probability $1/2,$ not
probability zero.

All this argument demonstrates, however, is a lack of imagination concerning
the interpretation of coherent superposition within an epistemic view. We
have already seen in Sec.~\ref{epistemicstates1} how in the toy theory one
can define some binary operations that are distinct from convex combination.
The possibility of representing coherent superposition and convex
combination differently within an epistemic view is what makes interference
understandable. This is made clear through the toy theory version of the
interference experiment discussed above. Recall from Eq.~(\ref%
{binaryoperationsmap}) of Sec.~\ref{epistemicstates1} that the toy theory
analogue of the coherent superposition $\sqrt{2}^{-1}\left( \left|
0\right\rangle +\left| 1\right\rangle \right)$ is $(1\vee 2) +_{1} (3\vee 4)$
which is simply $1\vee 3$. This is \emph{not} the equally weighted
probabilistic sum of the two epistemic states, which would be the epistemic
state $1\vee 2\vee 3\vee 4$. Thus, the analogue of the three experiments
are:
\begin{eqnarray*}
&&\text{(a) Prepare }
\begin{array}{l}
\includegraphics[width=15mm]{1or2.eps}%
\end{array}
\text{, then measure }
\begin{array}{l}
\includegraphics[width=15mm]{13vs24.eps}%
\end{array}
\\
&&\text{(b) Prepare }
\begin{array}{l}
\includegraphics[width=15mm]{3or4.eps}%
\end{array}
\text{, then measure }
\begin{array}{l}
\includegraphics[width=15mm]{13vs24.eps}%
\end{array}
\\
&&\text{(c) Prepare }
\begin{array}{l}
\includegraphics[width=15mm]{1or3.eps}%
\end{array}
\text{, then measure }
\begin{array}{l}
\includegraphics[width=15mm]{13vs24.eps}%
\end{array}
\text{.}
\end{eqnarray*}
It is straightforward to see that the probability distributions over the
outcomes are $(1/2,1/2)$ for (a), $(1/2,1/2)$ for (b), and $(1,0)$ for (c).
Thus, the empirical signature of interference is reproduced.


Interference phenomena have led interpreters of quantum theory to conclude
that whatever an equally weighted coherent superposition of two
possibilities might be, it is not the `or' of those possibilities nor the
`and' of those possibilities. This is certainly the case in the toy theory.
The coherent combination of a pair of disjoint pure epistemic states is
neither the `or' nor the `and' of those states, but rather a sampling of the
ontic states from each.

\section{Pairs of elementary systems}

\label{2systems}

\subsection{Epistemic states}

\label{epistemicstates2}

The simplest composite system is a pair of elementary systems. Since each
elementary system has four ontic states, the pair has sixteen ontic states.
We can represent the ontic states of the pair by conjunctions of the
possible ontic states of the constituents. Representing conjunction by `$%
\cdot $' (read as `and'), the sixteen possibilities are
\begin{eqnarray}
&&1\cdot 1,\text{ }1\cdot 2,\text{ }1\cdot 3,\text{ }1\cdot 4,2\cdot 1,\text{
}2\cdot 2,\text{ }2\cdot 3,\text{ }2\cdot 4,  \notag \\
&&3\cdot 1,\text{ }3\cdot 2,\text{ }3\cdot 3,\text{ }3\cdot 4,4\cdot 1,\text{
}4\cdot 2,\text{ }4\cdot 3,\text{ }4\cdot 4.
\end{eqnarray}
We can represent these graphically by a $4\times 4$ array of boxes, where
the rows represent the different ontic states of system $A,$ and the columns
represent the different ontic states of system $B.$ Specifically, we take
the box in the $j$th row from the bottom and $k$th column from the left to
represent the ontic state $j\cdot k.$%
\begin{equation}
\includegraphics[width=25mm]{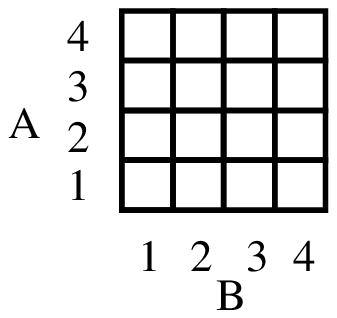}
\end{equation}
Since each system has two questions in a canonical set, the pair has four
questions in a canonical set. The knowledge balance principle ensures that
only two of these four questions may be answered in a state of maximal
knowledge. This corresponds to knowing the ontic state to be among four of
the sixteen possibilities. The pure epistemic states are therefore
disjunctions of four ontic states, for instance,
\begin{equation}
(1\cdot 3)\vee (1\cdot 4)\vee (2\cdot 3)\vee (2\cdot 4),  \label{example}
\end{equation}
which can be represented graphically by
\begin{equation}
\begin{array}{l}
\includegraphics[width=15mm]{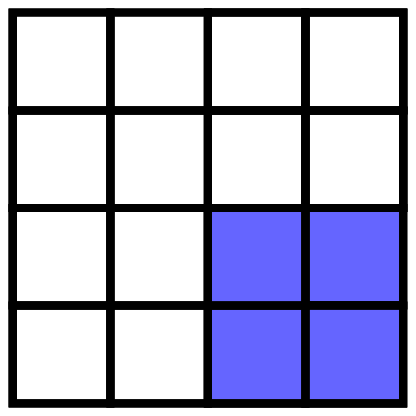}%
\end{array}%
\end{equation}
where we have dropped the labels on the rows and columns for convenience.

By applying the knowledge balance principle to each of the systems in the
pair individually, we obtain a further constraint: at most a single question
can be answered about the ontic state of each of the systems. Thus, an
epistemic state for $AB$ of the form
\begin{equation}
\begin{array}{l}
\includegraphics[width=15mm]{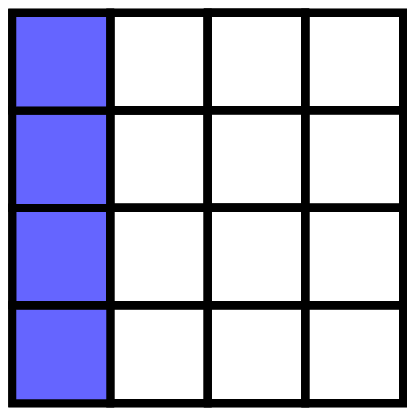}%
\end{array}%
\end{equation}
although satisfying the principle as it applies to the composite $AB$,
violates the principle as it applies to the system $B$ because the ontic
state of $B$ is known to be $1.$

The epistemic state for $AB$ of the form
\begin{equation}
\includegraphics[width=15mm]{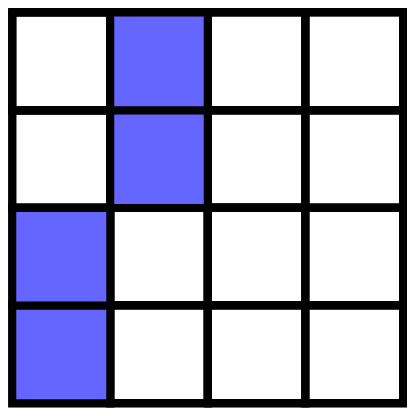}
\end{equation}
is also ruled out by application of the principle to the individual systems.
Here, it is not the marginals that are the problem. Rather, the problem is
that a reproducible measurement of $1\vee 2$ versus $3\vee 4$ on $A$, which
has outcome $1\vee 2$ for instance, allows one to rule out $3$ and $4$ as
possibilities for the ontic states of $A$ after the measurement, and, as
established earlier, causes the ontic state of $A$ to undergo an unknown
permutation: either ($12)(3)(4)$ or (1)(2)(3)(4). However, this leaves the
final epistemic state of $AB$ as
\begin{equation}
\includegraphics[width=15mm]{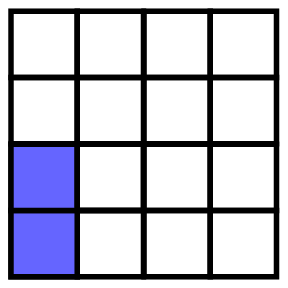}
\end{equation}
which corresponds to more knowledge about $AB\;$than is allowed by the
principle. We have here made use of the assumption that the transformation
that applies to $A$ is the same whether $A$ is correlated with $B$ or not,
since correlation is a feature of an observer's knowledge and therefore
cannot determine the nature of the physical transformation.

The full set of epistemic states that violate the knowledge balance
principle in some way are
\begin{eqnarray}
&&
\begin{array}{l}
\includegraphics[width=15mm]{2bitinvalid1.eps}%
\end{array}
\begin{array}{l}
\includegraphics[width=15mm]{2bitinvalid2.eps}%
\end{array}
\begin{array}{l}
\includegraphics[width=15mm]{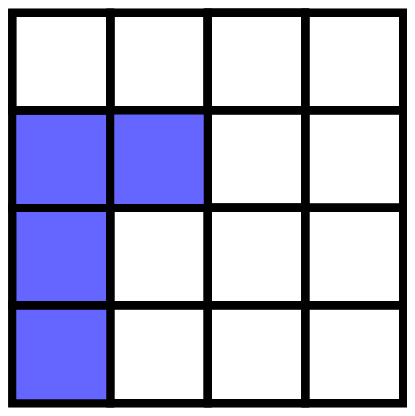}%
\end{array}
\begin{array}{l}
\includegraphics[width=15mm]{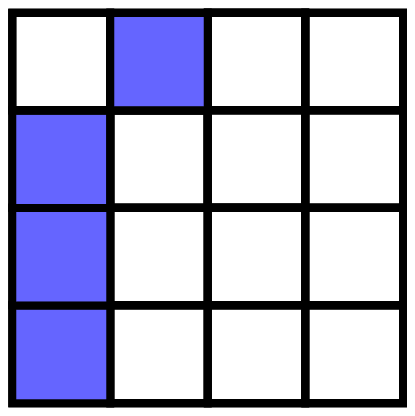}%
\end{array}
\notag \\
&&
\begin{array}{l}
\includegraphics[width=15mm]{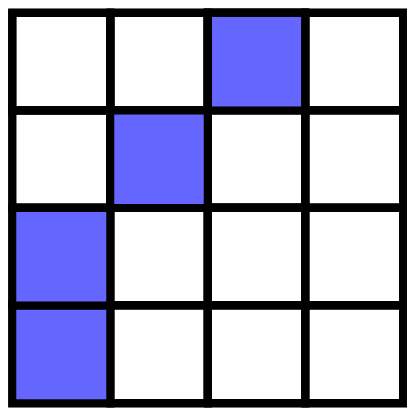}%
\end{array}
\begin{array}{l}
\includegraphics[width=15mm]{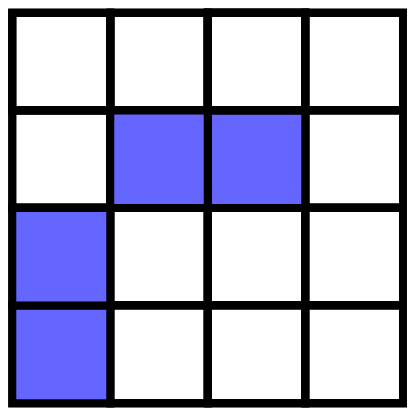}%
\end{array}
\begin{array}{l}
\includegraphics[width=15mm]{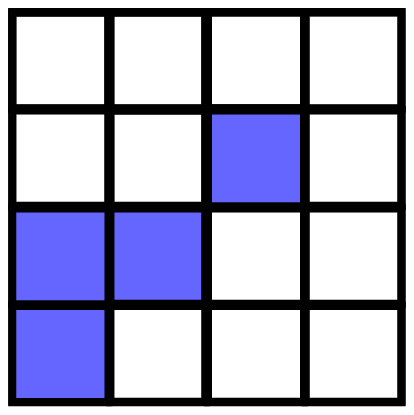}%
\end{array}
\begin{array}{l}
\includegraphics[width=15mm]{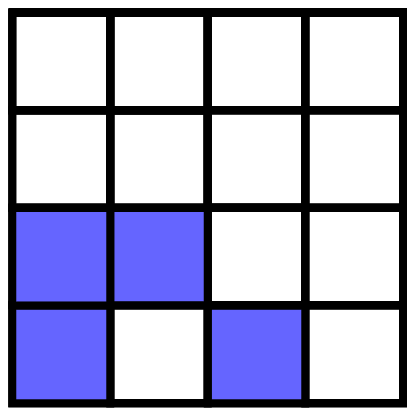}%
\end{array}
\label{violatingstates}
\end{eqnarray}
together with any epistemic state that can be obtained from one of these by
a permutation of $A$ and $B,$ or by any permutation of the rows and columns.

It follows that the valid states of maximal knowledge for a pair of systems
are of two types, represented as
\begin{equation}
\begin{array}{l}
\includegraphics[width=15mm]{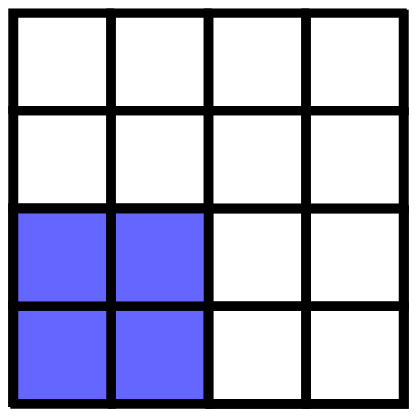}%
\end{array}
\begin{array}{l}
\includegraphics[width=15mm]{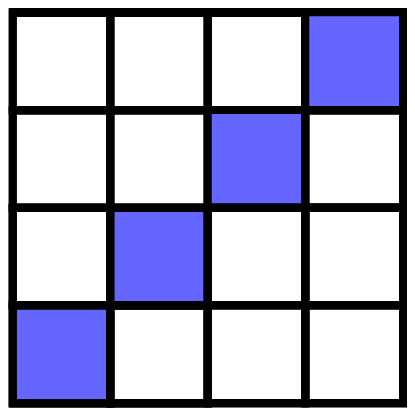}%
\end{array}%
\end{equation}
together with those that can be obtained from these by permutations of the
rows and columns. These are analogous in quantum theory to product states
and maximally entangled states respectively.

Epistemic states for composite systems can be classified according to
whether they describe correlations between the systems or not. An epistemic
state is said to describe correlations between a pair of systems if some
form of knowledge acquisition about one of the systems leads the bearer of
this epistemic state to refine their description of the other system. It is
clear that by these lights epistemic states of the first type are
uncorrelated while those of the second type are correlated.

The general form of the first type of epistemic state is
\begin{equation}
(a\vee b)\cdot (c\vee d),
\end{equation}
where $a,b,c,d\in \{1,2,3,4\}$ and $a\ne b,$ $c\ne d.$ These states are a
conjunction of states of maximal knowledge for each of the systems, and thus
satisfy the principle as it applies to the subsystems. Note that one can
distribute the conjunction over the disjunction to rewrite the epistemic
state as
\begin{equation}
(a\cdot c)\vee (a\cdot d)\vee (b\cdot c)\vee (b\cdot d),
\end{equation}
verifying that it is a disjunction of four ontic states and thus satisfies
the principle as it applies to the pair. Some examples of uncorrelated
epistemic states are
\begin{eqnarray}
(1\vee 2)\cdot (1\vee 2),  \notag \\
(1\vee 2)\cdot (2\vee 3),  \notag \\
(2\vee3)\cdot (1\vee 4),  \notag \\
(1\vee 3)\cdot (1\vee 3),
\end{eqnarray}
which are represented graphically as
\begin{equation}
\begin{array}{l}
\includegraphics[width=15mm]{2bit12and12.eps}%
\end{array}
\begin{array}{l}
\includegraphics[width=15mm]{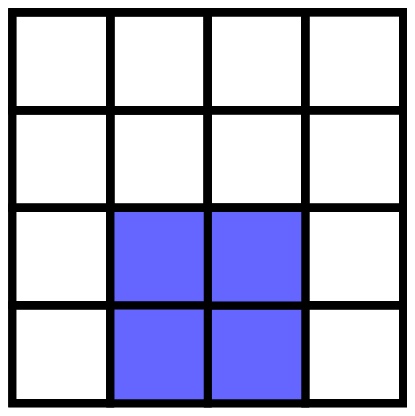}%
\end{array}
\begin{array}{l}
\includegraphics[width=15mm]{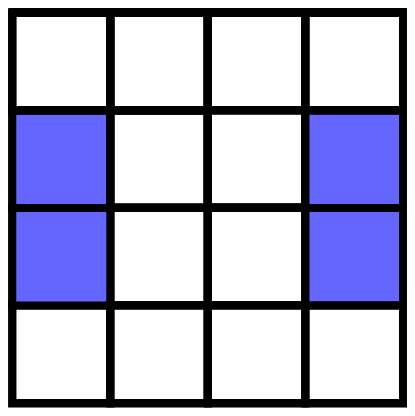}%
\end{array}
\begin{array}{l}
\includegraphics[width=15mm]{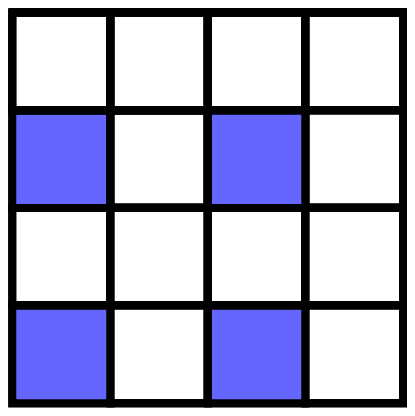}%
\end{array}
.
\end{equation}
By Eq.~(\ref{epistemicstatesmap}), these are analogous to the product states
$\left| 0\right\rangle $\strut $\left| 0\right\rangle $\strut , $\left|
0\right\rangle $\strut $\left| +i\right\rangle $\strut , $\left|
+i\right\rangle $\strut $\left| -i\right\rangle $\strut , and $\left|
+\right\rangle $\strut $\left| +\right\rangle $ respectively.

Since such an epistemic state is simply a ``product" of the marginals for $A$
and $B$, when a measurement on $A$ is implemented, only the marginal for $A$
is updated, and this occurs in precisely the manner described in Sec.~\ref%
{transformationaspect1}.
For instance, if the epistemic state for the composite is $(1\vee 2)\cdot
(2\vee3),$ and a measurement of $1\vee 4$ versus $2\vee 3$ on system $A$
finds the outcome $1\vee 4$, the final state is $(1\vee 4)\cdot (2\vee 3)$,
\begin{equation}
\begin{array}{l}
\includegraphics[width=15mm]{2bit12and23.eps}%
\end{array}
\rightarrow
\begin{array}{l}
\includegraphics[width=15mm]{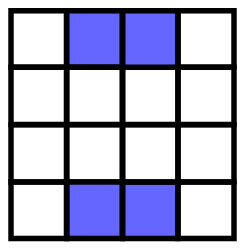}%
\end{array}
.
\end{equation}

The general form of the second type of allowed epistemic state is
\begin{equation}
(a\cdot e)\vee (b\cdot f)\vee (c\cdot g)\vee (d\cdot h),
\label{genericcorrelatedstate}
\end{equation}
where $a,b,c,d,e,f,g,h\in \{1,2,3,4\}$ and $a\ne b\ne $ $c\ne d,$ $e\ne f\ne
$ $g\ne h.$ Note that the marginal epistemic states for $A$ and $B$ are $%
1\vee 2\vee 3\vee 4$. Examples of such states are
\begin{eqnarray}
&&(1\cdot 1)\vee (2\cdot 2)\vee (3\cdot 3)\vee (4\cdot 4)  \notag \\
&&(1\cdot 2)\vee (2\cdot 3)\vee (3\cdot 4)\vee (4\cdot 1)  \notag \\
&&(1\cdot 4)\vee (2\cdot 3)\vee (3\cdot 1)\vee (4\cdot 2)  \notag \\
&&(1\cdot 4)\vee (2\cdot 1)\vee (3\cdot 3)\vee (4\cdot 2)
\end{eqnarray}
which are depicted as:
\begin{equation}
\begin{array}{l}
\includegraphics[width=15mm]{2bitcorrelated1.eps}%
\end{array}
\begin{array}{l}
\includegraphics[width=15mm]{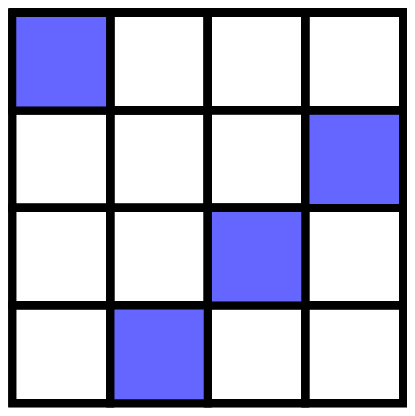}%
\end{array}
\begin{array}{l}
\includegraphics[width=15mm]{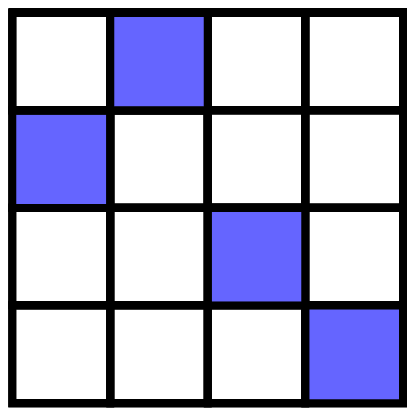}%
\end{array}
\begin{array}{l}
\includegraphics[width=15mm]{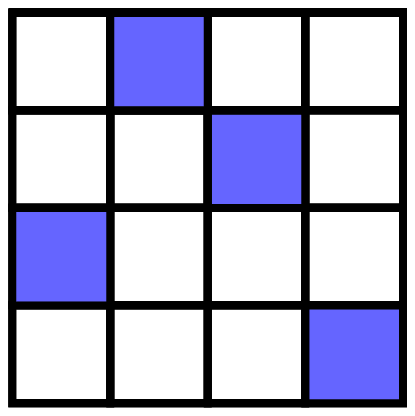}%
\end{array}
.
\end{equation}
For such epistemic states, nothing is known about the ontic states of the
individual systems, but everything is known about the relation between them.
In the first example, for instance, the two systems are known to be in the
same ontic state. In the second example, the ontic state of $B$ has an index
that is one greater (modulo 4) than the ontic state of $A$. In other words,
the ontic state of $B$ is related to the ontic state of $A$ by the
permutation (1234). In the third and fourth examples, the permutations are
(1423) and (142)(3) respectively. There is an epistemic state of this sort
for every permutation of the four ontic states, and thus 24 in all. These
are represented graphically by the 24 ways of filling only one box in every
row and column.

The following picture emerges. Unlike in classical theories, wherein one can
know the relation between two systems completely \emph{and} know their
individual ontic states, in the toy theory we have a trade-off. In a state
of maximal knowledge, \emph{either} one can know as much as is possible to
know about the individual ontic states of a pair of systems, in which case
one has an answer to a single question about each, yielding an uncorrelated
epistemic state, \emph{or} one can know as much as is possible to know about
the relation between the two systems, in which case one knows the answers to
two questions about their relation, yielding a correlated epistemic state.
It has been argued by Brukner, Zukowski and Zeilinger \cite{BZZ} (within the
context of a different interpretational approach) that this sort of account
captures the essence of entanglement.

It is worth noting that epistemic states of the second type are not only
correlated, they are \emph{perfectly} correlated, that is, for \emph{any}
form of knowledge acquisition about one of the systems, the description of
the other is refined. Further on, we shall consider epistemic states
describing imperfect correlations, for instance, in Eq.~(\ref%
{mixedcorrelatedstates}) of Sec.~\ref{mixedepistemicstates}.

It is useful to examine in detail how a perfectly correlated epistemic state
is updated if a reproducible measurement is implemented on one of the
subsystems. We describe this for a generic epistemic state of the form given
in Eq.~(\ref{genericcorrelatedstate}), and a generic measurement which
distinguishes $a\vee b$ from $c\vee d$. Upon obtaining the outcome $a\vee b$%
, the ontic states $c\cdot g$ and $d\cdot h$ for the composite are ruled
out. Thus one immediately sees that the marginal for $B$ after the
measurement will be $e\vee f$. Moreover, as discussed in Sec.~\ref%
{transformationaspect1}, the measurement causes system $A$ to undergo an
unknown permutation, namely, (a)(b)(c)(d) or (ab)(c)(d). The first case
yields $a\cdot e$ and $b\cdot f$ as possible final ontic states of the
composite, while the second case yields $b\cdot e$ and $a\cdot f$. The final
epistemic state is therefore the disjunction of these four possibilities,
which is simply $(a\vee b)\cdot (e\vee f)$. As an example, if the epistemic
state for $AB$ is initially $(1\cdot 4)\vee (2\cdot 3)\vee (3\cdot 1)\vee
(4\cdot 2)$ and a measurement of $2\vee 3$ versus $1\vee 4$ on $A$ finds the
outcome $1\vee 4$, the epistemic state is updated to $(1\vee 4)\cdot(2\vee 4)
$,
\begin{equation}
\begin{array}{l}
\includegraphics[width=15mm]{2bitcorrelated3.eps}%
\end{array}
\rightarrow
\begin{array}{l}
\includegraphics[width=15mm]{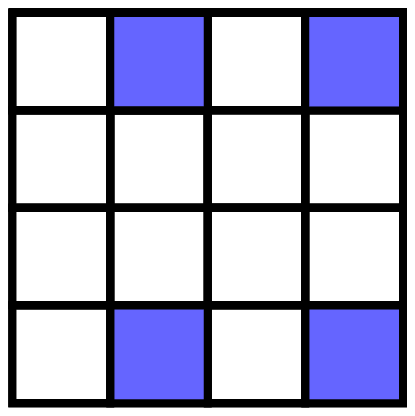}%
\end{array}
.
\end{equation}
The marginal for $B$ is updated from $1\vee 2\vee 3\vee 4$ to $2\vee 4$, so
there has been a refinement of the description of $B$ as a result of the
measurement on $A$.

\subsection{Remote steering}

\label{steering}

``Steering'' is the name given by Schr\"{o}dinger to the phenomenon that
lies at the heart of the Einstein-Podolsky-Rosen argument for the
incompleteness of quantum theory \cite{EPR}. We shall present the phenomenon
in a manner that is closer to the account given by Einstein in his
correspondence with Schr\"{o}dinger \cite{Einsteinletters} than to the
account found in the EPR paper. Alice and Bob each hold a qubit, denoted $A$
and $B$ respectively, and the pair $AB$ is described by the quantum state $%
\frac{1}{\sqrt{2}}\left( \left| 0\right\rangle \left| 0\right\rangle +\left|
1\right\rangle \left| 1\right\rangle \right) .$ Suppose Alice chooses to
measure the $\{\left| 0\right\rangle ,\left| 1\right\rangle \}$ basis (in a
reproducible way) on system $A.$ In this case, with probability 1/2 she
obtains the outcome $\left| 0\right\rangle $ and (following the standard
collapse rule) she updates the quantum state of the pair to $\left|
0\right\rangle \left| 0\right\rangle ,$%
\begin{equation}
\frac{1}{\sqrt{2}}\left( \left| 0\right\rangle \left| 0\right\rangle +\left|
1\right\rangle \left| 1\right\rangle \right) \rightarrow \left|
0\right\rangle \left| 0\right\rangle .  \label{rsq1}
\end{equation}
while with probability 1/2 she obtains the outcome $\left| 1\right\rangle $
and updates the quantum state of the pair to $\left| 1\right\rangle \left|
1\right\rangle ,$%
\begin{equation}
\frac{1}{\sqrt{2}}\left( \left| 0\right\rangle \left| 0\right\rangle +\left|
1\right\rangle \left| 1\right\rangle \right) \rightarrow \left|
1\right\rangle \left| 1\right\rangle .  \label{rsq2}
\end{equation}
On the other hand, if Alice chooses to measure the $\{\left| +\right\rangle
,\left| -\right\rangle \}$ basis on system $A$, then with probability 1/2
she obtains the outcome $\left| +\right\rangle $ and updates the quantum
state of the pair to $\left| +\right\rangle \left| +\right\rangle ,$%
\begin{equation}
\frac{1}{\sqrt{2}}\left( \left| 0\right\rangle \left| 0\right\rangle +\left|
1\right\rangle \left| 1\right\rangle \right) \rightarrow \left|
+\right\rangle \left| +\right\rangle ,  \label{rsq3}
\end{equation}
and with probability 1/2 she obtains the outcome $\left| -\right\rangle $
and updates the quantum state of the pair to $\left| -\right\rangle \left|
-\right\rangle ,$%
\begin{equation}
\frac{1}{\sqrt{2}}\left( \left| 0\right\rangle \left| 0\right\rangle +\left|
1\right\rangle \left| 1\right\rangle \right) \rightarrow \left|
-\right\rangle \left| -\right\rangle .  \label{rsq4}
\end{equation}
Note that for one choice of Alice's measurement, the final quantum state for
$B$ is either $\left| 0\right\rangle $ or $\left| 1\right\rangle $ whereas
for the other choice, it is either $\left| +\right\rangle $ or $\left|
-\right\rangle .$ In a 1935 paper discussing this phenomenon, Schr\"{o}%
dinger remarks (Ref. \cite{Schroedinger}, p. 555): ``It is rather
discomforting that the theory should allow a system to be steered or piloted
into one or the other type of state at the experimenter's mercy in spite of
his having no access to it.'' Indeed, if the quantum state is interpreted as
a state of reality, so that $\left| 0\right\rangle $,$\left| 1\right\rangle $%
,$\left| +\right\rangle $, and $\left| -\right\rangle $ are mutually
exclusive states of reality, then Alice's choice of measurement can directly
influence the reality in Bob's laboratory. If the collapse occurs
instantaneously, as is generally assumed, this would correspond to a
nonlocal influence. To be precise, it would lead to a failure of local
causality, in the sense defined by Bell \cite{Belllocalbeables}.

However, this example of the steering phenomenon does not imply a failure of
local causality if one adopts an epistemic view of quantum states.\footnote{%
Of course, a failure of locality \emph{is} implied by correlations that
violate Bell's inequalities \cite{Bell2}, and consequently there is nothing
analogous to such correlations in the toy theory. This will be discussed in
section \ref{phenomenathatdonotarise}.} Indeed, we now show that the
particular example of steering described above has a precise analogue in the
toy theory despite the fact that the latter is explicitly local. Here is how
it works. Alice and Bob each hold an elementary system, denoted $A$ and $B$
respectively, and Alice's epistemic state for the pair $AB$ is $(1\cdot
1)\vee (2\cdot 2)\vee (3\cdot 3)\vee (4\cdot 4).$ Suppose Alice implements
the reproducible measurement on $A$ that distinguishes $1\vee 2$ from $3\vee
4$. With probability $1/2$ she obtains the outcome $1\vee 2$ and, given the
results of the previous section, she must update her state of knowledge to $%
(1\vee 2)\cdot (1\vee 2).$ Graphically,
\begin{equation}
\begin{array}{l}
\includegraphics[width=15mm]{2bitcorrelated1.eps}%
\end{array}
\rightarrow
\begin{array}{l}
\includegraphics[width=15mm]{2bit12and12.eps}%
\end{array}
.  \label{rst1}
\end{equation}
On the other hand, if the outcome $3\vee 4$ occurs then Alice updates her
epistemic state for the pair to $(3\vee 4)\cdot (3\vee 4),$%
\begin{equation}
\begin{array}{l}
\includegraphics[width=15mm]{2bitcorrelated1.eps}%
\end{array}
\rightarrow
\begin{array}{l}
\includegraphics[width=15mm]{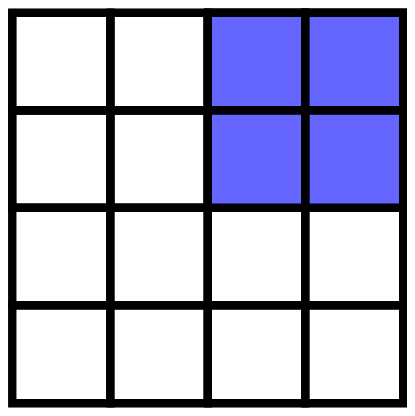}%
\end{array}
.  \label{rst2}
\end{equation}

Alice could also choose to implement the measurement that distinguishes $%
1\vee 3$ from $2\vee 4.$ She again obtains both outcomes with equal
probability. If the outcome is $1\vee 3,$ she updates her epistemic state
for the pair to $(1\vee 3)\cdot (1\vee 3)$%
\begin{equation}
\begin{array}{l}
\includegraphics[width=15mm]{2bitcorrelated1.eps}%
\end{array}
\rightarrow
\begin{array}{l}
\includegraphics[width=15mm]{2bit13and13.eps}%
\end{array}
,  \label{rst3}
\end{equation}
while if the outcome is $2\vee 4$, she updates her epistemic state to $%
(2\vee 4)\cdot (2\vee 4)$%
\begin{equation}
\begin{array}{l}
\includegraphics[width=15mm]{2bitcorrelated1.eps}%
\end{array}
\rightarrow
\begin{array}{l}
\includegraphics[width=15mm]{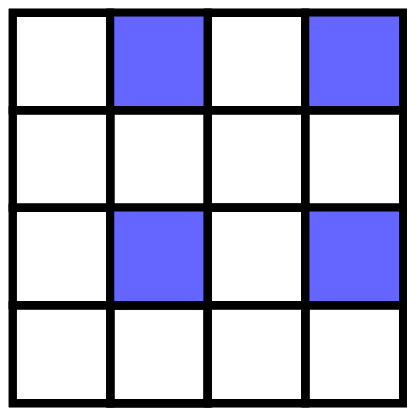}%
\end{array}
.  \label{rst4}
\end{equation}
Note that the right hand sides of Eqs.~(\ref{rst1})-(\ref{rst4}) are
precisely analogous to those of Eqs.~(\ref{rsq1})-(\ref{rsq4}) under the
mapping of Eq.~(\ref{epistemicstatesmap}).

The important point to note about the steering phenomenon in the toy theory
is that the choice of measurement at $A$ does not change the ontic state at $%
B.$ The measurement \emph{does} sometimes lead to a disturbance, but this is
a disturbance to the ontic state of $A.$ The only change associated with $B$
is to Alice's \emph{knowledge} of $B.$ Suppose, for instance, that the ontic
state of $AB$ was initially $1\cdot 1.$ Alice only knows that it is $(1\cdot
1)$ or $(2\cdot 2)$ or $(3\cdot 3)$ or $(4\cdot 4),$ and therefore initially
assigns the marginal $1\vee 2\vee 3\vee 4$ to $B.$ If she measures $1\vee 2$
versus $3\vee 4$ on $A,$ she will obtain the outcome $1\vee 2$ (by virtue of
$A$ being in ontic state 1), and she will update her marginal for $B$ to $%
1\vee 2.$ If, on the other hand, she measures $1\vee 3$ versus $2\vee 4,$
then she will obtain the outcome $1\vee 3$ (by virtue of $A$ being in ontic
state 1), and she will update her marginal for $B$ to $1\vee 3.$ In both
cases, $B$ remains in the ontic state $1$ throughout. Alice has simply
narrowed down the possibilities in two different ways.

\subsection{Epistemic states of nonmaximal knowledge}

\label{mixedepistemicstates}

One way to have nonmaximal knowledge of a pair of systems is to know nothing
about their ontic state. This corresponds to the epistemic state ($1\vee
2\vee 3\vee 4)\cdot $($1\vee 2\vee 3\vee 4),$ depicted as
\begin{equation}
\begin{array}{l}
\includegraphics[width=15mm]{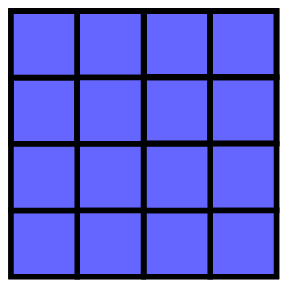}%
\end{array}
.
\end{equation}
\strut It is analogous to the completely mixed state for two qubits, $%
I/2\otimes I/2.$

In the case of a single elementary system, we found that knowing nothing was
the \emph{only} way to have nonmaximal knowledge. In the case of two
elementary systems, however, there are other possibilities. Since there are
four questions in the canonical set, one could know the answer to just one
of these, rather than two or none. This corresponds to an ontic base with
eight elements. These epistemic states are also highly constrained by the
knowledge balance principle. Their marginals must be valid epistemic states
for the individual subsystems, and they must be mapped to valid epistemic
states under the update rule for measurements on one of the subsystems. Some
examples of epistemic states of nonmaximal knowledge that contain eight
ontic states but still violate the principle in some way are
\begin{equation*}
\begin{array}{l}
\includegraphics[width=15mm]{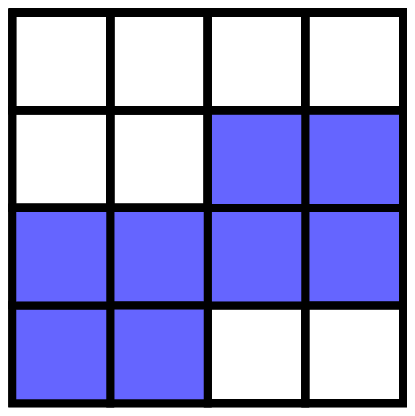}%
\end{array}
\begin{array}{l}
\includegraphics[width=15mm]{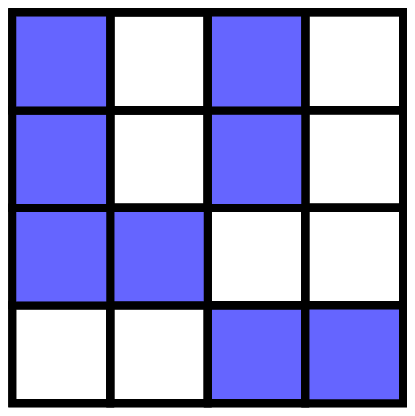}%
\end{array}
\begin{array}{l}
\includegraphics[width=15mm]{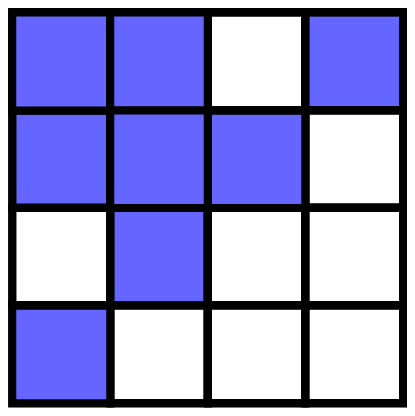}%
\end{array}
\begin{array}{l}
\includegraphics[width=15mm]{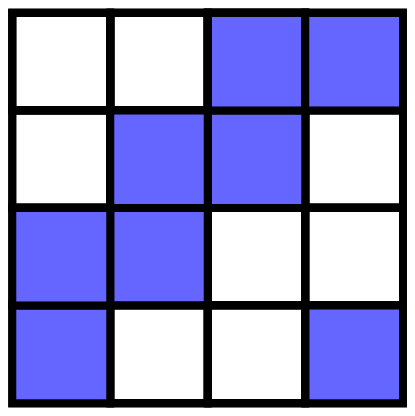}%
\end{array}%
\end{equation*}

The epistemic states of nonmaximal knowledge that abide by the principle are
again found to be of two types. The first type is essentially a conjunction
of a pure epistemic state for one system and a mixed epistemic state for the
other. Examples are $(3\vee 4)\cdot (1\vee 2\vee 3\vee 4)$ and $(1\vee 2\vee
3\vee 4)\cdot (1\vee 3),$ which are graphically depicted as
\begin{equation}
\begin{array}{l}
\includegraphics[width=15mm]{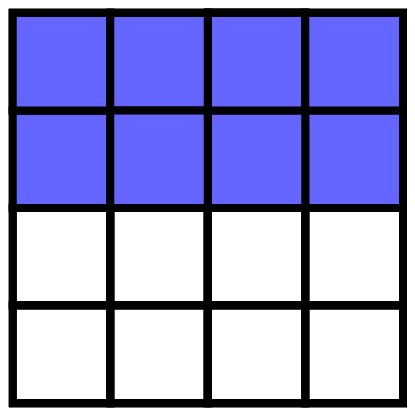}%
\end{array}
\begin{array}{l}
\includegraphics[width=15mm]{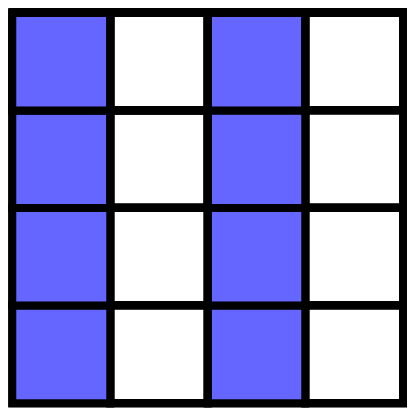}%
\end{array}
\text{ ,}
\end{equation}
and which are analogous to the density operators $\left| 1\right\rangle
\left\langle 1\right| \otimes I/2$ and $I/2\otimes \left| +\right\rangle
\left\langle +\right| $ respectively. These are uncorrelated states. The
second type of state is more interesting. Examples are $[(1\vee 2)\cdot
(1\vee 2)]\vee [(3\vee 4)\cdot (3\vee 4)]$ and $[(1\vee 3)\cdot (2\vee
4)]\vee [(2\vee 4)\cdot (1\vee 3)],$ which are depicted as
\begin{equation}
\begin{array}{l}
\includegraphics[width=15mm]{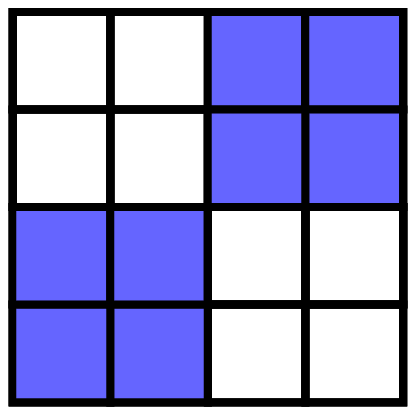}%
\end{array}
\begin{array}{l}
\includegraphics[width=15mm]{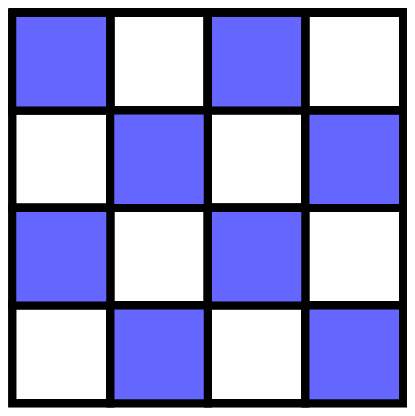}%
\end{array}
,  \label{mixedcorrelatedstates}
\end{equation}
which are analogous to the density operators $\frac{1}{2}\left|
0\right\rangle \left\langle 0\right| \otimes \left| 0\right\rangle
\left\langle 0\right| +\frac{1}{2}\left| 1\right\rangle \left\langle
1\right| \otimes \left| 1\right\rangle \left\langle 1\right| $ and $\frac{1}{%
2}\left| +\right\rangle \left\langle +\right| \otimes \left| -\right\rangle
\left\langle -\right| +\frac{1}{2}\left| -\right\rangle \left\langle
-\right| \otimes \left| +\right\rangle \left\langle +\right| .$ These are
correlated states. Measurements upon one system require an update of the
epistemic state of the other. For instance, if the initial epistemic state
is $[(1\vee 2)\cdot (1\vee 2)]\vee [(3\vee 4)\cdot (3\vee 4)]$ and a
measurement of $1\vee 2$ versus $3\vee 4$ is implemented on system $A$, then
the final epistemic state of the pair is $(1\vee 2)\cdot (1\vee 2)$ if the
outcome $1\vee 2$ is obtained, and $(3\vee 4)\cdot (3\vee 4)$ if the outcome
$3\vee 4$ is obtained. Note however that other measurements, such as a
measurement of $1\vee 3$ versus $2\vee 4,$ do not lead to an update of the
marginal of the nonmeasured system. Thus, the correlation is not perfect, in
the sense defined in Sec.~\ref{epistemicstates2}. The same sort of thing
occurs for the density operator $\frac{1}{2}\left| 0\right\rangle
\left\langle 0\right| \otimes \left| 0\right\rangle \left\langle 0\right| +%
\frac{1}{2}\left| 1\right\rangle \left\langle 1\right| \otimes \left|
1\right\rangle \left\langle 1\right| .$ There is correlation for
measurements in the $\{\left| 0\right\rangle ,\left| 1\right\rangle \}$
basis, but none for measurements in the $\{\left| +\right\rangle ,\left|
-\right\rangle \}$ basis. The existence of a distinction between epistemic
states exhibiting perfect correlations and those exhibiting imperfect
correlations is analogous to the existence of a distinction, in quantum
theory, between states that are said to be quantum correlated, or entangled,
and those that are said to be merely \emph{classically} correlated.\footnote{%
Again, this is not to say that perfect correlations in the toy theory have
\emph{all} the features of quantum correlations. In particular, they do not
violate any Bell inequality.}

\strut Note that states of nonmaximal knowledge are mixed states. Indeed,
they may be viewed as convex combinations of pure states in several
different ways. For instance,
\begin{eqnarray}
\begin{array}{l}
\includegraphics[width=15mm]{2bitmixed3.eps}%
\end{array}
&=&
\begin{array}{l}
\includegraphics[width=15mm]{2bit12and12.eps}%
\end{array}
+_{\text{cx}}
\begin{array}{l}
\includegraphics[width=15mm]{2bit34and34.eps}%
\end{array}
\notag \\
&=&
\begin{array}{l}
\includegraphics[width=15mm]{2bitcorrelated1.eps}%
\end{array}
+_{\text{cx}}
\begin{array}{l}
\includegraphics[width=15mm]{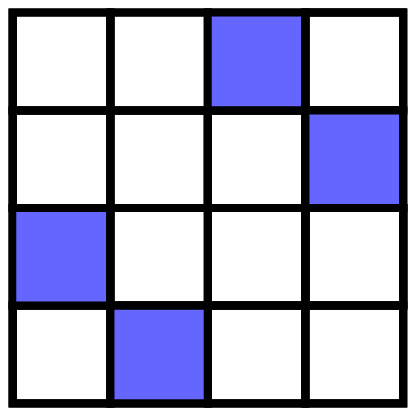}%
\end{array}
\notag \\
&=&
\begin{array}{l}
\includegraphics[width=15mm]{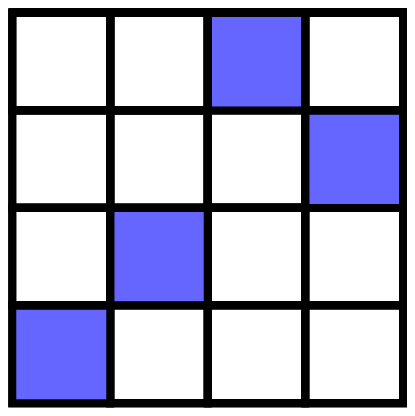}%
\end{array}
+_{\text{cx}}
\begin{array}{l}
\includegraphics[width=15mm]{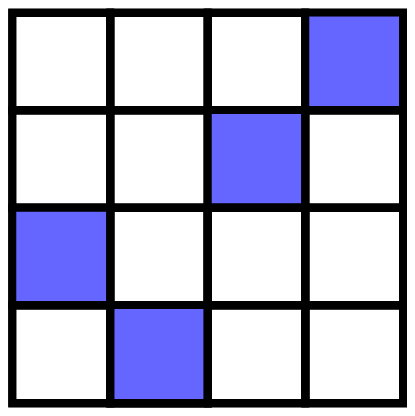}%
\end{array}%
.  \label{multiple decompositions}
\end{eqnarray}

Coherent binary operations on the pure epistemic states of a pair of systems
could also be defined, but we shall not do so here. Note that our
definitions of disjointness and compatibility and of the fidelity between
epistemic states, presented in the context of a single elementary system in
Sec.~\ref{epistemicstates1}, are applicable for composite systems as well.

\subsection{Transformations}

\label{transformations2}

The transformations that can be performed upon a pair of systems is a subset
of the set of permutations of the sixteen ontic states. It is a subset
because some permutations take valid epistemic states to invalid ones. For
instance, the permutation
\begin{equation}
\begin{array}{l}
\includegraphics[width=15mm]{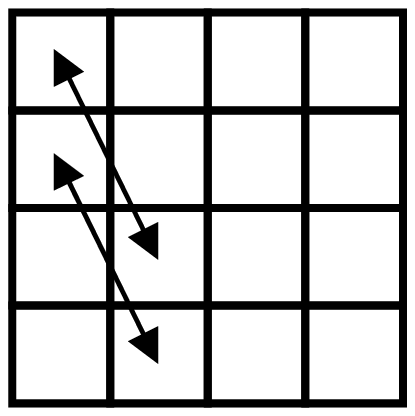}%
\end{array}%
\end{equation}

is invalid because it leads to the transition
\begin{equation}
\begin{array}{l}
\includegraphics[width=15mm]{2bit12and12.eps}%
\end{array}
\rightarrow
\begin{array}{l}
\includegraphics[width=15mm]{2bitinvalid1.eps}%
\end{array}%
.
\end{equation}

Independent permutations of each subsystem's ontic states are among the
subset of allowed permutations of the composite's ontic states. For
instance, the permutation $(12)(3)(4)$ on system $A$ yields
\begin{equation}
\begin{array}{l}
\includegraphics[width=17mm]{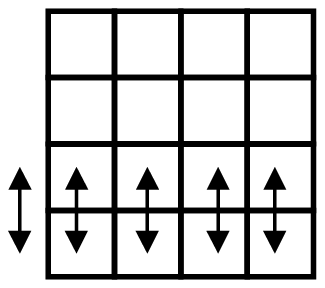}%
\end{array}%
,
\end{equation}%
and the permutation (12)(3)(4) on $A$ and (13)(2)(4) on $B$ yields
\begin{equation}
\begin{array}{l}
\includegraphics[width=17mm]{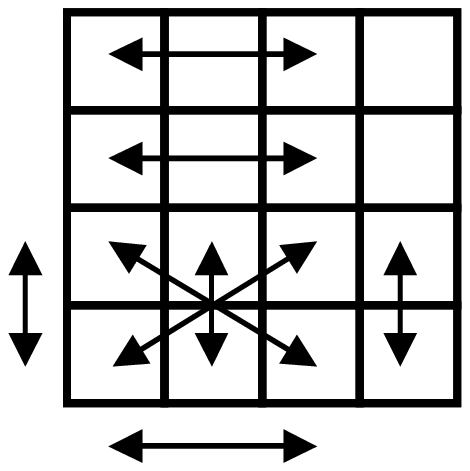}%
\end{array}%
.
\end{equation}%
Clearly, such local permutations cannot change the degree of correlation
between the systems: uncorrelated states are transformed into uncorrelated
states and correlated states are transformed into correlated states. These
permutations are analogous to local unitary operations in quantum theory,
which do not change the degree of entanglement. Other permutations \emph{can
}alter the degree of correlation, and are thus analogous to entangling
operations in quantum theory. One such permutation is
\begin{equation}
\begin{array}{l}
\includegraphics[width=15mm]{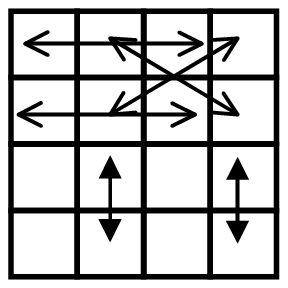}%
\end{array}%
\end{equation}%
which yields the transition
\begin{equation}
\begin{array}{l}
\includegraphics[width=15mm]{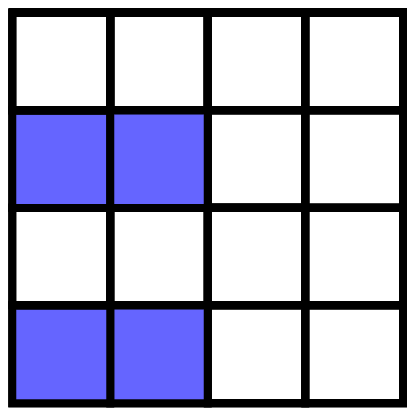}%
\end{array}%
\rightarrow
\begin{array}{l}
\includegraphics[width=15mm]{2bitcorrelated1.eps}%
\end{array}%
.
\end{equation}%
It is analogous to the controlled NOT operation for a pair qubits \cite%
{NielsenChuang}.

\subsection{No cloning}

\label{nocloning}

\strut Given the nature of transformations for a pair of
elementary systems, it is possible to prove the existence of a
no-cloning theorem. We begin by reviewing this theorem in the
context of the ontic view of quantum states~
\cite{WoottersZurek,Dieks}. A cloning process for a set of states
$\{\left| \psi _{i}\right\rangle \}$ is defined as a
transformation satisfying
\begin{equation}
\left| \psi _{i}\right\rangle \left| \chi \right\rangle \rightarrow \left|
\psi _{i}\right\rangle \left| \psi _{i}\right\rangle  \label{defncloning}
\end{equation}
for all $\left| \psi _{i}\right\rangle ,$ where $\left| \chi \right\rangle $
is an arbitrar\.{y} fixed state. The idea is that the quantum state of $A$
is unknown and the goal is to implement a transformation that leaves system $%
B$ in this unknown state.

The simplest case to consider is when the set contains two states $\{\left|
\psi _{1}\right\rangle ,$ $\left| \psi _{2}\right\rangle \}$. If $\left|
\psi _{1}\right\rangle $ and $\left| \psi _{2}\right\rangle $ are
nonorthogonal states, then the cloning process is impossible because it does
not preserve inner products, and so cannot be a unitary map. For instance, a
cloning process for the set $\{\left| 1\right\rangle ,\left| +\right\rangle
\}$ satisfies
\begin{eqnarray}
\left| 1\right\rangle \left| 0\right\rangle &\rightarrow &\left|
1\right\rangle \left| 1\right\rangle  \notag \\
\left| +\right\rangle \left| 0\right\rangle &\rightarrow &\left|
+\right\rangle \left| +\right\rangle ,
\end{eqnarray}
where we have taken the arbitrary initial state of system $B$ to be $\left|
0\right\rangle .$ The inner product squared between the two possible initial
states is $\left| \left\langle 1|+\right\rangle \left\langle
0|0\right\rangle \right| ^{2}=1/2,$ while the inner product squared between
the two possible final states is $\left| \left\langle 1|+\right\rangle
\left\langle 1|+\right\rangle \right| ^{2}=1/4.$

If one adopts an epistemic view of quantum states, then the question of
whether cloning is possible or not is a question of whether the epistemic
state that pertains to one system can be made to be also applicable to
another. It is \emph{not} the question of whether the \emph{ontic} state of
a system can be duplicated in another. Here is the manner in which it is
defined in the toy theory. A cloning process for a set of epistemic states $%
\{(a_{i}\vee b_{i})\}$ is defined as a transformation satisfying
\begin{equation}
(a_{i}\vee b_{i})\cdot (c\vee d)\rightarrow (a_{i}\vee b_{i})\cdot
(a_{i}\vee b_{i})
\end{equation}
for all epistemic states $a_{i}\vee b_{i}$ in the set, where $c\vee d$ is an
arbitrary initial epistemic state for $B$. The cloning process cannot be
implemented for nondisjoint epistemic states because it leads to a decrease
in the classical fidelity (defined in Sec.~\ref{epistemicstates1}) and
because this fidelity is preserved under permutations. This is easily
illustrated by an example. The analogue of the cloning process for the set $%
\{\left| 1\right\rangle ,\left| +\right\rangle \}$ is the cloning process
for the set $\{3\vee 4,1\vee 3\}$. We require
\begin{eqnarray}
(3\vee 4)\cdot (1\vee 2) &\rightarrow &(3\vee 4)\cdot (3\vee 4)  \notag \\
(1\vee 3)\cdot (1\vee 2) &\rightarrow &(1\vee 3)\cdot (1\vee 3).
\end{eqnarray}
where we have taken the arbitrary initial epistemic state for $B$ to be $%
1\vee 2.$ Graphically, this is depicted as
\begin{eqnarray}
\begin{array}{l}
\includegraphics[width=15mm]{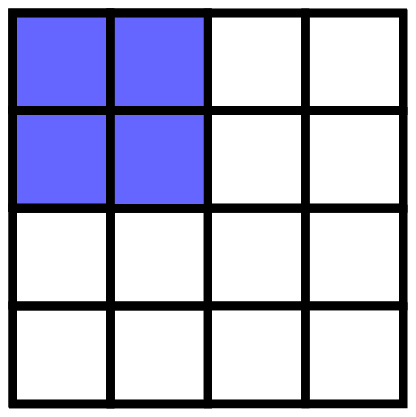}%
\end{array}
&\rightarrow &
\begin{array}{l}
\includegraphics[width=15mm]{2bit34and34.eps}%
\end{array}
\notag \\
\begin{array}{l}
\includegraphics[width=15mm]{2bit13and12.eps}%
\end{array}
&\rightarrow &
\begin{array}{l}
\includegraphics[width=15mm]{2bit13and13.eps}%
\end{array}
.
\end{eqnarray}
Imagine that the upper and lower diagrams are superimposed on top of one
another. It is then easy to see that there are two ontic states in the
overlap of the two possible initial epistemic states, namely, the ontic
states $3\cdot 1$ and $3\cdot 2$, whereas there is only one ontic state in
the overlap of the two possible final epistemic states, namely, $3\cdot 3$.
However, a permutation of the ontic states can only change the places
wherein the two epistemic states overlap, not the \emph{number} of places
where they overlap. Thus, the cloning process is not a permutation and
therefore is impossible in the toy theory \footnote{%
In this case, the fidelity between the initial epistemic states is 1/2
whereas between the final epistemic states it is 1/4.}

As it turns out, one does not actually need a restriction on knowledge to
obtain a no-cloning theorem. By defining cloning in terms of epistemic
states rather than ontic states, one obtains a no-cloning theorem for sets
of non-disjoint epistemic states, even in classical theories~{%
Fuchscloning,classicalnocloning}. A restriction on knowledge \emph{is }%
necessary however in order to have \emph{pure} states that are nondisjoint,
which is necessary if there is to be a no-cloning theorem for \emph{pure}
states. In this sense, the toy theory is more analogous to quantum theory,
vis-a-vis cloning possibilities, than any classical theory.

\subsection{No broadcasting}

\label{nobroadcasting}

Broadcasting is a process wherein one's state of knowledge about a system is
duplicated in the marginals of a pair of systems while allowing that these
systems might become correlated \cite{broadcasting}. This differs from
cloning insofar as the latter does not allow for such correlation. A
broadcasting process for a set of density operators $\{\rho _{i}\}$ has the
form
\begin{equation}
\rho _{i}\otimes \sigma \rightarrow W_{i}  \label{defnbroadcasting}
\end{equation}
where $W_{i}$ is a density operator for the composite $AB$ that has
marginals
\begin{equation}
\text{Tr}_{A}(W_{i})=\text{Tr}_{B}(W_{i})=\rho _{i}.
\label{constraintonmarginals}
\end{equation}
Broadcasting is only possible in quantum theory for a set of commuting
density operators \cite{broadcasting}.

The simplest case to consider is when the set $\{\rho _{i}\}$ contains only
pure states. Since nonorthogonal pure states do not commute, these cannot be
broadcast. However, one does not need the result of Ref.~\cite{broadcasting}
to see this. It follows immediately from the fact that any quantum state for
$AB$ with pure marginals is uncorrelated. That is, if $\rho _{i}$ is a pure
density operator, then the only way to satisfy Eq.~(\ref%
{constraintonmarginals}) is to have $W_{i}=\rho _{i}\otimes \rho _{i}.$ This
implies that the only way to duplicate a pure state of a system in the
marginals of a pair of systems is if the pair ends up in a product state.
But this is simply cloning, and cloning of nonorthogonal pure states is
impossible.

It may seem that the no-broadcasting theorem for pure quantum states tells
us nothing that wasn't already contained in the no-cloning theorem. However,
the former does capture something that the latter does not, namely, that
pure states can never arise as the marginals of a correlated state. Although
this is mathematically obvious given the formalism of quantum theory, it is
a conceptually significant fact in the context of the epistemic view, since
pure states are states of incomplete knowledge within the epistemic view,
and it is natural to expect states of incomplete knowledge to arise as the
marginals of a correlated state. Indeed, in a classical theory any state of
incomplete knowledge can arise as the marginal of a correlated state, and
consequently a broadcasting process exists for any set of epistemic states,
even though a cloning process need not. Specifically, one can broadcast any
set of epistemic states classically as follows. Measure the ontic state of
system $A$ (which can be done classically), prepare $B$ in this ontic state,
then forget the outcome of the measurement. The result is that the marginals
for $A$ and $B$ will be whatever the initial epistemic state for $A$ was,
and the two systems will also be known to be perfectly correlated.

Thus, at first sight, the no broadcasting theorem may seem to be at odds
with the view that quantum states are states of incomplete knowledge.
However, the toy theory provides an enlightening example of how broadcasting
of epistemic states may be ruled out. First note that the classical protocol
for achieving broadcasting does not work in a toy theory universe since one
cannot measure the ontic state of a system. The fact that \emph{no} protocol
can achieve broadcasting follows from the fact that in the toy theory, as in
quantum theory, pure epistemic states never arise as the marginals of
correlated states. This implies that a broadcasting process for pure states
is simply a cloning process, and as we saw in the previous section, such a
process is impossible in the toy theory. This proof has the same structure
as the one we provided for quantum theory. In this case, however, we can
identify the conceptual underpinnings of the fact that pure epistemic states
never arise as the marginals of correlated states.

Recall that the pure epistemic states in the toy theory are states of
maximal knowledge. Thus, if every system is described by a pure epistemic
state, one has maximal knowledge about each system. One cannot also have
knowledge of the relations among the systems (that is, a correlated
epistemic state), since this would exceed what is allowed by the knowledge
balance principle. For example, if the marginal epistemic states for a pair
of elementary systems are $a\vee b$ and $e\vee f$ respectively, then the
only possible epistemic state for the pair is $(a\vee b)\cdot (e\vee f)$
which is an uncorrelated state.

Simply assuming that maximal information is incomplete is not sufficient to
conclude that broadcasting of pure states will be impossible. For this, it
needs to be the case that having maximal knowledge of $A$ and maximal
knowledge of $B$ constitutes having maximal knowledge of the composite $AB.$
The knowledge balance principle ensures that this is the case in the toy
theory.

No-broadcasting for \emph{mixed} epistemic states also admits an analogue in
the toy theory, but we do not consider it here.

\subsection{Measurements}

\label{measurements2}

We now consider the measurements that may be performed upon a pair of
systems. Every partitioning of the set of sixteen ontic states into four
disjoint pure epistemic states yields a maximally informative measurement.
If all of these correspond to uncorrelated epistemic states, we have a
measurement such as
\begin{equation}
\includegraphics[width=15mm]{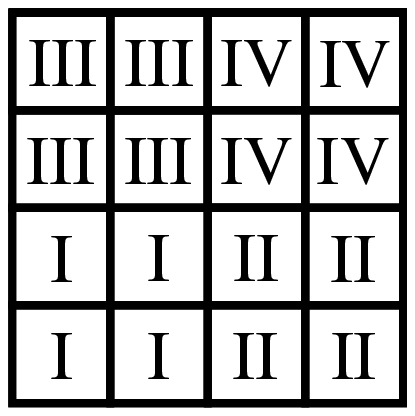}  \label{productmeas1}
\end{equation}
where the different roman numerals represent the different outcomes. This is
simply a conjunction of a measurement upon the first system and a
measurement upon the second, in this case a measurement of \{$1\vee 2$, $%
3\vee 4\}$ on both. We can represent the measurement on the pair by the
partitioning
\begin{equation}
\{S_{I},S_{II},S_{III},S_{IV}\},
\end{equation}
where
\begin{eqnarray}
S_{I} &=&(1\vee 2)\cdot (1\vee 2),  \notag \\
S_{II} &=&(1\vee 2)\cdot (3\vee 4),  \notag \\
S_{III} &=&(3\vee 4)\cdot (1\vee 2),  \notag \\
S_{IV} &=&(3\vee 4)\cdot (3\vee 4).
\end{eqnarray}
This is analogous to the product basis
\begin{equation}
\{\left| 0\right\rangle \left| 0\right\rangle ,\left| 0\right\rangle \left|
1\right\rangle ,\left| 1\right\rangle \left| 0\right\rangle ,\left|
1\right\rangle \left| 1\right\rangle \}
\end{equation}
in quantum theory. We call measurements of this type \emph{product
measurements. }

Other examples of product measurements can be obtained by permuting the rows
and columns in the above example, for instance
\begin{equation}
\begin{array}{l}
\includegraphics[width=15mm]{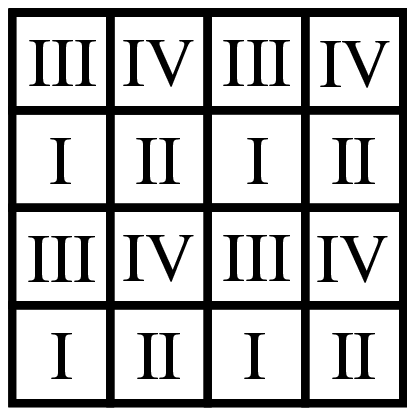}%
\end{array}
\begin{array}{l}
\includegraphics[width=15mm]{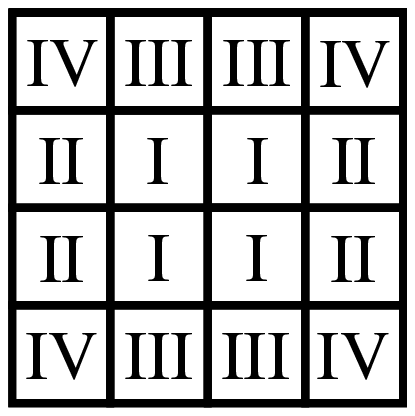}%
\end{array}
\begin{array}{l}
\includegraphics[width=15mm]{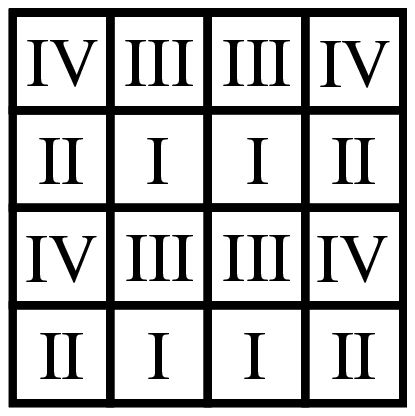}%
\end{array}%
\end{equation}
which are analogous to the bases
\begin{eqnarray}
&&\{\left| +\right\rangle \left| +\right\rangle ,\left| +\right\rangle
\left| -\right\rangle ,\left| -\right\rangle \left| +\right\rangle ,\left|
-\right\rangle \left| -\right\rangle \},  \notag \\
&&\{\left| +i\right\rangle \left| +i\right\rangle ,\left| +i\right\rangle
\left| -i\right\rangle ,\left| -i\right\rangle \left| +i\right\rangle
,\left| -i\right\rangle \left| -i\right\rangle \},  \notag \\
&&\{\left| +\right\rangle \left| +i\right\rangle ,\left| +\right\rangle
\left| -i\right\rangle ,\left| -\right\rangle \left| +i\right\rangle ,\left|
-\right\rangle \left| -i\right\rangle \},
\end{eqnarray}
respectively. Another form that a product measurement can take is:
\begin{equation*}
\begin{array}{l}
\includegraphics[width=15mm]{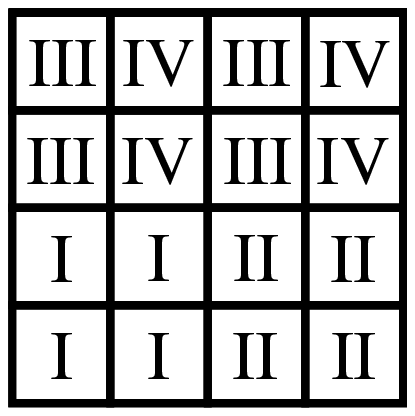}%
\end{array}
\end{equation*}
which is analogous to the product basis
\begin{equation}
\{\left| 0\right\rangle \left| 0\right\rangle ,\left| 0\right\rangle \left|
1\right\rangle ,\left| 1\right\rangle \left| +\right\rangle ,\left|
1\right\rangle \left| -\right\rangle \}
\end{equation}
in quantum theory.

If the disjoint epistemic states are all perfectly correlated, then we have
a measurement such as
\begin{equation}
\begin{array}{l}
\includegraphics[width=15mm]{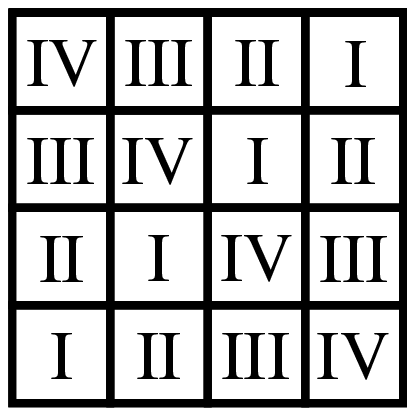}%
\end{array}
.  \label{Bell basis analogue}
\end{equation}
This example is analogous to the Bell basis
\begin{equation}
\{\left| \Phi ^{+}\right\rangle ,\left| \Phi ^{-}\right\rangle ,\left| \Psi
^{+}\right\rangle ,\left| \Psi ^{-}\right\rangle \},
\end{equation}
where
\begin{eqnarray}
\left| \Phi ^{\pm }\right\rangle &=& \sqrt{2}^{-1} (\left| 0\right\rangle
\left| 0\right\rangle \pm \left| 1\right\rangle \left| 1\right\rangle) ,
\notag \\
\left| \Psi ^{\pm }\right\rangle &=& \sqrt{2}^{-1} (\left| 0\right\rangle
\left| 1\right\rangle \pm \left| 1\right\rangle \left| 0\right\rangle) .
\label{Bell basis}
\end{eqnarray}
Other examples of measurements composed entirely of correlated epistemic
states include:
\begin{equation}
\begin{array}{l}
\includegraphics[width=15mm]{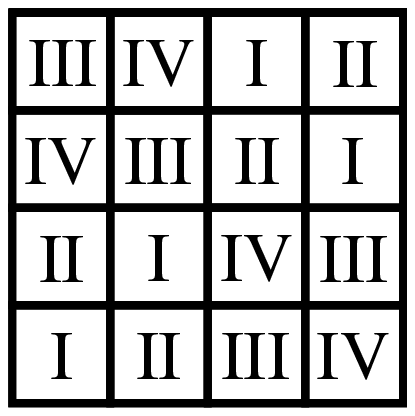}%
\end{array}
\begin{array}{l}
\includegraphics[width=15mm]{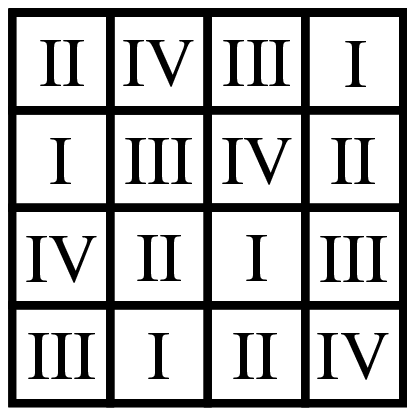}%
\end{array}
\begin{array}{l}
\includegraphics[width=15mm]{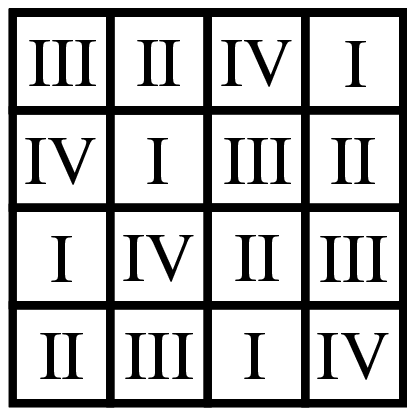}%
\end{array}
.
\end{equation}

There also exist measurements that are composed of some uncorrelated and
some correlated epistemic states, for instance:
\begin{equation}
\begin{array}{l}
\includegraphics[width=15mm]{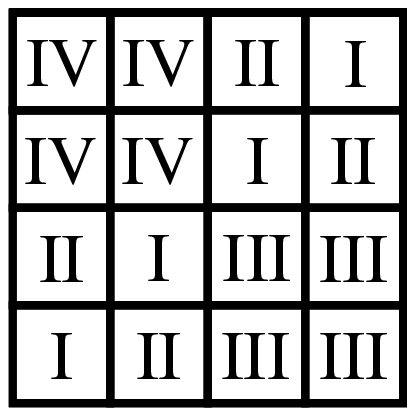}%
\end{array}%
\end{equation}
which is analogous to a measurement of the basis
\begin{equation}
\{\left| \Phi ^{+}\right\rangle ,\left| \Phi ^{-}\right\rangle ,\left|
0\right\rangle \left| 1\right\rangle ,\left| 1\right\rangle \left|
0\right\rangle \}.
\end{equation}

We call measurements that contain correlated epistemic states \emph{joint}
measurements since they cannot be implemented by separate measurements on
the individual systems. Note that joint measurements can only be implemented
directly if the systems are not spatially separated.

\subsection{Mutually unbiased measurements}

\label{MUBs}

In quantum theory, two bases are said to be \emph{mutually unbiased} if all
the pairwise fidelities between elements from the two bases have the same
value. Thus, bases $\{\left| \psi _{i}\right\rangle \}$ and $\{\left| \chi
_{j}\right\rangle \}$ are mutually unbiased if $\left| \left\langle \psi
_{i}|\chi _{j}\right\rangle \right| ^{2}$ is independent of $i$ and $j.$ The
number of mutually unbiased bases (MUBs) that can be constructed depends on
the dimensionality $d$ of the Hilbert space. For $d$ a power of a prime,
there are $d+1$ MUBs \cite{WoottersMUBs}.

For a single qubit, the number of MUBs that can be constructed is three. An
example of such a triplet of MUBs is
\begin{equation}
\{\left| 0\right\rangle ,\left| 1\right\rangle \},\{\left| +\right\rangle
,\left| -\right\rangle \},\{\left| +i\right\rangle ,\left| -i\right\rangle
\}.
\end{equation}
For a pair of qubits, one can construct five MUBs, an example being
\begin{eqnarray}
&&\{\left| 0\right\rangle \left| 0\right\rangle ,\left| 0\right\rangle
\left| 1\right\rangle ,\left| 1\right\rangle \left| 0\right\rangle ,\left|
1\right\rangle \left| 1\right\rangle \},  \notag \\
&&\{\left| +\right\rangle \left| +\right\rangle ,\left| -\right\rangle
\left| -\right\rangle ,\left| +\right\rangle \left| -\right\rangle ,\left|
-\right\rangle \left| +\right\rangle \},  \notag \\
&&\{\left| -i\right\rangle \left| -i\right\rangle ,\left| +i\right\rangle
\left| +i\right\rangle ,\left| -i\right\rangle \left| +i\right\rangle
,\left| +i\right\rangle \left| -i\right\rangle \},  \notag \\
&&\{I\otimes U\left| \Phi ^{+}\right\rangle ,I\otimes U\left| \Phi
^{-}\right\rangle ,I\otimes U\left| \Psi ^{+}\right\rangle ,I\otimes U\left|
\Psi ^{-}\right\rangle \}  \notag \\
&&\{I\otimes V\left| \Phi ^{+}\right\rangle ,I\otimes V\left| \Phi
^{-}\right\rangle ,I\otimes V\left| \Psi ^{+}\right\rangle ,I\otimes V\left|
\Psi ^{-}\right\rangle \}  \notag \\
\end{eqnarray}
where $U$ is the unitary map that corresponds to a clockwise rotation by $%
120^{\circ }$ about the $\hat{x}+\hat{y}+\hat{z}$ axis in the Bloch sphere,
and $V=U^{-1}$ \cite{WoottersWignerfunctions}.

As discussed previously, the analogue in the toy theory of a basis of states
is a set of disjoint epistemic states that yield a partitioning of the full
set of ontic states. We call two such partitionings \emph{mutually unbiased}
if all pairwise classical fidelities (defined in Sec.~\ref{epistemicstates1}%
) between elements from the two partitionings have the same value. For a
pair of pure epistemic states, the classical fidelity is proportional to the
number of ontic states they have in common. it follows that the number of
mutually unbiased partitionings (MUPs) for a single elementary system is
three:
\begin{equation}
\begin{array}{l}
\includegraphics[width=15mm]{12vs34.eps}%
\end{array}
\begin{array}{l}
\includegraphics[width=15mm]{13vs24.eps}%
\end{array}
\begin{array}{l}
\includegraphics[width=15mm]{23vs14.eps}%
\end{array}
.
\end{equation}
There exist sets of five MUPs for a pair of elementary systems, an example
being the set:
\begin{equation}
\begin{array}{l}
\includegraphics[width=15mm]{2bitproductmeas1.eps}%
\end{array}
\begin{array}{l}
\includegraphics[width=15mm]{2bitproductmeas2.eps}%
\end{array}
\begin{array}{l}
\includegraphics[width=15mm]{2bitproductmeas3.eps}%
\end{array}
\begin{array}{l}
\includegraphics[width=15mm]{2bitjointmeas2.eps}%
\end{array}
\begin{array}{l}
\includegraphics[width=15mm]{2bitjointmeas3.eps}%
\end{array}
.
\end{equation}
We conjecture that the number of MUPs for any number of elementary systems
is equal to the number of MUBs for the same number of qubits.

\subsection{Dense coding}

\label{densecoding}

By transmitting a single qubit from Alice to Bob, the most classical
information that can be communicated is one classical bit. This is a
consequence of Holevo's theorem. However, if Alice and Bob initially share
an entangled pair of qubits, then they can communicate two bits of classical
information by transmitting a single qubit. This is known as dense coding
\cite{densecoding}.

The phenomenon is surprising because it is unclear how adding a resource of
entanglement can possibly increase the capacity for communication given that
the distribution of the resource may occur at a time prior to Alice even
deciding which message she wishes to send, and need not involve any
transmission from Alice to Bob; they may both simply receive their half of
the entangled pair from a third party. The puzzle is sufficiently acute that
some have suggested that the additional bit of information travels backwards
in time through the channel that established the entanglement. A different
sort of resolution of the puzzle is suggested by the analogue of dense
coding in the toy theory. In order to see the extent of the analogy, we
begin by presenting the quantum protocol.

A pair of qubits, $A$ and $B$, described by the entangled state $\left| \Phi
^{+}\right\rangle =\sqrt{2}^{-1}(\left| 0\right\rangle \left| 0\right\rangle
+\left| 1\right\rangle \left| 1\right\rangle ),$ are distributed to Alice
and Bob ($A$ to Alice and $B$ to Bob). Depending on which of four messages, $%
00,$ 01, 10 or 11, Alice wishes to communicate to Bob, she implements one of
four transformations on $A$ corresponding to unitary operators $I,\sigma
_{z},\sigma _{x},$ and $i\sigma _{y}$ (where $\sigma _{x},\sigma _{y},$ and $%
\sigma _{z}$ are the Pauli operators \cite{NielsenChuang}). These
transformations map $\left| \Phi ^{+}\right\rangle $ to the four Bell
states, $\left| \Phi ^{+}\right\rangle ,\left| \Phi ^{-}\right\rangle
,\left| \Psi ^{+}\right\rangle ,$ and $\left| \Psi ^{-}\right\rangle $
respectively. Since these are orthogonal, they can be distinguished with
certainty. Thus, if Alice sends qubit $A$ to Bob, he holds the pair and can
perform a measurement of the Bell basis to determine which of the four
messages Alice wished to communicate. In this way, Alice has succeeded in
communicating two bits of information to Bob.

In the toy theory, it is also true that the transmission of a single
elementary system (without a shared resource of correlation) can only
communicate a single classical bit. The reason is as follows. Although a
single elementary system has four ontic states, allowing it to carry two
bits of classical information, Alice cannot prepare the system to be in
precisely one of these ontic states, nor can Bob measure which of the four
ontic states describes the system. The best Alice can do is to choose which
state of incomplete knowledge describes the system after her preparation
procedure. Thus, she could encode one bit of classical information by
choosing to perform one or the other of two preparations associated with the
epistemic states
\begin{equation}
\begin{array}{l}
\includegraphics[width=15mm]{1or2.eps}%
\end{array}
\text{and}
\begin{array}{l}
\includegraphics[width=15mm]{3or4.eps}%
\end{array}
.
\end{equation}
Bob can distinguish which preparation was implemented by subjecting the
system to the measurement of the form
\begin{equation}
\begin{array}{l}
\includegraphics[width=15mm]{12vs34.eps}%
\end{array}
.
\end{equation}
It should be clear that one classical bit is the most that Alice can
communicate to Bob in this way.

On the other hand, if Alice and Bob initially each hold one half of a pair
of elementary systems that are correlated, then Alice can communicate two
bits to Bob. Here is a protocol that achieves this. Suppose that initially
Alice holds an elementary system $A$ and Bob holds an elementary system $B,$
and these are known to be described by the epistemic state
\begin{equation}
\begin{array}{l}
\includegraphics[width=15mm]{2bitcorrelated1.eps}%
\end{array}
.
\end{equation}
Alice can, depending on which of four messages she wishes to send, perform
one of four permutations on $A,$ namely, $(1)(2)(3)(4),$ (12)(34), (13)(24),
or (14)(23), graphically,
\begin{equation}
\begin{array}{l}
\includegraphics[width=15mm]{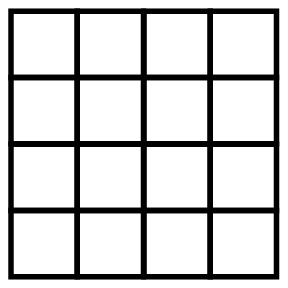}%
\end{array}
\begin{array}{l}
\includegraphics[width=15mm]{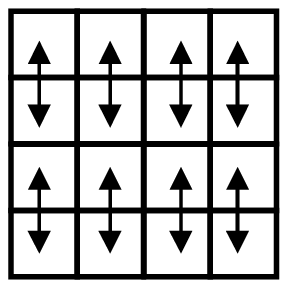}%
\end{array}
\begin{array}{l}
\includegraphics[width=15mm]{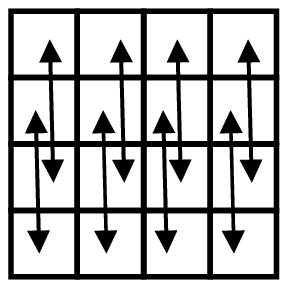}%
\end{array}
\begin{array}{l}
\includegraphics[width=15mm]{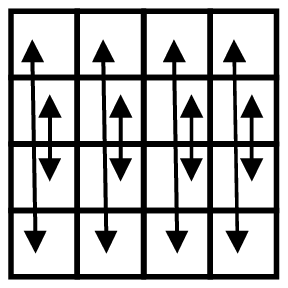}%
\end{array}
.
\end{equation}
These map the initial epistemic state to the four epistemic states
\begin{equation}
\begin{array}{l}
\includegraphics[width=15mm]{2bitcorrelated1.eps}%
\end{array}
\begin{array}{l}
\includegraphics[width=15mm]{2bitcorrelated8.eps}%
\end{array}
\begin{array}{l}
\includegraphics[width=15mm]{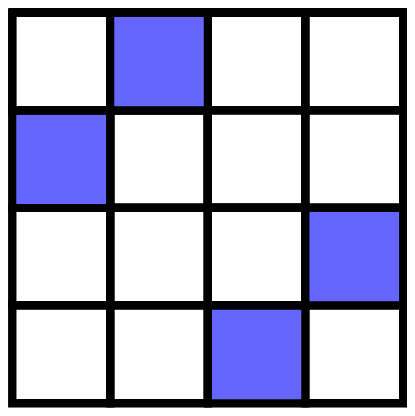}%
\end{array}
\begin{array}{l}
\includegraphics[width=15mm]{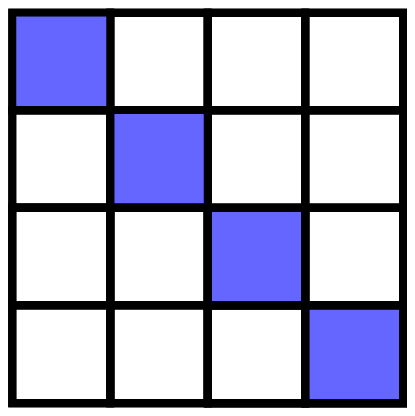}%
\end{array}
.
\end{equation}
There is a measurement that distinguishes these four epistemic states,
namely,
\begin{equation}
\begin{array}{l}
\includegraphics[width=15mm]{2bitBellmeas.eps}%
\end{array}
.
\end{equation}
Thus, if Alice sends the system $A$ to Bob, he can implement this
measurement and determine which of the four messages Alice wished to
communicate.

One can summarize these facts about the toy theory as follows. Every
elementary system has the inherent capability of encoding two bits of
classical information. However, the knowledge balance principle imposes a
restriction that prevents Alice and Bob from making use of this capacity
unless they initially share correlated systems. In a toy theory universe,
one \emph{cannot} come to know which of four possible ontic states describe
a single system, because one cannot learn two bits of information about a
single system. However, one \emph{can} come to know which of four possible
relations hold between two systems, because one can learn two bits of
information about a \emph{pair} of systems. Moreover, one can fix which of
these four relations holds by acting on just one of the systems.

Note that the toy theory yields an interesting new perspective on how to
compare quantum and classical information theories: rather than comparing a
single qubit to a single classical bit, as is conventionally done, the toy
theory suggests that it is more appropriate to compare a single qubit to
\emph{two }classical bits.

\subsection{\strut Nonmaximally informative measurements}

\label{nonmaximalmeasurements}

In addition to the measurements considered in Sec.~\ref{measurements2},
there are measurements that are not maximally informative. These do not
answer as many questions as are allowed by the knowledge balance principle.
An example of a \emph{product} measurement that is nonmaximally informative
is one that is trivial for one of the systems. For instance, the measurement
that is trivial on $B$ and distinguishes $1\vee 2$ from $3\vee 4$ on $A$ is
depicted by
\begin{equation}
\begin{array}{l}
\includegraphics[width=15mm]{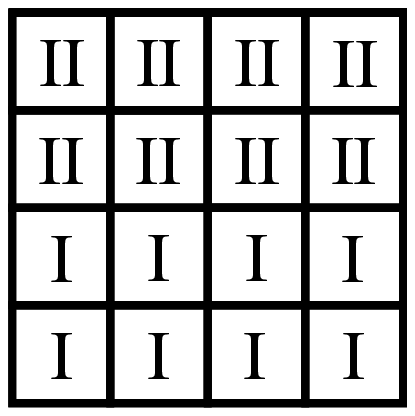}%
\end{array}
.
\end{equation}

\emph{Joint} measurements can also fail to be maximally informative. For
instance, the measurement
\begin{equation}
\begin{array}{l}
\includegraphics[width=15mm]{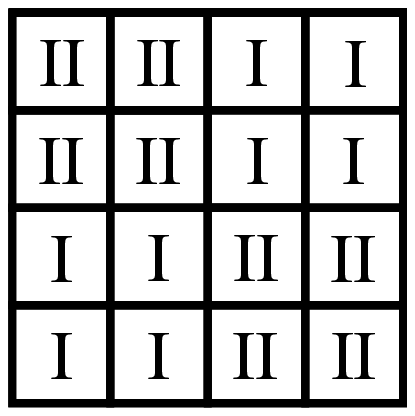}%
\end{array}
\label{non-conjunctive non-maximal}
\end{equation}
yields information about the relation between the two systems, but is not as
informative as it could be. Indeed, it can be obtained by coarse-graining of
the outcomes of the measurement described by Eq.~(\ref{productmeas1}) or the
one described by Eq.~(\ref{Bell basis analogue}).

\subsection{Measurement update rule}

\label{transformationaspect2}

The transformation associated with measurements that act upon a \emph{single}
elementary system has been described in Sec.~\ref{transformationaspect1}.
The transformation associated with a product measurement is simply a
conjunction of such transformations on the individual subsystems. For joint
measurements, we must see what the principle dictates. If the measurement is
maximally informative, then in order for it to be repeatable, the updated
epistemic state must assign zero probability to all the ontic states that
are inconsistent with the outcome that occurred. But no more ontic states
can receive probability zero without violating the principle. If the set of
ontic states consistent with the outcome of the measurement is $S,$ then the
updated epistemic state must have $S$ as its ontic base$.$ This must be true
regardless of the initial epistemic state. This implies that an unknown
permutation must occur as the result of the measurement, specifically, a
permutation drawn uniformly from any set that has the property of
randomizing the elements of $S$ (the set of all permutations of the elements
of $S$, for instance, has this property).

For instance, if the initial state is $(2\vee 3)\cdot (1\vee 2),$ and a
reproducible measurement of the form of Eq.~(\ref{Bell basis analogue})
(analogous to the Bell basis) finds the outcome $\mathrm{I}$, then we have
\begin{equation}
\begin{array}{l}
\includegraphics[width=15mm]{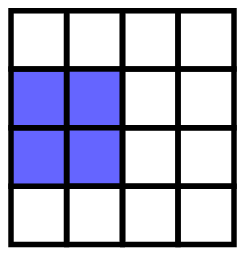}%
\end{array}
\rightarrow
\begin{array}{l}
\includegraphics[width=15mm]{2bitcorrelated1.eps}%
\end{array}
.
\end{equation}

The situation is more complicated for joint measurements that fail to be
maximally informative. Suppose, for instance, that the initial state is $%
(2\vee 3)\cdot (1\vee 2),$ and that a reproducible measurement of the form
of Eq.~(\ref{non-conjunctive non-maximal}) finds the outcome $\mathrm{I}$.
There are many update rules that are consistent with the reproducibility of
the measurement. For instance,
\begin{eqnarray}
\begin{array}{l}
\includegraphics[width=15mm]{2bit23and12.eps}%
\end{array}
&\rightarrow &
\begin{array}{l}
\includegraphics[width=15mm]{2bitmixed3.eps}%
\end{array}
\\
\begin{array}{l}
\includegraphics[width=15mm]{2bit23and12.eps}%
\end{array}
&\rightarrow &
\begin{array}{l}
\includegraphics[width=15mm]{2bit12and12.eps}%
\end{array}
\label{secondrule} \\
\begin{array}{l}
\includegraphics[width=15mm]{2bit23and12.eps}%
\end{array}
&\rightarrow &
\begin{array}{l}
\includegraphics[width=15mm]{2bitcorrelated8.eps}%
\end{array}
.
\end{eqnarray}
Indeed, any epistemic state appearing in Eq.~(\ref{multiple decompositions})
could be the final epistemic state while still yielding reproducibility.

It turns out that the update rule is not uniquely defined in this case. This
is completely analogous to quantum theory, wherein a reproducible
measurement that is not maximally informative can be associated with many
different maps. Which map applies depends on how the measurement is
implemented. One update rule, however, is particularly common. This is the
one wherein the final quantum state is the projection of the initial quantum
state into the subspace associated with the outcome that occurs. We can
define an analogous update rule in the toy theory: the final epistemic state
is the one with the highest classical fidelity with the initial epistemic
state. In this form, the analogy to the quantum update rule is apparent,
since the quantum state that is the projection of the initial quantum state
into the subspace associated with the outcome is the element of that
subspace that has the maximal inner product with the initial quantum state.
In the example provided above, this particular update rule corresponds to
the second rule we depicted, that is, Eq.~(\ref{secondrule}).

\section{Triplets of elementary systems}

\label{3systems}

\subsection{Epistemic states}

\label{epistemicstates3}

For three elementary systems, each of which has four ontic states, there are
$64$ ontic states in all. We can represent these by a $4\times 4\times 4$
grid of boxes, with the three systems labelled by $A$, $B$ and $C$.
\begin{equation}
\includegraphics[width=35mm]{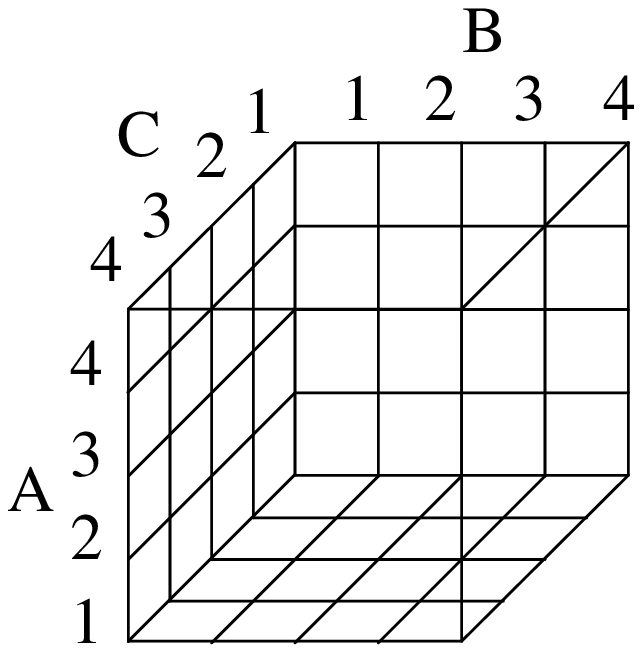}
\end{equation}
There are six yes/no questions in a canonical set for the three systems. In
a state of maximal knowledge, three questions are answered and three are
unanswered, which implies that a state of maximal knowledge contains eight
ontic states. Any pair of systems in the triplet must also abide by the
principle, so that the marginal distributions for the pairs must all be of
one of the forms described in Sec.~\ref{epistemicstates2}.

The pure epistemic states that are allowed by the knowledge balance
principle are of three types: (1) no correlations between any of the
systems, (2) correlations between one pair of the systems, and (3)
correlations between all three systems. We shall see that these are
analogous, respectively, to product states, products of a Bell state and a
pure state, and the so-called Greenberger-Horne-Zeilinger (GHZ) states \cite%
{GHZ}.

The uncorrelated epistemic states are of the form
\begin{equation}  \label{1v2a1v2a1v2}
(1\vee 2)\cdot (1\vee 2)\cdot (1\vee 2)
\end{equation}
to within local permutations. The marginals over any pair of systems are
pure uncorrelated epistemic states. For instance, the marginal on $AB$ is
simply $(1\vee 2)\cdot (1\vee 2).$ We can represent the epistemic states
graphically as collections of solid coloured $1\times 1\times 1$ blocks in
our $4\times 4\times 4$ grid, and the marginals as the shadows of these
blocks. For instance, the example of Eq.~(\ref{1v2a1v2a1v2}) is represented
graphically as follows:
\begin{equation}
\begin{array}{l}
\includegraphics[width=50mm]{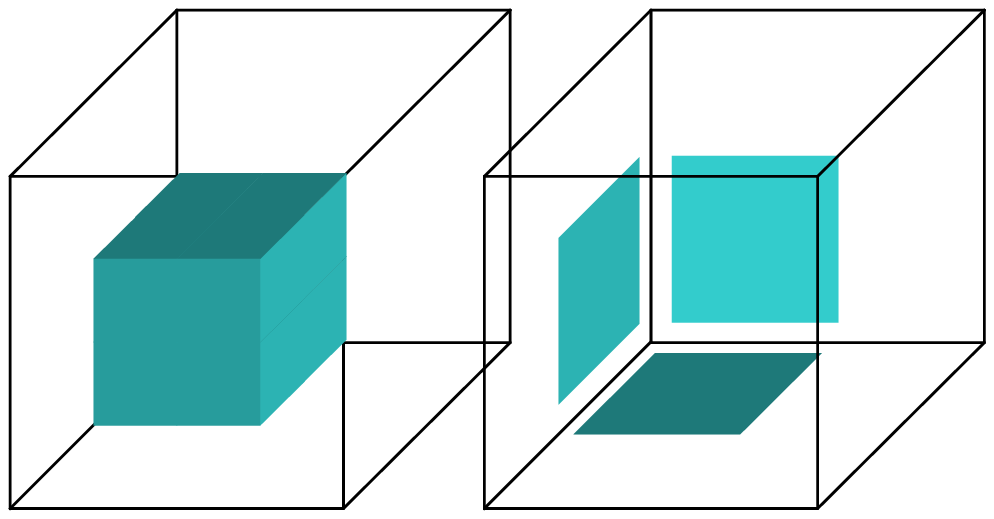}%
\end{array}
.
\end{equation}
This is analogous to the quantum state $\left| 0\right\rangle \left|
0\right\rangle \left| 0\right\rangle .$

The pair-correlated epistemic states are of the form
\begin{equation}
\left[ (1\cdot 1)\vee (2\cdot 2)\vee (3\cdot 3)\vee (4\cdot 4)\right] \cdot
(1\vee 2),
\end{equation}
to within local permutations.
This is represented graphically as
\begin{equation}
\begin{array}{l}
\includegraphics[width=50mm]{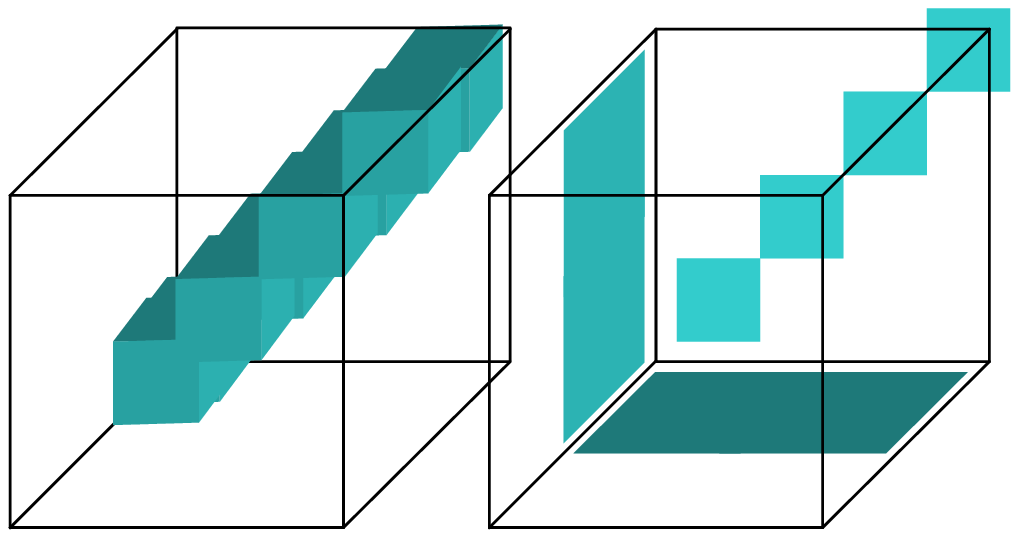}%
\end{array}
,
\end{equation}
and is analogous to $\frac{1}{\sqrt{2}}(\left| 0\right\rangle \left|
0\right\rangle +\left| 1\right\rangle \left| 1\right\rangle )\left|
0\right\rangle .$

The triplet-correlated epistemic states have the form
\begin{eqnarray}
&&(1\cdot 1\cdot 1)\vee (1\cdot 2\cdot 2)\vee (2\cdot 1\cdot 2)\vee (2\cdot
2\cdot 1)  \notag \\
&&\vee (3\cdot 3\cdot 3)\vee (3\cdot 4\cdot 4)\vee (4\cdot 3\cdot 4)\vee
(4\cdot 4\cdot 3)  \notag \\
\end{eqnarray}
to within local permutations. The marginals over every pair of elementary
systems are correlated mixed states. For the particular example we have
provided, they are all of the form $\left[ (1\vee 2)\cdot (1\vee 2)\right]
\vee \left[ (3\vee 4)\cdot (3\vee 4)\right] .$ This is represented
graphically as follows
\begin{equation}
\begin{array}{l}
\includegraphics[width=50mm]{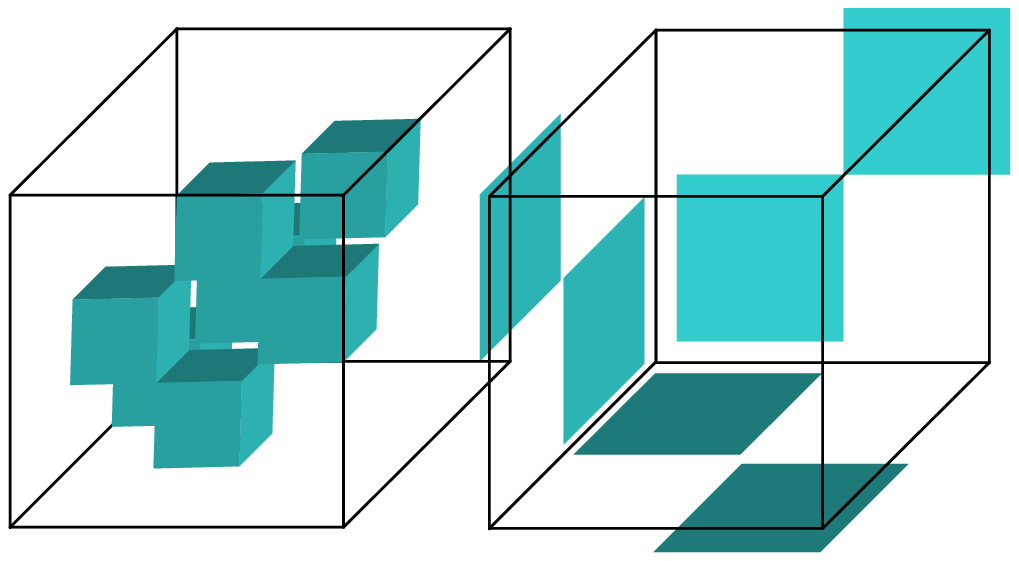}%
\end{array}
.
\end{equation}
This epistemic state is analogous to the GHZ state for three qubits of the
form $\frac{1}{\sqrt{2}}(\left| 0\right\rangle \left| 0\right\rangle \left|
0\right\rangle +\left| 1\right\rangle \left| 1\right\rangle \left|
1\right\rangle )$ which has marginals $\frac{1}{2}\left| 00\right\rangle
\left\langle 00\right| +\frac{1}{2}\left| 11\right\rangle \left\langle
11\right| $ over every pair of subsystems.

\subsection{The monogamy of pure entanglement}

\label{monogamy}

In quantum theory, a system can be pure entangled with only \emph{one} other
system. The reason is that if $A$ and $B$ are pure entangled, then the
reduced density operator over $AB$ is a pure state. However, for the
composite $AB$ to be entangled with another system, the reduced density
operator of $AB$ must be mixed. Consequently, there is no entanglement
between $AB$ and any other system, and thus no entanglement between $A$ and
any other system besides $B.$ This feature of pure state entanglement is
sometimes referred to as the \emph{monogamy }of entanglement \cite%
{CoffmanKunduWootters}.

From the epistemic perspective, the monogamy of pure entanglement is a
monogamy of perfect correlations. In both classical theories and the toy
theory, a pair of systems are perfectly correlated if one knows the precise
relation between their ontic states. Perfect correlations are monogamous if
a system can only be perfectly correlated with one other.


Classical statistical theories are polygamous when it comes to perfect
correlations. For instance, it is possible to know that three systems, $A,B$
and $C,$ are in precisely the same ontic state. In this case, $A$ is
perfectly correlated with $B$ and perfectly correlated with $C.$

The toy theory, however, forbids such polygamy. We demonstrate this in the
case of three elementary systems by supposing the contrary and deriving a
contradiction with the knowledge balance principle. Suppose three elementary
systems, $A,$ $B$ and $C,$ are all pairwise perfectly correlated. This would
imply that for every ontic state of $A$ there was associated a unique ontic
state of $B$ and a unique ontic state of $C.$ For instance, one way for $A,B$
and $C$ to be perfectly correlated would be if they were known to be in
precisely the same ontic state, that is, if the epistemic state was ($1\cdot
1\cdot 1)\vee (2\cdot 2\cdot 2)\vee (3\cdot 3\cdot 3)\vee (4\cdot 4\cdot 4).$
This epistemic state and its marginals are represented graphically as
follows:
\begin{equation}
\begin{array}{l}
\includegraphics[width=50mm]{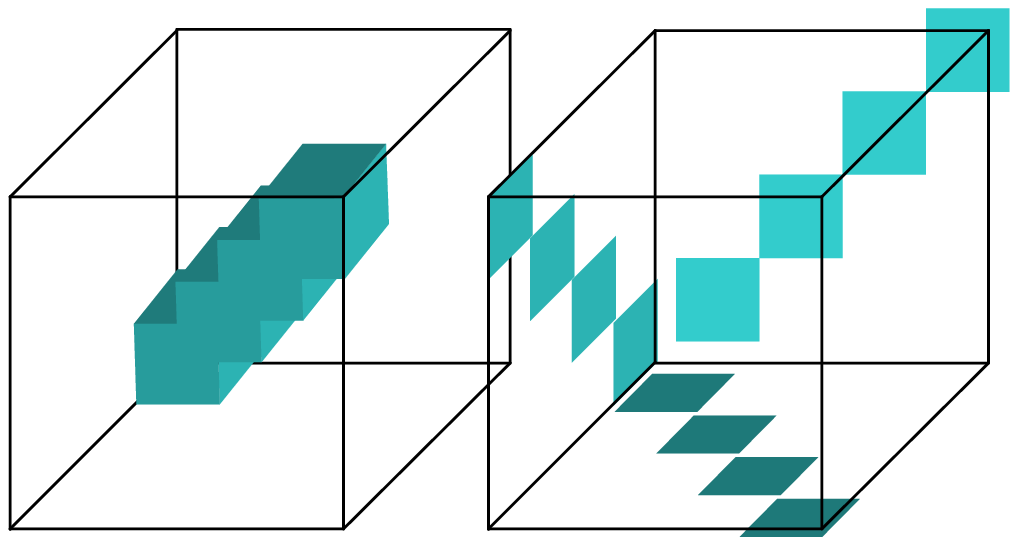}%
\end{array}
.
\end{equation}
But this is not one of the three valid forms of epistemic state for a
triplet of elementary systems. The problem is that it contains only four
ontic states rather than eight, which is the minimum number that is allowed
by the knowledge balance principle.

\subsection{Teleportation}

\label{teleportation}

A teleportation protocol in quantum theory makes use of a pair of qubits
that are maximally entangled and a classical channel in order to transfer
the applicability of an unknown quantum state from a qubit in Alice's
possession to one in Bob's possession \cite{teleportation}. We shall begin
by providing a standard account of how teleportation works, which assumes an
ontic view of quantum states.

A pair of qubits, denoted $A$ and $B,$ are prepared in the quantum state $%
\left| \Phi ^{+}\right\rangle =\sqrt{2}^{-1}\left( \left| 0\right\rangle
\left| 0\right\rangle +\left| 1\right\rangle \left| 1\right\rangle \right) ,$
after which $A$ is given to Alice and $B$ to Bob. A third party, Victor,
prepares another system, denoted $A^{\prime }$, in the quantum state $\left|
\psi \right\rangle ,$ and passes it to Alice. The identity of $A^{\prime }$%
's quantum state is unknown to Alice and Bob. Their task is to implement a
protocol that leaves $B$ in the quantum state $\left| \psi \right\rangle .$
The initial quantum state of $A^{\prime }AB$ is
\begin{equation}
\left| \psi \right\rangle \left| \Phi ^{+}\right\rangle
\end{equation}
It turns out that this can be rewritten as follows
\begin{eqnarray}
&&\frac{1}{2}\left| \Phi ^{+}\right\rangle \left| \psi \right\rangle +\frac{1%
}{2}(I\otimes \sigma _{z})\left| \Phi ^{+}\right\rangle \sigma _{z}\left|
\psi \right\rangle  \notag \\
&&+\frac{1}{2}(I\otimes \sigma _{x})\left| \Phi ^{+}\right\rangle \sigma
_{x}\left| \psi \right\rangle +\frac{1}{2}(I\otimes i\sigma _{y})\left| \Phi
^{+}\right\rangle i\sigma _{y}\left| \psi \right\rangle .  \notag \\
\end{eqnarray}
Note that $(I\otimes \sigma _{z})\left| \Phi ^{+}\right\rangle =\left| \Phi
^{-}\right\rangle ,$ $(I\otimes \sigma _{x})\left| \Phi ^{+}\right\rangle
=\left| \Psi ^{+}\right\rangle ,$ and $(I\otimes i\sigma _{y})\left| \Phi
^{+}\right\rangle =\left| \Psi ^{-}\right\rangle ,$ so that the states for $%
A^{\prime }A$ in this decomposition are just the elements of the Bell basis.
If Alice measures the Bell basis on $A^{\prime }A$ and obtains the outcome
associated with the unitary operator $U$, where $U\in \{I,\sigma _{z},\sigma
_{x},i\sigma _{y}\},$ then the quantum state of $A^{\prime }AB$ is updated
to
\begin{equation}
(I\otimes U)\left| \Phi ^{+}\right\rangle U\left| \psi \right\rangle .
\end{equation}
If she classically communicates to Bob the identity of $U$ -- only two bits
of information are required to do so -- then Bob can apply the inverse of $U$
to $B$ to leave $A^{\prime }AB$ in the state
\begin{equation}
(I\otimes U)\left| \Phi ^{+}\right\rangle \left| \psi \right\rangle .
\end{equation}
Thus, at the end of the protocol the quantum state of $B$ is $\left| \psi
\right\rangle ,$ as required, and $A^{\prime }A$ is left in one of the Bell
states. The protocol succeeds regardless of the identity of $\left| \psi
\right\rangle ,$ so Alice and Bob need not know its identity. Note that if
system $A^{\prime }$ is entangled with a fourth system, $C,$ then the
quantum state of $A^{\prime }C$ is transferred to $BC,$ which is known as
\emph{entanglement swapping}.

Teleportation is often thought to be surprising because it takes an infinite
amount of information to completely specify a quantum state, but somehow
this state can be transferred from one system to another given the
transmission of only two bits of classical information. This fact is only
surprising, however, if one takes on ontic view of the quantum state. \strut
From the perspective that quantum states are states of incomplete knowledge,
teleportation is a protocol wherein someone's \emph{knowledge} about the
system $A^{\prime }$ becomes applicable to the system $B,$ and, as we shall
see, a transfer of the applicability of a state of knowledge from $A$ to $B$
requires much less communication from Alice to Bob. We demonstrate this
first in the context of a classical theory, and then in the context of the
toy theory, where there is a strong analogue to the quantum protocol.

In a classical theory, a transfer of the applicability of a state of
knowledge is easily achieved. Suppose Victor describes system $A^{\prime }$
by some probability distribution $p(x)$ over its ontic states, and suppose
that Alice and Bob do not know the nature of this distribution. Nonetheless,
Alice can simply measure the ontic state of system $A^{\prime },$ then
communicate this information to Bob, and Bob can prepare system $B$ to be in
this particular ontic state. Assuming that Victor knows that they have
implemented this protocol, but does not know the outcome of Alice's
measurement, he will assign the marginal distribution $p(x)$ to $B.$
However, teleportation requires more than just getting the marginal
distribution for $B$ to reflect the initial marginal distribution for $%
A^{\prime }$ -- the correlations of $A^{\prime }$ to other systems must also
be reproduced. Since Victor initially describes $A^{\prime }$ as
uncorrelated with all other systems, he should, in the end, describe $B$ as
uncorrelated with all other systems. The protocol we have described does not
quite achieve teleportation because Victor ends up describing $B$ as
perfectly correlated with another system, namely, $A^{\prime }$. However,
this problem is easily fixed: Alice can simply randomize the ontic state of $%
A^{\prime }$ at the end of the protocol.

Note that this protocol only requires Alice to communicate to Bob an amount
of information that is sufficient to specify the ontic state of $A^{\prime }$%
, and this is in general much less than is required to specify Victor's
epistemic state (for instance, there might be a finite number of ontic
states, but an infinite number of epistemic states). Note also that this
classical protocol succeeds without Alice and Bob requiring any resource of
classical correlations.

In the toy theory, this protocol does not work since Alice cannot measure
the precise ontic state of $A^{\prime }$. Nonetheless, teleportation can be
achieved if Alice and Bob initially share correlated systems. Here is how it
works. Suppose Alice holds an elementary system $A$ and Bob holds an
elementary system $B$, and the pair is described by the epistemic state $%
(1\cdot 1)\vee (2\cdot 2)\vee (3\cdot 3)\vee (4\cdot 4)$ (analogous to $%
\left| \Phi ^{+}\right\rangle ).$ It is known that $A$ and $B$ are in the
same ontic state, but it is not known what this state is. A third party,
Victor, sends to Alice a system $A^{\prime },$ which he describes by the
epistemic state $a\vee b,$ where the identity of $a$ and $b$ are unknown to
Alice and Bob (analogous to the unknown state $\left| \psi \right\rangle )$.
Victor's initial epistemic state for $A^{\prime }AB$ is
\begin{equation}
(a\vee b)\cdot \left( (1\cdot 1)\vee (2\cdot 2)\vee (3\cdot 3)\vee (4\cdot
4)\right) .
\end{equation}
Although Alice cannot determine which of the four possible ontic states
applies to $A^{\prime },$ she can determine which of four relations hold
between $A^{\prime }$ and $A.$ For instance, she can determine whether the
permutation that relates $A$ to $A^{\prime }$ is (1)(2)(3)(4), (12)(34),
(13)(24), or (14)(23). This is simply the measurement of Eq.~(\ref{Bell
basis analogue})~(analogous to the Bell basis), applied to $A^{\prime }A.$

Suppose the permutation relating $A$ to $A^{\prime }$ is found to be $P.$
Since the permutation that related $B$ to $A$ prior to the measurement was
identity, one can conclude that the permutation that related $B$ to $%
A^{\prime }$ prior to the measurement was $P.$ Since Victor's state of
knowledge about the initial ontic state of $A^{\prime }$ (where by "initial"
we mean prior to the measurement) is $a\vee b$, it follows that, upon
learning the outcome of Alice's measurement, his state of knowledge about
the initial ontic state of $B$ is $P[a]\vee P[b]$ where $P[a]$ is the image
of $a$ under the permutation $P.$ Victor knows that Alice's measurement does
not cause a physical disturbance to $B,$ so his state of knowledge about its
\emph{final} ontic state (where by "final" we mean \emph{after }the
measurement) is also $P[a]\vee P[b].$ On the other hand, $A^{\prime }$ and $A
$ \emph{do} suffer an unknown permutation due to Alice's measurement, which
causes the epistemic state for the pair to be updated to $(1\cdot P[1])\vee
(2\cdot P[2])\vee (3\cdot P[3])\vee (4\cdot P[4]),$ the state appropriate
for finding $A$ to be related to $A^{\prime }$ by the permutation $P.$ Thus,
after Alice's measurement, Victor's epistemic state for $A^{\prime }AB$ is
\begin{equation*}
((1\cdot P[1])\vee (2\cdot P[2])\vee (3\cdot P[3])\vee (4\cdot P[4]))\cdot
(P[a]\vee P[b]).
\end{equation*}

To complete the teleportation protocol, Alice communicates the outcome of
her measurement, the permutation $P,$ to Bob$.$ Since there are four
possible outcomes, this requires Alice to communicate two bits of
information to Bob. Upon learning $P,$ Bob applies its inverse to $A^{\prime
}$. Thus, Victor's epistemic state at the end of the protocol is
\begin{equation*}
((1\cdot P[1])\vee (2\cdot P[2])\vee (3\cdot P[3])\vee (4\cdot P[4]))\cdot
(a\vee b).
\end{equation*}
The epistemic state $a\vee b,$ which was applicable to $A^{\prime }$ at the
start of the protocol, is now applicable to $B.$ The epistemic state for the
pair $A^{\prime }A$ is left as one of the four correlated epistemic states
that are analogous to the Bell states. This is the analogue of
teleportation. Had Victor initially known $A^{\prime }$ to have a particular
correlation with a fourth system, $C,$ then at the end of the protocol, he
would judge $B$ to have this correlation with $C.$ This is the analogue of
entanglement swapping.

It should be noted that even if Victor does not learn the outcome of Alice's
measurement, at the end of the protocol he still describes $B$ by the
epistemic state he initially assigned to $A^{\prime }$ (since he knows that
Bob will implement the inverse of $P,$ regardless of the identity of $P$).
Note also that we could have chosen a different initial correlated epistemic
state for $AB$, or a different basis of correlated epistemic states for
Alice's measurement on $A^{\prime }A$ (for instance, the basis associated
with the four permutations $(1)(2)(3)(4)$, $(1234)$, $(13)(24)$ and $(1432)$%
) and teleportation could still be achieved. These freedoms are analogous to
freedoms that are present in the quantum protocol.

In the toy theory, even though it takes more than two bits of information to
specify which of the six possible epistemic states applies (the analogue of
the continuum of quantum states), the applicability of an unknown epistemic
state can clearly be transferred from one system to another using only two
bits of information. This is precisely what is achieved by the protocol we
have described. In fact, a transfer of the applicability of a description
from one place to another doesn't require \emph{any} communication between
those locations. Suppose that Alice refrains from sending Bob the two bits
of information specifying the outcome of her measurement, but does send this
information to Victor. It is still the case that in one quarter of the
trials, namely those where Alice finds the ontic states of $A^{\prime }$ and
$A$ to be identical, the applicability of Victor's epistemic state is
transferred from system $A^{\prime }$ to system $B.$ The way in which the
applicability of an epistemic state is transferred from one system to
another is not by information transmission between the systems, but by
information transmission to the individual who is describing the systems.

This toy version of a teleportation protocol is essentially the
one provided by Hardy \cite{Hardydisentangling}, modulo the
arbitrariness in the choice of the set of permutations. Note,
however, that his goal was to distinguish teleportation from
nonlocality, not to provide an argument for the epistemic view of
quantum states. His point, that not every phenomenon involving
entanglement involves nonlocality, is reinforced by other examples
we have considered here, such as remote steering, dense coding and
the monogamy of entanglement.

\section{Further analogues}

\label{otheranalogues}

There are some more analogies between the toy theory and quantum theory
which we have opted not to present in detail. Nonetheless, it is worth
pointing out some of these, lest the phenomena in question be mistaken as
uniquely quantum.

The first such phenomenon is:

\begin{itemize}
\item The existence of unsharp measurements
\end{itemize}

In quantum theory, measurements on a system are typically associated with
projective-valued measures (PVMs) or, equivalently, Hermitian operators on
the system's Hilbert space. These are known as sharp measurements. There are
other sorts of measurements on a system, called unsharp measurements, which
are associated with positive-operator valued measures (POVMs) on the
system's Hilbert space. They may arise by a convex combination of sharp
measurements, or by coupling the system to an auxiliary system (called an
\emph{ancilla}) and performing a sharp measurement on the composite \cite%
{NielsenChuang}. In the toy theory, one can also contemplate certain convex
combinations of measurements, and one can implement effective measurements
on a system by coupling to an ancilla and measuring the composite. These
constitute unsharp measurements in the toy theory.

Similarly, we have:

\begin{itemize}
\item The existence of irreversible transformations
\end{itemize}

A transformation on a system in quantum theory is reversible if it is
associated with a unitary map on the space of density operators for that
system. An arbitrary transformation, however, is associated with a
completely positive trace-preserving linear map \cite{NielsenChuang}, which
can be non-unitary. These can arise as a result of a convex combination of
unitary maps, or by coupling the system to an ancilla and applying a
reversible transformation to the pair. Again, operations of these sorts are
allowed in the toy theory, and so irreversible transformations arise there
as well.

The main features of state discrimination tasks in quantum theory \cite%
{Cheflesreview} are also reproduced in the toy theory. For instance, we have

\begin{itemize}
\item No deterministic error-free discrimination of nonorthogonal states

\item The possibility of \emph{in}deterministic error-free discrimination of
nonorthogonal states (also known as unambiguous discrimination)
\end{itemize}

We also have

\begin{itemize}
\item No information gain without disturbance in discrimination of
nonorthogonal states
\end{itemize}

The latter phenomenon accounts for the possibility of key distribution in
quantum theory \cite{infogaintradeoff,BB84}. It follows that one expects key
distribution to be possible in the toy theory as well.

The toy theory also contains analogues of a few recently discovered
phenomena involving product bases, namely:

\begin{itemize}
\item The existence of locally indistinguishable product bases \cite%
{NLwoutentanglement}

\item The existence of unextendible product bases (that is, product bases
for which no additional product state can be found that is orthogonal to
every element of the basis) \cite{UPBs}
\end{itemize}

The first phenomena is sometimes referred to as ``nonlocality without
entanglement'' (in fact, this was the title of Ref.~\cite{NLwoutentanglement}%
). This description is perhaps inappropriate given that the toy theory is
explicitly local (in Bell's sense) and yet reproduces this phenomena.

Another interesting feature of quantum theory is:

\begin{itemize}
\item The fact that for every outcome of a maximally informative
measurement, there is a unique quantum state that yields that outcome with
certainty.
\end{itemize}

To be specific, in a measurement associated with the basis $\{\left| \psi
_{i}\right\rangle \},$ the only state that yields outcome $i$ with certainty
is $\left| \psi _{i}\right\rangle .$ This feature cannot be captured within
an epistemic approach if one allows for arbitrary probability distributions
over the ontic states. The reason is that within such an approach,
measurements correspond to a partitioning of the ontic states into disjoint
sets, and a particular outcome of a measurement is only certain to occur if
the epistemic state prior to the measurement has its ontic base within the
ontic base of that outcome. But if arbitrary distributions are allowed, then
there will be many epistemic states with the same ontic base. In the toy
theory, on the other hand, there are no two distinct epistemic states with
the same ontic base, since only uniform distributions are allowed. As a
result, there is a one-to-one correspondence between the outcomes of
maximally informative measurements and the pure epistemic states.

We have not exhausted the list of quantum phenomena that have analogues in
the toy theory, however the point should be clear: the toy theory captures a
good deal of quantum theory.

\section{Phenomena that are not reproduced}

\label{phenomenathatdonotarise}

There is a certain satisfaction in being able to reproduce quantum phenomena
in a theory that admits a simple interpretation. Nonetheless, what is even
more interesting is to identify the quantum phenomena that \emph{can't }be
reproduced by the toy theory, since these now present the greatest challenge
to the proponent of the epistemic view, and since these provide the best
clues for determining what other conceptual ingredients, besides the idea
that maximal information is incomplete, are at play in quantum theory.

Here are some features of quantum theory that are absent from the toy theory:

\begin{itemize}
\item Contextuality (i.e. the existence of a Kochen-Specker theorem \cite%
{Bell,KochenSpecker})

\item Nonlocality (i.e. the existence of a Bell theorem \cite{Bell2})

\item The continuum of quantum states, measurements, and transformations

\item The fact that convex combination and coherent superposition are full
rather than partial binary operations on the space of quantum states

\item The fact that two levels of a fundamentally three-level system behave
like a fundamentally two-level system

\item The possibility of an exponential speed-up relative to classical
computation, assuming certain computational problems are classically hard.
\end{itemize}

We shall consider each of these in turn.

\textbf{Contextuality and nonlocality. }The Kochen-Specker theorem \cite%
{Bell,KochenSpecker} and Bell's theorem \cite{Bell2} state that any hidden
variable theory that is local or noncontextual cannot reproduce all the
predictions of quantum theory. The toy theory is, by construction, a local
and noncontextual hidden variable theory. Thus, it cannot possibly capture
all of quantum theory. In the face of these no-go theorems, a proponent of
the epistemic view is forced to accept alternative possibilities for the
nature of the ontic states to which our knowledge pertains in quantum
theory. It is here that the novel conceptual ingredients are required. Note
that since nonlocality is an instance of contextuality \cite{Mermin}, the
latter can be considered as the more fundamental of the two phenomena.
Indeed, if quantum theory can be derived from a principle asserting that
maximal information is incomplete and some other conceptual ingredient, then
contextuality may be our best clue as to what this other conceptual
ingredient must be.

\textbf{Continuum of states, measurements and transformations. }
The finite cardinality of epistemic states, reproducible
measurements and reversible transformations in the toy theory is
due to the fact that these are associated respectively with
uniform distributions over, partitionings of, and permutations of
a finite set of ontic states.

Of course, by allowing non-uniform probability distributions over
the ontic states, measurements whose outcomes are determined only
probabilistically by the ontic states, and probabilistic
combinations of permutations, one could have a continuum of
distinct epistemic states, measurements and transformations over a
finite number of ontic states.

As it turns out, however, such a theory cannot reproduce the
predictions of quantum theory. The proof is as follows. For every
pair of pure quantum states, one can find a measurement and an
outcome of this measurement such that the first quantum state
assigns zero probability to this outcome while the second assigns
to it a non-zero probability. This implies that the first quantum
state does not contain in its ontic base any state that is in the
ontic base of the measurement outcome, while the second quantum
state does. It follows that every pure quantum state has an ontic
base that is unique to it, equivalently, every pure quantum state
picks out a unique subset of the ontic states. Since there are a
continuum of pure quantum states, there must be a continuum of
distinct subsets of the ontic states, which is only possible if
the full set
of ontic states is a continuum. This proof is due to Hardy \cite%
{Hardydiscomfort}. Since in practice one cannot verify that the number of
distinct pure quantum states is really a continuum as opposed to being very
large but finite, all one can strictly conclude is that there must be a very
large number of ontic states.

Given these considerations, one is immediately led to the idea of modifying
the toy theory to allow for a continuum of ontic states. In this case, there
would be an infinite number of questions in the canonical set. However, if
one were to keep the knowledge balance principle intact, this would imply
that a single elementary system was capable of encoding an infinite number
of classical bits, in contrast to the single classical bit that can be
encoded in a qubit. Thus, if this variant of the toy theory is to be
analogous to quantum theory, it must also involve some modification of the
foundational principle; we must consider other ways to guarantee that
knowledge is incomplete. An obvious choice is to assume that for $N$
systems, only $N$ of the infinite number of questions in a canonical set can
be answered. (There would obviously be a great imbalance of knowledge in
this case, since one's ignorance would always far exceed one's knowledge.)
This choice, however, has significant problems. The most notable is the fact
that there are an infinite number of mutually unbiased partitionings of an
infinite set and therefore such a theory would have an infinite number of
mutually unbiased measurements. By contrast, in quantum theory there are
only three mutually unbiased bases for a qubit, and five for a pair of
qubits. Other options for modifying the foundational principle are required
here.

\textbf{Binary operations on epistemic states. }We have seen that there are
two types of binary operations defined for epistemic states in the toy
theory, analogous to convex combinations and coherent superpositions of
quantum states. However, these operations are partial; they are not defined
for every pair of epistemic states.

It might therefore seem desirable to close the set of epistemic states in
the toy theory under convex combination with arbitrary probability
distributions. In this case, the set of allowed epistemic states for a
single elementary system would have the shape of an octahedron in the Bloch
sphere picture. Hardy's toy theory, for instance, has this feature \cite%
{Hardydisentangling}. Such a variant of our toy theory has also
been considered by Halvorson \cite{Halvorson}. However, there is
an important sense in which such a theory is \emph{less} analogous
to quantum theory than the one presented in this paper. The toy
theory shares with quantum theory the feature that every mixed
state has multiple convex decompositions into pure states, whereas
in this modified version, there are many mixed states that have
unique decompositions. Similarly, in the toy theory, as in quantum
theory, every mixed state has a ``purification'' -- a correlated
state between the system of interest and another of equal size
such that the marginal over the system of interest is equal to the
mixed state in question -- whereas in the modified version, there
are many mixed states that do not.

The problem with the modified theory is that although convex
combination has been extended to a full binary operation rather
than a partial one, the coherent binary operations have not been
so extended. Moreover, although one has allowed arbitrary weights
in the convex combinations, one has not allowed the analogue of
arbitrary amplitudes and phases for the coherent binary
operations. It is likely that a better analogy with quantum theory
can be obtained only if \emph{both} operations are generalized.
Unfortunately, it is unclear how to do so in a conceptually
well-motivated way.

\textbf{Embedding two-level systems in three-level systems. }The
toy theory we have described does not contain anything analogous
to a three-level quantum system (called a ``qutrit" in quantum
information theory). Nonetheless, a variant of the toy theory
does. One simply needs to change the measure of knowledge to one
that refers to ternary questions (having three possible answers)
rather than binary questions. We can then introduce canonical sets
of ternary questions, and measure knowledge in terms of these. The
knowledge balance principle then dictates that in a state of
maximal knowledge, the maximum number of ternary questions for
which the answer is known must equal the number for which the
answer is unknown. The simplest possible system one can consider
is completely specified by the answers to a pair of ternary
questions and thus has nine ontic states. In a state of maximal
knowledge one has the answer to one of these questions, which
corresponds to knowing that the system is in one of three ontic
states.

Although this variant of the toy theory does a good job of reproducing
quantum phenomena involving qutrits, it cannot be combined, in any obvious
way, with the original toy theory. For instance, two disjoint epistemic
states of a toy qutrit are not isomorphic to two disjoint epistemic states
of a toy qubit, since the former involve six ontic states, and the latter
four. This is in contrast to quantum theory, where two levels of a
fundamentally three-level system are isomorphic to a fundamentally two-level
system \footnote{%
The existence of such an isomorphism is one of the axioms in Hardy's
axiomatization of quantum theory \cite{Hardyaxioms}.}.

Similarly, a pair of qubits is described in the same way as a fundamentally
4-level system in quantum theory. We could define a variant of the toy
theory involving tertiary questions (having four answers), which would yield
an analogue of a fundamentally 4-level system, but this theory would be
distinct from the original toy theory applied to a pair of elementary
systems. For instance, in the original toy theory a pair of systems must
satisfy the knowledge balance principle at the level of the pair and at the
level of the individual systems, but nothing analogous to the latter
constraint occurs in the variant involving tertiary questions.

\textbf{Exponential speed-ups in computation. }If it is indeed the case that
quantum computers offer an exponential speed-up over classical computers for
certain computational problems \cite{NielsenChuang} (we currently do not
have a \emph{proof} that these problems are in fact difficult for a
classical computer), then such a speed-up would be a feature of quantum
theory that is not reproduced by the toy theory. This is clear since the toy
theory can be efficiently simulated classically. $N$ elementary system of
the toy theory can be modelled by $2N$ classical bits and every operation in
the toy theory has a counterpart in the classical model since the toy theory
involves a \emph{restriction}, relative to the full classical model,\emph{\ }%
on the permissible preparations, transformations and measurements.
Thus, if quantum theory \emph{does} offer an algorithmic speed-up,
this is likely to be connected is some way to the other phenomena
that the toy theory fails to reproduce, such as the contextuality
and nonlocality of quantum theory. In this vein, note that some
quantum information-processing tasks that offer an advantage over
their classical counterparts have already been
shown to have such a connection, specifically, random access codes \cite%
{Galvao} and communication complexity problems \cite{BZPZ}.

A distinction between those quantum phenomena that are due to maximal
information being incomplete and those quantum phenomena that arise from
some other conceptual ingredient is likely to be very useful in the field of
quantum information theory, where there is currently a paucity of intuitions
regarding what sorts of information-processing tasks can be implemented more
successfully in a quantum universe than in a classical universe.

\section{Related work}

\label{relatedwork}

Kirkpatrick has considered a model of a system with two variables wherein it
is assumed that the measurement of one variable causes a randomization in
the value of the other \cite{Kirkpatrick}. This model exhibits
noncommutativity of measurements as well as an analogue of interference. The
manner in which these phenomena arise for a single elementary system in the
toy theory is no different. Kirkpatrick does not, however, consider the
possibility of transformations nor the case of multiple systems. \emph{\ }%
Our conclusions are also quite different. While Kirkpatrick emphasizes the
classicality of his model, we have tried to focus on the toy theory's
innovation relative to a classical theory, namely, that maximal information
is incomplete.

Hardy has introduced a toy theory very similar to the one described here
\cite{Hardydisentangling}. The elementary systems within his theory also
have four ontic states. Hardy postulates restrictions on the sorts of
measurements that are possible, and a disturbance upon measurement that
randomizes the ontic state among the possibilities consistent with the
measurement outcome. He also postulates that permutations of the ontic
states of a single system are possible transformations. As possible states
of knowledge he allows for any probability distribution over the ontic
states.

In its treatment of a single system, Hardy's toy theory is essentially the
same as the one presented here, except for the fact that the set of
epistemic states in his case are the convex hull of the ones we consider (an
octahedron on the Bloch sphere). Some of the disadvantages associated with
adopting this set of epistemic states were discussed in Sec.~\ref%
{phenomenathatdonotarise}. For multiple systems, the differences
between Hardy's theory and the one presented here are more
significant. Specifically, the set of measurements allowed in
Hardy's theory is larger than the set picked out by the knowledge
balance principle, and joint transformations on several systems
are not considered. As a result, Hardy's theory is less analogous
to quantum theory than the one presented here. Note however that
Hardy invented his theory for the purpose of demonstrating the
possibility of a local theory that exhibits teleportation, and for
this it is quite sufficient.

Smolin has constructed several toy models involving ``lockboxes'' \cite%
{Smolin}. The motivation for his work is to reproduce certain
information-theoretic phenomena which have been suggested as postulates for
quantum theory, specifically, no superluminal signalling, no broadcasting,
no bit commitment and key distribution. One of Smolin's models, involving
pairs of lockboxes, succeeds in this task. It assumes, however, that every
pair of lockboxes bears a unique label and this assumption has recently been
criticized as unphysical \cite{HalvorsonBub}.

There is an interesting connection between Smolin's theory and our own. By
abandoning the assumption of unique labels, and by formulating Smolin's
model in a different manner, one obtains a variant of the toy theory.\emph{\
}Suppose that every elementary system (a single lockbox in Smolin's
terminology) has two possible ontic states, and thus only a single yes/no
question that can be asked of it. Now assume that the answer to this
question is always unknown. Denoting the two ontic states by $1$ and $2,$ it
follows that the only valid epistemic state for a single system is $1\vee 2,$
and there are no non-trivial measurements. However, permuting $1$ and $2$
does not increase one's knowledge and is therefore an allowed
transformation. For a pair of such systems, there are four possible ontic
states, and thus two yes/no questions that can be asked of the pair. Assume
that one can know the answer to \emph{one} of these questions. Recalling
that the marginals on the individual systems must be $1\vee 2,$ it follows
that the only valid epistemic states for the pair are ($1\cdot 1)\vee
(2\cdot 2)$ and ($1\cdot 2)\vee (1\cdot 2).$ This corresponds to knowing
that the ontic states of the two systems are the same, or knowing that they
are different. The only possible measurement on the pair is the one that
determines whether the ontic states of the two are the same or different.
Note that a permutation on either system takes one epistemic state to the
other. It is this last feature which is critical for establishing the
impossibility of bit commitment.

Because of the assumption of unique labels for pairs, Smolin's model did not
incorporate the possibility of correlation between more than two systems. By
the lights of our reformulation however, it is natural to assume that for
three systems (and three yes/no questions) one could still only have the
answer to a single question, while for four systems, one could have the
answer to two, and so forth. Although the resulting theory will not provide
as good an analogy to quantum theory as does our toy theory, it would be
interesting to explore the differences, since this is likely to shed light
on how much work is being done by the assumption of a \emph{balance} of
knowledge and ignorance and how much is being done by the assumption of
maximal knowledge being incomplete.

\strut The above models all resemble the toy theory insofar as
they are local noncontextual hidden variable theories. They do
not, however, share the foundational principle from which the toy
theory was derived. By contrast, Zeilinger has advocated an
approach to quantum theory which is operational, denying any
hidden ontic states, but which adopts a similar foundational
principle \cite{Zeilinger}. Zeilinger's principle is that $N$
elementary systems represent the truth values of $N$ propositions.
The propositions to which Zeilinger is referring are propositions
stating the outcomes of measurements on the system, rather than
propositions about the ontic state of the system. In particular,
these propositions concern the outcomes of measurements associated
with a set of mutually unbiased bases (Zeilinger calls these
``mutually complementary measurements''). Note that the structure
of the set of measurements in quantum theory is taken for granted
in this approach; the existence of a particular number of mutually
unbiased bases for an elementary system is assumed rather than
derived. Had one assumed that there was only a single measurement
for every elementary system, then Zeilinger's principle would be
consistent with knowing the truth values for all propositions
pertaining to a system and would therefore yield a classical
theory. In the toy theory, the ratio of the number of known
propositions to the total number of propositions which pertain to
a system is fixed by the assumption of a balance between knowledge
and ignorance.

Finally, Wootters has recently introduced a representation of the quantum
states of $N$ qubits as real functions on a discrete space of $4^{N}$
elements \cite{WoottersWignerfunctions}. This is a generalized Wigner
function representation of the quantum states. Since these functions can be
negative, they cannot be interpreted as epistemic states. Nonetheless, this
approach is likely to facilitate the comparison of quantum theory to the toy
theory.

\section{Conclusions}

\label{conclusions}

We have considered the consequences of a principle of equality between
knowledge and ignorance to the structure of the set of possible states of
knowledge. We have examined the manner in which such states of knowledge may
be decomposed into convex sums, decomposed into ``coherent'' sums,
transformed, inverted, updated, remotely ``steered", cloned, broadcast,
teleported, and so forth. In all of these respects we have found that they
resemble quantum states. This is strongly suggestive that quantum states
should be interpreted as states of incomplete knowledge.

The toy theory contains almost no \emph{physics. }The motional
degree of freedom was assumed classical, and there were no masses
or charges or forces or fields or Hamiltonians anywhere in the
theory. Although this is a shortcoming from the perspective of
obtaining an empirically adequate theory, it helps make the case
for the epistemic view. Specifically, it supports the idea that a
great number of quantum phenomena, and in particular all the
phenomena that the toy theory reproduces, have nothing to do with
physics, but rather concern only the manipulation of our
information about the world.~\footnote{This idea has also been
defended by Fuchs~\cite{Fuchssamizdat}} Since the spectra of atoms
are not reproduced in the toy theory, these might well be
indicative of some real physics, but no-cloning and quantum
teleportation, for instance, are probably not.

The following questions for future research suggest themselves:

\begin{itemize}
\item Can we derive the knowledge balance principle from a physical
principle governing the interactions between systems, treating observers as
physical systems?
\end{itemize}

Most scientific realists seek a universal physical theory, wherein
apparatuses and observers are physical systems like any other rather than
unanalyzed primitives that appear in the axioms of the theory. Thus, even if
one could derive quantum theory from a set of axioms that included a
principle of maximal information being incomplete, the question of whether
and how this principle could be justified by some \emph{physical} principle,
governing all systems, including observers, would be left open. It may be
useful to begin by attempting to answer this question in the context of the
toy theory, rather than in the context of quantum theory.

\begin{itemize}
\item What are the ontic states of which quantum states are states of
knowledge?
\end{itemize}

Within the context of the research program outlined here, this
question captures the central mystery of quantum theory.
Contextuality and nonlocality imply that there must be some
modification, relative to classical theories, of our conception of
reality if we are to interpret quantum states as states of
incomplete knowledge about this reality. Specifically, there
cannot be local systems with attributes that are measured in a
noncontextual way. Many who adopt an epistemic interpretation of
the quantum state abandon the notion that the knowledge
represented by the quantum state is knowledge of a pre-existing
reality. Rather, it is assumed that the quantum state can only
represent someone's knowledge about the outcomes of future
measurements, or, more generally, the outcomes of future
interventions into the world, for instance, whether or not there
will be an audible click in a certain detector~\cite{Fuchs}.
However, a proponent of the epistemic view is not forced to this
conclusion. Noncontextual hidden variables and the outcomes of
future interventions do not exhaust the possibilities for the
sample space over which states of knowledge could be defined. We
feel that the most promising avenue for the epistemic program is
to investigate these other possibilities.

\begin{itemize}
\item Is there a second principle that can capture the missing quantum
phenomena?
\end{itemize}

\strut \strut A principle stating that maximal knowledge is incomplete
knowledge is likely to serve as a foundational principle in a simple
axiomatization of quantum theory. This is the claim that we argue is made
plausible by the strength of the analogy between the toy theory and quantum
theory. Nonetheless, this principle is insufficient for deriving quantum
theory. It is intriguing to speculate that we are lacking just one
additional conceptual ingredient, just one extra principle about reality,
from which all the phenomena of quantum theory, including contextuality and
nonlocality, might be derived. To find a plausible candidate for a second
such principle, it may be useful to adopt a similar strategy to the one used
here to argue for the first principle: do not attempt to derive all of
quantum theory, but rather focus on the more modest goal of reproducing a
variety of quantum phenomena, even if only qualitatively and in the context
of some incomplete and unphysical theory. In particular, attempt to
reproduce those phenomena that the toy theory fails to reproduce. Armed with
a conceptual innovation that captures the essence of the missing quantum
phenomena, a path to quantum theory might suggest itself.

\begin{acknowledgements}
I am indebted to Joseph Emerson and Chris Fuchs for opening my eyes to the epistemic view. I am also thankful to Stephen Bartlett, Chris Fuchs, Lucien Hardy, Terry Rudolph, Peter Turner and Antony Valentini for many helpful
comments.
\end{acknowledgements}

\appendix

\section{Differences to restricted quantum}

\label{whytoyisnotrestriction}

The strong similarity of the toy theory to quantum theory might
lead one to believe that the epistemic states, measurements and
transformations that apply to $N$ elementary systems in the toy
theory are simply subsets of the states, measurements and
transformations that apply to $N$ qubits in quantum theory. This
is not the case however. First, there is the fact that the
coherent binary operations in the toy theory are not precisely
analogous to coherent superpositions in quantum theory, as
described in Sec.~\ref{epistemicstates1}. Second, there is the
fact that the set of transformations in the toy theory includes
permutations analogous to anti-unitary maps, which do not arise in
a restricted version of quantum theory. A third fact is that the
nature of the correlations for mutually unbiased measurements is
different in the two theories, as we now demonstrate.

Suppose a pair of qubits is described by one of the four Bell states $\left|
\Phi ^{+}\right\rangle ,\left| \Phi ^{-}\right\rangle ,\left| \Psi
^{+}\right\rangle ,$ or $\left| \Psi ^{-}\right\rangle ,$ and that one of
three mutually unbiased measurements are implemented on each qubit: $%
\{\left| 0\right\rangle ,\left| 1\right\rangle \}$ on each qubit, $\{\left|
+\right\rangle ,\left| -\right\rangle \}$ on each qubit, or $\{\left|
+i\right\rangle ,\left| -i\right\rangle \}$ on each qubit. For each state
and each possible measurement, one obtains either correlation between the
outcomes (the same outcome for each qubit), or anti-correlation (different
outcomes for the two qubits). The results are summarized in Table \ref%
{correlationsquantum}, where `C' denotes correlation and `A' denotes
anti-correlation. One notes that in all cases there are an odd number of
anti-correlations.

\begin{table}[h]
\begin{tabular}{|c|c|c|c|}
\hline
& $\{\left| 0\right\rangle ,\left| 1\right\rangle \}$ & $\{\left|
+\right\rangle ,\left| -\right\rangle \}$ & $\{\left| +i\right\rangle
,\left| -i\right\rangle \}$ \\ \hline
$\left| \Phi ^{+}\right\rangle $ & C & C & A \\ \hline
$\left| \Phi ^{-}\right\rangle $ & C & A & C \\ \hline
$\left| \Psi ^{+}\right\rangle $ & A & C & C \\ \hline
$\left| \Psi ^{-}\right\rangle $ & A & A & A \\ \hline
\end{tabular}%
\caption{Correlations (C) and anti-correlations (A) for different
measurements on each qubit of a pair prepared in one of the Bell
states.} \label{correlationsquantum}
\end{table}

We can consider the analogous experiment in the toy theory. A pair
of elementary systems are described by one of four pure correlated
epistemic states (heading the rows of Table
\ref{correlationstoy}), and one of three mutually
unbiased measurements is implemented on each system (heading the columns of Table \ref%
{correlationstoy}). Again, one finds either correlated or anti-correlated
outcomes, however, the number of anti-correlations is always even. Since one
cannot achieve an even number of anti-correlations for any quantum state, it
is clear that the toy theory for $N$ elementary systems is not simply a
restricted version of quantum theory for $N$ qubits.

\begin{table}[h]
\begin{tabular}{|c|c|c|c|}
\hline
& $%
\begin{array}{l}
\includegraphics[width=15mm]{12vs34.eps}%
\end{array}
$ & $%
\begin{array}{l}
\includegraphics[width=15mm]{13vs24.eps}%
\end{array}
$ & $%
\begin{array}{l}
\includegraphics[width=15mm]{23vs14.eps}%
\end{array}
$ \\ \hline
$%
\begin{array}{l}
\includegraphics[width=15mm]{2bitcorrelated1.eps}%
\end{array}
$ & C & C & C \\ \hline
$%
\begin{array}{l}
\includegraphics[width=15mm]{2bitcorrelated8.eps}%
\end{array}
$ & C & A & A \\ \hline
$%
\begin{array}{l}
\includegraphics[width=15mm]{2bitcorrelated7.eps}%
\end{array}
$ & A & C & A \\ \hline
$%
\begin{array}{l}
\includegraphics[width=15mm]{2bitcorrelated9.eps}%
\end{array}
$ & A & A & C \\ \hline
\end{tabular}%
\caption{Correlations (C) and anti-correlations (A) for mutually unbiased
measurements given correlated epistemic states analogous to the Bell states.}
\label{correlationstoy}
\end{table}

\section{Relevance to quantum axiomatics}

\label{relevancetoaxiomatics}

There has recently been much interest in the possibility of deriving some or
all of the quantum formalism from information-theoretic axioms. Fuchs has
popularized the question \cite{Fuchs}, and it has been addressed in many
recent articles \cite{Fuchs,Halvorson,CBH,Smolin,HalvorsonBub}. The toy
theory shows that many of the information-theoretic effects one finds in
quantum theory are not unique to the latter, and this has important
consequences for some proposed axiomatizations.

For instance, it is likely that in the toy theory key distribution \cite%
{BB84} is possible, as discussed in Sec.~\ref{otheranalogues}. Moreover, it
is likely that arbitrarily concealing and arbitrarily binding bit commitment
\cite{bitcommitment,SpekkensRudolphPRA} is not possible in the toy theory.
For instance, the fact that there is an analogue of remote steering, as
demonstrated in Sec.~\ref{steering}, shows that an analogue of the BB84
protocol for bit commitment \cite{BB84} will not be secure against Alice. We
have not \emph{rigorously} established the possibility of key distribution
and the impossibility of bit commitment, since to do so properly is a
non-trivial task. Nonetheless, our results strongly suggest the falsity of
an informal conjecture that the possibility of key distribution and the
impossibility of bit commitment together imply quantum theory \cite%
{FuchsBrassard}.

Recently, Clifton, Bub and Halvorson (CBH) \cite{CBH} have shown that within
the context of a C$^{*}$ algebraic framework, one can derive quantum theory
from three information-theoretic postulates: the impossibility of
superluminal information transfer through measurements, the impossibility of
broadcasting, and the impossibility of bit commitment.

As we have shown, broadcasting is impossible in the toy theory, and since
the theory is explicitly local, there is clearly no superluminal information
transfer through measurement. Moreover, as discussed above, it is very
likely that bit commitment is impossible in the toy theory. These facts do
not, however, challenge the CBH characterization theorem since the toy
theory does not fall within the C$^{*}$ algebraic framework. For instance,
convex combination is only a partial binary operation within the toy theory
and is not defined for arbitrary probability distributions, features that
are required within the C$^{*}$ algebraic framework \cite{Halvorson}. Two
possibilities suggest themselves: either the assumption of a C$^{*}$
algebraic framework rules out physically reasonable theories, or a closer
examination of those features of the toy theory which cause it to fall
outside this framework will show that it is not physically reasonable after
all. Similar conclusions can be drawn form the work of Smolin \cite{Smolin}.

Although the toy theory might ultimately be a set-back for the CBH approach
insofar as it leads one to question the innocence of the assumption of a C$%
^{*}$ algebraic framework, the fact that it is derived from a simple
information-theoretic principle, the knowledge balance principle, and the
fact that it is so close in spirit to quantum theory suggests that the
prospects for an axiomatization of quantum theory that is predominantly
information-theoretic are actually quite good.

\end{document}